\documentclass[a4paper,11pt]{article}
\usepackage[utf8]{inputenc}
\usepackage[english]{babel}
\usepackage{amsmath, amssymb,graphicx}
\usepackage{braket}
\usepackage{yfonts}
\usepackage{caption}
\usepackage{subcaption}
\usepackage{ulem}
\pdfoutput=1
\usepackage{jheppub}
\newcommand{\be}{\begin{equation}}
\newcommand{\ee}{\end{equation}}
\newcommand{\bea}{\begin{eqnarray}}
\newcommand{\eea}{\end{eqnarray}}
\newcommand{\nn}{\nonumber}

\newcommand{\fb}{\mathfrak{b}}
\newcommand{\fB}{\mathfrak{B}}

\newcommand{\dd}{{\mathrm{d}}}

\definecolor{darkraspberry}{rgb}{0.53, 0.15, 0.34}

\definecolor{darkblue}{rgb}{0., 0, 1}

\definecolor{dgreen}{rgb}{0.,0.6,0.}

\author[a]{Irina Ya. Aref'eva,}
\author[b,c]{Anastasia A. Golubtsova}
\author[d]{and Eric Gourgoulhon}

\affiliation[a]{Steklov Mathematical Institute, Russian Academy of Sciences,\\Gubkina str. 8, 119991, Moscow, Russia}
\affiliation[b]{Bogoliubov Laboratory of Theoretical Physics, JINR,\\141980  Dubna, Moscow region, Russia}
\affiliation[c]{Dubna State University,\\Universitetskaya str. 19, Dubna, 141980, Russia}
\affiliation[d]{\,Laboratoire Univers et Th\'{e}ories, Observatoire de Paris, CNRS,
Universit\'{e} PSL, Universit\'{e} de Paris,
5 place Jules Janssen, 92190 Meudon, France}

\emailAdd{arefeva@mi-ras.ru}
\emailAdd{golubtsova@theor.jinr.ru}
\emailAdd{eric.gourgoulhon@obspm.fr}

\title{Holographic drag force in 5d Kerr-AdS black hole}

\abstract{
We consider the 5d Kerr-AdS black hole as a gravity dual to rotating quark-gluon plasma.
In the holographic prescription we calculate the drag force acting on a heavy quark. According to the holographic approach a heavy quark can be considered through the string in the gravity dual. We study the dynamics of the string for the Kerr-AdS backgrounds with one non-zero rotational parameter and two non-zero rotational parameters that are equal in magnitude. For the case of one non-zero rotational parameter we find good agreement with the prediction from the 4d case considered by  \href{http://arxiv.org/abs/1012.3800}{{\tt arXiv:1012.3800}}.}

\keywords{black holes,  Kerr-AdS metric,  holography and quark-gluon plasma, drag force}

\begin{document}
\maketitle
\flushbottom
\newpage
\section{Intoduction}
\label{sec:intro}
Starting with Landau's works \cite{Landau}, the hydrodynamic description of the quark-gluon plasma (QGP) is the main theoretical method for linking  the QGP models with experimental results. Experimental data, as a rule, are distributions by energy, momenta, spin polarization,  and types of  particles, etc. Parameters of the  QGP fluid are primary important and are fitted to describe experimental data.
After several tests it is believed that  the QGP  fluid satisfies  the relativistic ideal hydrodynamics equations, in the simplest cases such as the Bjorken  \cite{Bjorken} or Gubser's ones \cite{Gubser}. Vorticity is an important characteristic of fluids \cite{Liang:2004ph,Becattini:2007sr,1602.06580,1603.06117,Pang:2016igs}.
Recent experiments by the STAR collaboration at the RHIC facility \cite{STARcoll,STARcoll2,STARcoll3,STARcoll4}
indicate that the QGP fluid produced in peripheral relativistic heavy-ion collisions is in fact a rotating fluid.
It was proposed to use the average polarization of $ \Lambda $ hyperons to experimentally estimate the fluid vorticity in heavy-ion collisions.
The global spin polarization of $\Lambda$/$\overline{\Lambda}$ hyperons  related with vorticity
  was observed at RHIC \cite{STARcoll,STARcoll2,STARcoll3,STARcoll4} (see \cite{Dey:2020ogs,Taya:2020sej,1606.03808} for more details).  The angular momentum $J$  is shown to be in the range of $10^3 \hbar-10^5 \hbar$ \cite{1602.06580,1603.06117}.  Rotation is also important for neutron stars \cite{1602.01081}.

On the other hand, a powerful method for the theoretical study of QGP is the holographic approach, which is closely related to the hydrodynamic method \cite{PSS,Baier:2007ix,Bhattacharyya:2008jc}. As was established by Bhattacharyya et al. \cite{BLLMS:2007,BLMMS:2008}, the rotating fluid models are associated with rotating AdS (Kerr-AdS) black holes.  A description of dynamics of fluid flows through the conformal Navier-Stokes equations with transport coefficients were holographically computed in the 5d Kerr-AdS background  and the dual field stress-energy tensor was calculated in \cite{BLLMS:2007,BLMMS:2008}. In the paper \cite{Bantilan:2018vjv} a solution of relativistic ideal hydrodynamics that describes rotation around two axes and parametrized by two parameters, the first one  is related with non-vanishing angular momentum for off-center collisions, while the second one parametrizes  inhomogeneities of incoming nuclei, is constructed and its relevance to experimental data is shown. Recently, the effects of rotation on the hydrodynamic quasinormal modes of spinning black hole were studied in \cite{Garbiso:2020puw}.

The holographic duality relates the properties of black holes in an AdS spacetime to the thermal properties of dual conformal field theories at strong coupling in a spacetime of smaller dimension. The five-dimensional rotating black holes with AdS asymptotics were discovered in \cite{HHT}. 5d Kerr-AdS metrics  due to $SO(4)$-symmetry  are characterized by two rotation parameters that are associated with the number of Casimirs for $ SO (4) $ and are preserved independently.  Kerr-AdS black holes have  been studied  in the framework of holographic duality \cite{HHT}. {In this case CFT duals to Kerr-AdS black holes are CFTs in a 4d rotating Einstein universe\footnote{After \cite{HHT}, a number of interesting facts concerning the Kerr-AdS/CFT correspondence were found out  \cite{Berman}-\cite{AJ},  \cite{Gibbons:2004ai}.}.
Application of the Kerr-AdS/CFT duality to QGP  is justified at the high energies reached at LHC and RHIC, since in these regions the restoration of conformal invariance takes place.
Note, that in dual conformal theories operators are characterized by two quantum numbers, dimension and spin, the last of which has two independent parts if the theory is defined on
$\mathbb{R}\times \mathbb{S}^3$.

Within holographic applications Kerr-AdS black holes were discussed for both high-energy physics and condensed matter phenomena \cite{BLLMS:2007}-\cite{McInnes:2018hid},\cite{kn:sonner}-\cite{CLP:2018}.

 One of the well-tested holographic predictions for QGP are related with energy loss calculations \cite{GubserDF}-\cite{0906.1890}, in particular, jet suppression calculations \cite{HoloJet}. To model  QGP phase transition these calculations were made  for non-conformal
isotropic \cite{1603.01254}  and anisotropic  \cite{1202.3696,1312.7474,DF-aniz-IA} backgrounds,
some of them taking into account the motion of the QGP fluid \cite{LR,1505.07379}.
In the context of recent interest to a rotational vorticity and the efforts to get this quantity experimentally, it is  of interest to perform a holographic calculation of energy loss in a rotating fluid. This is the main goal of this paper.

For our analysis we follow up the holographic prescription  \cite{GubserDF,HKKKY},
where the motion of a heavy quark in QGP is dual to an open string with an endpoint at the boundary of the  background under consideration. In our case the  string is stretched down from the Kerr-AdS space boundary to the black hole horizon.  The dynamics of the string is described by the Nambu-Goto action.
To solve the corresponding equations we generalize an approach proposed for the more simple
Kerr-${\rm AdS}_4$ case of \cite{NAS}. This approach  consists in expanding an ansatz for solutions to string equations of motion in series by a rotating parameter. We write string equations of motion in
Kerr-${\rm AdS}_5$, expand them on rotations parameters and solve the linearized version of these equations of motion. Then we  find  the components of conjugate momenta. We perform  calculations  both in Boyer-Lindquist (rotating-at-infinity frame) and AdS
(static-at-infinity frame) coordinate systems.
In this paper we consider explicitly two cases:  a) rotational parameters are non-zero and equal in absolute value; b) one of rotational parameters vanishes. The case of non-zero and non-equal rotational parameters will be the subject of separated considerations.
For the extremal black holes (with merging horizons) that are related to the case with equal rotational  parameters we observe  vanishing of the drag force.

The paper is organized as follows.  Section \ref{SU} consists of several subsections that contain the description of main tools using in this paper. In Section \ref{DFH} we briefly remind the main steps of the drag force calculations.  In Section \ref{hydro} we remind how the drag forces appear in hydrodynamic equations. In Section \ref{DF-WL} we present the relation between drag forces and special Wilson loops, and notice the difference of this connection between a spherical symmetric case considered here and a planar AdS case.  In Section \ref{Kerr5} we describe thermodynamics of the 5d Kerr-AdS black holes. In Section \ref{HDF} we find the drag force  for a fixed quark studying string dynamics in a 5d Kerr-AdS black hole with one rotational parameter.
 In Section \ref {HDF1r} we find solutions to linearized string equations of motion.   Assuming that the rotational parameter is small we find the conjugate momenta and the leading term for the drag force. Then in Section \ref{HEL1r} we calculate corrections to the thermal quark mass at rest. In Section \ref{HDF2r} we discuss the case with two rotational parameters that are equal in magnitude. We perform computations of the drag force in Boyer-Lindquist and global AdS coordinates in Sections \ref{HDF2rBL} and  \ref{GC}, respectively.  In Section \ref{Sum} we conclude with a summary of the presented results as well with outline of future directions mainly related with NICA, which requires the calculations in the deformed Kerr-AdS metric.  In the appendix, we keep supplementary relations, that are useful for the main calculations.

\section{Setup}\label{SU}
\subsection{Drag forces in QGP as resistance to string moving in holographic space}\label{DFH}

The drag force is a force acting opposite to the relative motion of the heavy quark moving with respect to a surrounding quark-gluon plasma.
Drag forces define the energy loss in QGP.  This is an important property and
there is a large literature dedicated to this subject  (see \cite{GubserDF}-\cite{0906.1890} and refs. therein).
Following the dictionary of the gauge/gravity duality the heavy quark is associated to an endpoint of a relativistic string suspended from the boundary of the Kerr-AdS background  into the interior \cite{GubserDF,HKKKY}. Here we briefly describe this approach for an arbitrary curved background. In the holographic approach to investigate quark dynamics one has to study the string motion described by the Nambu-Goto action
\be\label{NGA}
S_{NG} = - \frac{1}{2\pi \alpha'} \int d\sigma^{0}d\sigma^{1}\sqrt{|g|},
\ee
where $g = \det g_{\alpha \beta}$ is the determinant of the induced metric which is defined
through the 5d spacetime metric $G_{\mu\nu}$ (in Sect.~3 and 4 this corresponds to the 5d Kerr-AdS black hole metric)  by
\be\label{NGAIM}
 g_{\alpha\beta} = G_{\mu\nu}\partial_{\alpha}X^{\mu}\partial_{\beta}X^{\nu},
\ee
$X^{\mu}$ are the embedding functions of the string worldsheet in the spacetime, we also assume that $X^{\mu} = X^{\mu}(\sigma)$.
 The equations of motion have the form
\be\label{maineqC1}
\frac{1}{\sqrt{-g}} \partial_\alpha \Big(\sqrt{-g}\, G_{\mu\nu} \,
 \partial^{\alpha} X^\nu\Big)-\frac12 \partial _\mu G_{\rho\nu}\partial_\alpha X^\rho\, \partial^{\alpha}X^\nu=0.\ee
The  conserved currents are defined as variational derivatives on $\partial _\alpha X^\mu$
   \be\label{conjmom}
   \pi^\alpha _\mu \equiv -2\pi \alpha' \frac{\delta S}{\delta \partial_{\alpha} X^{\mu}} =\sqrt{-g}\,g^{\alpha\beta}G_{\mu\nu}(X) \,\partial_\beta X^\nu.
\ee
Note, that this current is related to the translational invariance.

In (\ref{NGA})-(\ref{NGAIM}) we define $\sigma^{\alpha}$ with $\alpha = 0,1$ as the string worldsheet coordinates. So for the conjugated momenta  $\pi^{\alpha}_{\mu}$  one can write
\be
\partial_{\alpha} \pi^{\alpha}_{\mu} = 0, \quad \partial_{\alpha}\pi^{\alpha}_{\mu} = \frac{\partial \pi^{\sigma^{0}}_{\mu}}{\partial \sigma^{0}} + \frac{\partial \pi^{\sigma^{1}}_{\mu}}{\partial \sigma^{1}}.
\ee
The corresponding charge reads
\bea
\int_{M}\partial_{\alpha}\pi^{\alpha}_{\mu}d\sigma^{2}&= &\int_{M}\left(\frac{\partial \pi^{\sigma^{0}}_{\mu}}{\partial \sigma^{0}} + \frac{\partial \pi^{\sigma^{1}}_{\mu}}{\partial \sigma^{1}}\right)d\sigma^{0}d\sigma^{1} \nonumber \\
&=&\int_{\partial_{M}} -\pi^{\sigma^{1}}_{\mu}d\sigma^{0}+\int_{\partial_{M}}\pi^{\sigma^{0}}_{\mu}d\sigma^{1} = \int_{\partial M} \vec{\pi}_{\mu}d\vec{\sigma},
\eea
where $M$ is the two-dimensional worldsheet manifold and $\partial M$ is its boundary.
The associated conserved charge is the total momentum in the $\mu$-direction
\be
p_{\mu} =\int d\Sigma_{\alpha}\pi^{\alpha}_{\mu},
\ee
where $\Sigma_{\alpha}$ is a cross-sectional surface on the worldsheet.
Taking into account $\Sigma_{\sigma^{1}} =d\sigma^{0}\sqrt{-g_{00}}\hat{n}_{\sigma^{1}}$ with
\be
\hat{n}^\alpha =\Big(-\cfrac{g_{10}}{\sqrt{-g\,g_{11}}}, \displaystyle{\sqrt{\cfrac{g_{11}}{-g}}}\Big),
\ee
 the time-independent  force on the string is
\be\label{dragfp}
\frac{\partial p_{\mu}}{\partial \sigma^{0}} = - \frac{1}{2\pi\alpha'}\pi^{\sigma^{1}}_{\mu}.
\ee
Therefore, to find the components of the drag force (\ref{dragfp}) we need to calculate the conjugate momenta of the string (\ref{conjmom}).

The total energy of the string is given by
\be\label{enstrdef}
E  =  -\frac{1}{2\pi\alpha'} \int dr \pi^{\sigma^{0}}_{t},
\ee
where the conjugate momentum $\pi^{\sigma^{0}}_{t}$ is calculated using (\ref{conjmom}). In what follows we simplify notations:  $\pi^{\sigma^{0}}_{t}=\pi^{0}_{t}$,  $\pi^{\sigma^{1}}_{t}=\pi^{r}_{t}$.

Drag forces in the holographic approach
 have been calculated for several isotropic models including models describing the quark confinement/deconfinement  and  chiral symmetry  breaking  phase transitions
 \cite{GubserDF,HKKKY,0605235,0906.1890,1505.07379,1507.06556}.
Drag forces in anisotropic QGP using the suspended  string are calculated in \cite{1202.3696,1202.4436,1312.7474,1410.7040,1412.8433,1711.08943,DF-aniz-IA} including plasma with magnetic field \cite{1605.06061,1606.01598,1901.09304,Zhang:2019jfq,2012.05758}.

\subsection{Drag forces via hydrodynamic equations}\label{hydro}

Here we show that the drag force can be considered as a relativistic pressure gradient force in the fluid. Originally this was demonstrated in \cite{NAS} for the 3d case.
Here we consider a rotating fluid in a 4-dimensional spacetime $\mathbb{R}\times \mathbb{S}^{3}$
that has the metric
\be\label{metricsp}
ds^{2} =-dT^{2} + \frac{1}{\ell^2}\left(d\Theta^{2} + \sin^{2}\Theta d\Phi^{2} + \cos^{2}\Theta d\Psi^{2}\right).
\ee
The metric has the following non-zero Christoffel symbols
\be\label{Chrsymb}
\Gamma^{2}_{\,\,\,33} =-\frac{1}{2}\sin(2\Theta) , \quad \Gamma^{2}_{\,\,\,44} = \frac{1}{2}\sin(2\Theta),\quad \Gamma^{3}_{\,\,\,32} =\cot{\Theta}, \quad \Gamma^{4}_{\,\,\,42} = - \tan{\Theta}.
\ee
The stress-energy tensor in hydrostationary equilibrium reads
\be
T^{AB} = (\rho+ P)u^A u^B + Pg^{AB},
\ee
where $g^{AB}$ are components of the metric (\ref{metricsp}), $A,B =1,\ldots,4$, $\rho$ is the density, $P$ is the pressure and $u^{A}$ is the velocity field:
\be\label{velfcontr}
u^{A} = \frac{1}{\sqrt{1-\ell^{-2}\Omega^2_{\phi}\sin^2\Theta -\ell^{-2}\Omega^2_{\psi}\cos^{2}\Theta}}(1,0,\Omega_{\phi},\Omega_{\psi}),
\ee
so $u^{A}u_{A} = -1$, $\Omega_{\phi} = a \ell^2$, $\Omega_{\psi} = b\ell^2$.

Correspondingly, taking into account $u_{A} = g_{AB}u^{B}$ we also have
\be\label{velfco}
 u_{A} = \frac{1}{\sqrt{1-\ell^{-2}\Omega^2_{\phi}\sin^2\Theta -\ell^{-2}\Omega^2_{\psi}\cos^{2}\Theta}}(-1,0,\ell^{-2}\Omega_{\phi}\sin^{2}\Theta,\ell^{-2}\Omega_{\psi}\cos^{2}\Theta).
\ee

Let's put $\ell^{-2}\Omega^2_{\Phi}\sin^2\Theta +\ell^{-2}\Omega^2_{\Psi}\cos^{2}\Theta = v^{2}$,  $(1- v^{2})^{-1/2} = \gamma$ and following \cite{BLLMS:2007} we get
\begin{equation}
T^{AB} =\gamma^{2} \left(
\begin{array}{cccc}
\rho+Pv^{2}& 0 &(\rho+P)\Omega_{\phi} & (\rho+P)\Omega_{\psi}\\
0  & P\ell^{2}\gamma^{-2} & 0 & 0 \\
(\rho+P)\Omega_{\phi}& 0 &  \rho\Omega^{2}_{\phi}+ P(\ell^{2}\csc^{2}\Theta - \Omega^{2}_{\psi}\cot^{2}\Theta) &   (\rho+P)\Omega_{\phi}\Omega_{\psi}\\
 (\rho+P)\Omega_{\psi} & 0 & (\rho+P)\Omega_{\phi}\Omega_{\psi}   &  \rho\Omega^{2}_{\psi}+ P(\ell^2\sec^{2}\Theta-\Omega_{\phi}^{2} \tan^{2}\Theta)
\end{array}
\right),
\end{equation}
the same as  eq.(30) from \cite{BLLMS:2007}. Taking $\rho = 3P$ and $3P =3 \tau^4\gamma^4$ we obtain
\begin{equation}
T^{AB} =\gamma^{6}\tau^{4} \left(
\begin{array}{cccc}
(3+v^{2})& 0 & 4\Omega_{\phi} & 4\Omega_{\psi}\\
0  & \ell^{2}(1 - v^2) & 0 & 0 \\
4\Omega_{\phi}& 0 &  3\Omega^{2}_{\phi}+\ell^{2}\csc^{2}\Theta - \Omega^{2}_{\psi}\cot^{2}\Theta &   4\Omega_{\phi}\Omega_{\psi}\\
 4\Omega_{\psi} & 0 & 4\Omega_{\phi}\Omega_{\psi}   &  3\Omega^{2}_{\psi}+\ell^2\sec^{2}\Theta-\Omega_{\phi}^{2} \tan^{2}\Theta
\end{array}
\right),
\end{equation}
which coincides with eq. (34) of  \cite{BLLMS:2007} and eq.~(11) of \cite{Bantilan:2018vjv}.

The corresponding conservation law of the stress-energy tensor is \be\label{nablaTAB}  \nabla_A T^{AB} =0.\ee
One can rewrite the conservation law projecting (\ref{nablaTAB}) onto the direction  orthogonal to the velocity field
\be
\left(g_{BC} +u_{B}u_C\right)\nabla_{A}T^{AC} =0,
\ee
that can be rewritten as
\be \label{ener_mom_cons_ortho}
(g_{BC}+u_{B}u_C)\nabla^C P +(\rho+P)u^{C}\nabla_{C}u_{B}=0.
 \ee
 Owing to rotational symmetry, $\rho$ and $P$ depend on $\Theta$ only :
 \be \rho= \rho(\Theta),\quad P= P(\Theta),\ee
so that Eq.~(\ref{ener_mom_cons_ortho}) reduces to
 \be\label{derpressure}
\partial_{\Theta} P = (\rho +  P)u^{3}\Gamma^{3}_{\,\,\,3 2}u_{3} + (\rho +  P)u^{4}\Gamma^{4}_{\,\,\,4 2}u_{4}.
  \ee
Substituting (\ref{Chrsymb}),  (\ref{velfcontr}), (\ref{velfco}) into (\ref{derpressure}) we get
\be
\partial_{\Theta}P =  (\rho +  P)\frac{\cot{\Theta}\ell^{-2}\Omega^{2}_{\phi}\sin^{2}\Theta}{1-\ell^{-2}\Omega^2_{\phi}\sin^2\Theta -\ell^{-2}\Omega^2_{\psi}\cos^{2}\Theta} - (\rho +  P)\frac{\tan{\Theta}\ell^{-2}\Omega^{2}_{\psi}\cos^{2}\Theta}{1-\ell^{-2}\Omega^2_{\phi}\sin^2\Theta -\ell^{-2}\Omega^2_{\psi}\cos^{2}\Theta},
\ee
which can be recast as
 \be\label{pressure}
\partial_{\Theta}P =  (\rho +  P)\frac{(\Omega^{2}_{\phi} - \Omega^{2}_{\psi})\sin(2\Theta)}{2\ell^2(1-\ell^{-2}\Omega^2_{\phi}\sin^2\Theta -\ell^{-2}\Omega^2_{\psi}\cos^{2}\Theta)}.
\ee
In particular, considering the contribution from $\Omega_{\phi}$ only, we get
\be\label{hydromain}
\partial_{\Theta}P =  (\rho +  P)\frac{\Omega^{2}_{\phi}\sin(2\Theta)}{2\ell^2(1-\ell^{-2}\Omega^2_{\phi}\sin^2\Theta)},
\ee
which is in agreement with our result for the drag force from a 5d Kerr-AdS black hole with one non-zero rotational parameter and the prediction from \cite{NAS}.
If $\Omega_{\phi} = \Omega_{\psi}$ we get
\be
\partial_{\Theta}P= 0.
\ee
Below we will show that this matches with the holographic calculations for the case with two equal rotational parameters.

Now taking into account
\be
\rho + P =sT.
\ee
the entropy density can be written as
\be
s = \left(\frac{\pi T}{\ell}\right)^{3}\frac{1}{4G_{5}(1-a^2)},\quad s\sim  \left(\frac{\pi T}{\ell}\right)^{3}\frac{1}{4G_{5}}
\ee
where the Newton's constant equals $G_{5} = \frac{\pi}{2N^2_{c}\ell^3}$. A relativistic pressure gradient force is given by
\be
\frac{d p_{\theta}}{dt}= -3m_{\rm rest}\frac{\partial_{\theta}P}{sT}.
\ee
So we can write
\be
\frac{d p_{\theta}}{dt}= -3m_{\rm rest}\frac{2}{\pi^{2}T^{4}N^{2}_{c}}\partial_{\theta}P,
\ee
which is in good agreement with the calculation for the drag force considering a curved string in the 5d Kerr-AdS with one non-zero rotational parameter presented in the next section. We note that $m_{\rm rest}$ is related with of a cut-off  at the point $r_{m}$  (where $r_{m}$ is a location of a quark).

\subsection{Drag forces and special Wilson loops}\label{DF-WL}
It has been  noticed \cite{0606049,1707.05045} that holographic calculations of the drag forces for an isotropic matter are directly related with calculations of spatial Wilson loops.   This relation is also inherited  for non-isotropic case including full anisotropic case \cite{DF-aniz-IA,2012.05758}.

Holographic consideration of drag forces in spherically symmetric backgrounds  brings with it a new feature  that we would like to note before we are going to special calculations. First of all, we note that, by analogy with the case of a flat boundary, it is natural to expect the connection between the technique of suspended strings and the calculation of the spatial Wilson loops.
In spherical symmetric backgrounds we deal with  boundaries that have ${\mathbb R}^1\times \mathbb{S}^3$ topology. In these cases it is more natural to deal not with rectangular special Wilson loops, but with circular ones. Circular Wilson loops have been considered in holographic approach \cite{0512150} and refs. therein, however there are obvious obstacles to deal with these objects on the lattice.

In Fig.~\ref{Fig:RS} we present the geometry of a suspended string in the ${\mathbb R}^1\times \mathbb{S}^2$ case, with $\mathbb{S}^2$ parametrized by two spherical angles
$(\theta,\phi)$
(in addition, it can be assumed that there is a third angle $\psi$, used for the parameterization of $\mathbb{S}^3$ but it is suppressed in Fig.~\ref{Fig:RS}).

\begin{figure}[h!]
  \centering
 \qquad\qquad\qquad \qquad\includegraphics[scale=0.3]{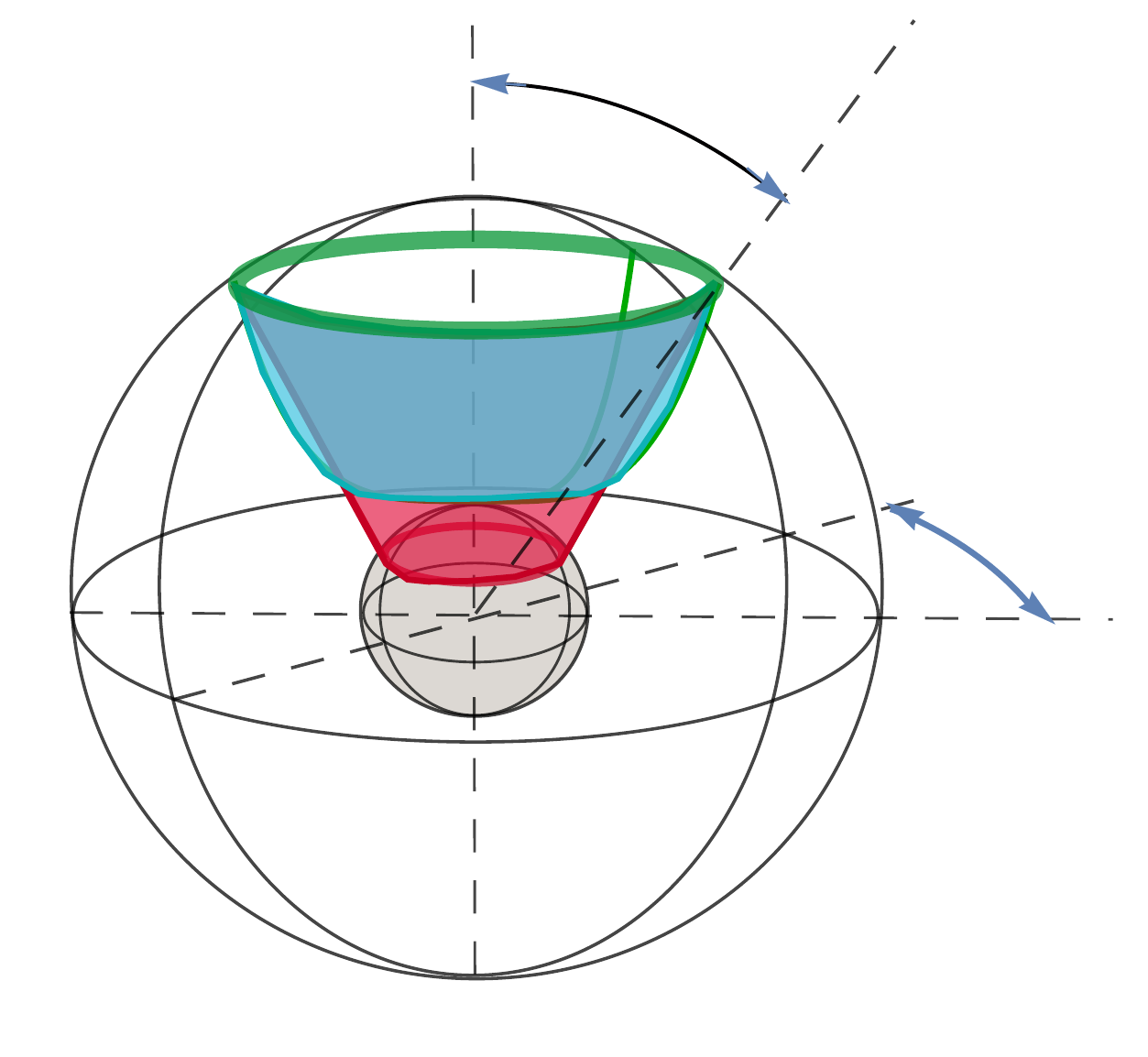}
   \begin{picture}(50,150)\put(-28,80){$\phi$}
\put(-90,65){$r_h$}
\put(-80,150){$\theta$}
  \end{picture} \\A\\  \includegraphics[scale=0.3]{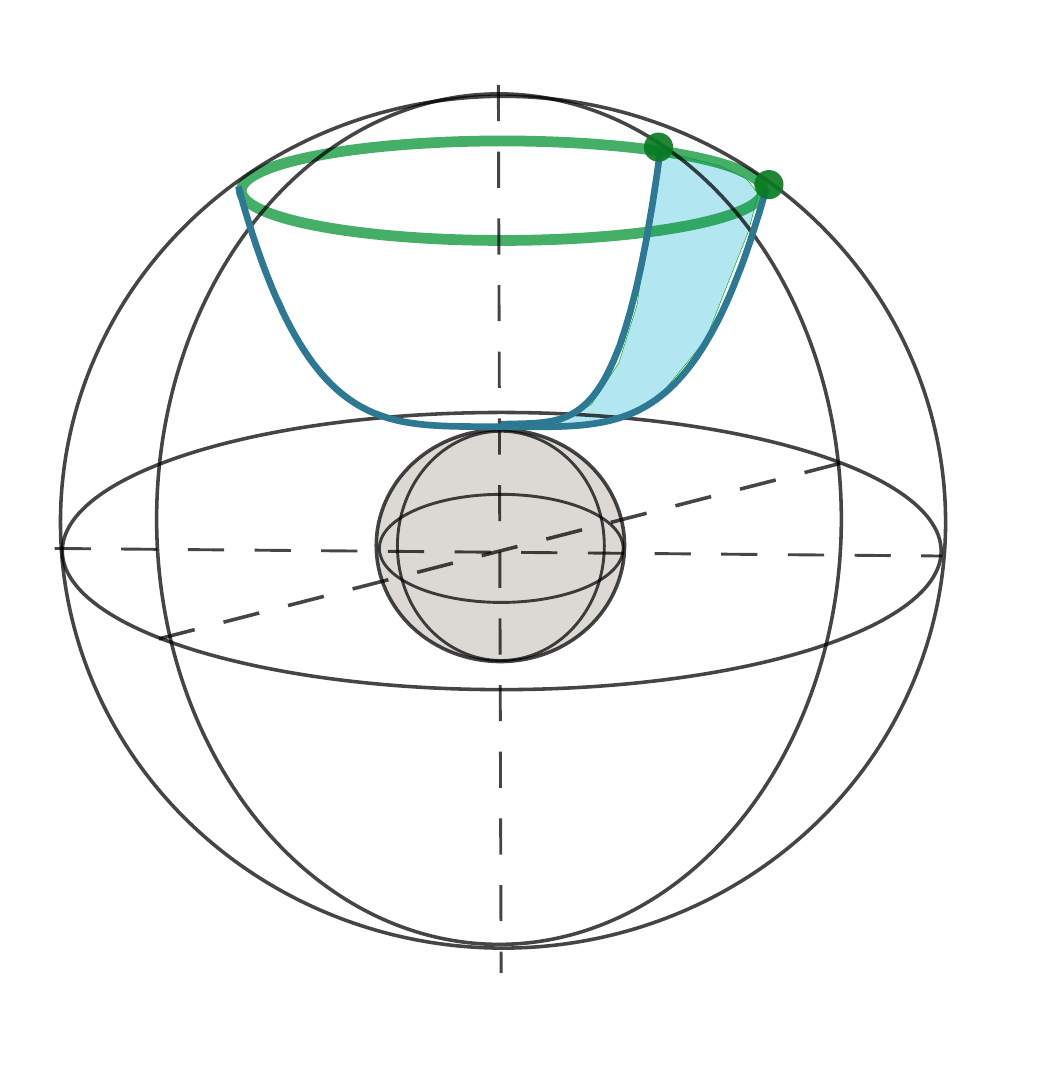}  \qquad\qquad\includegraphics[scale=0.3]{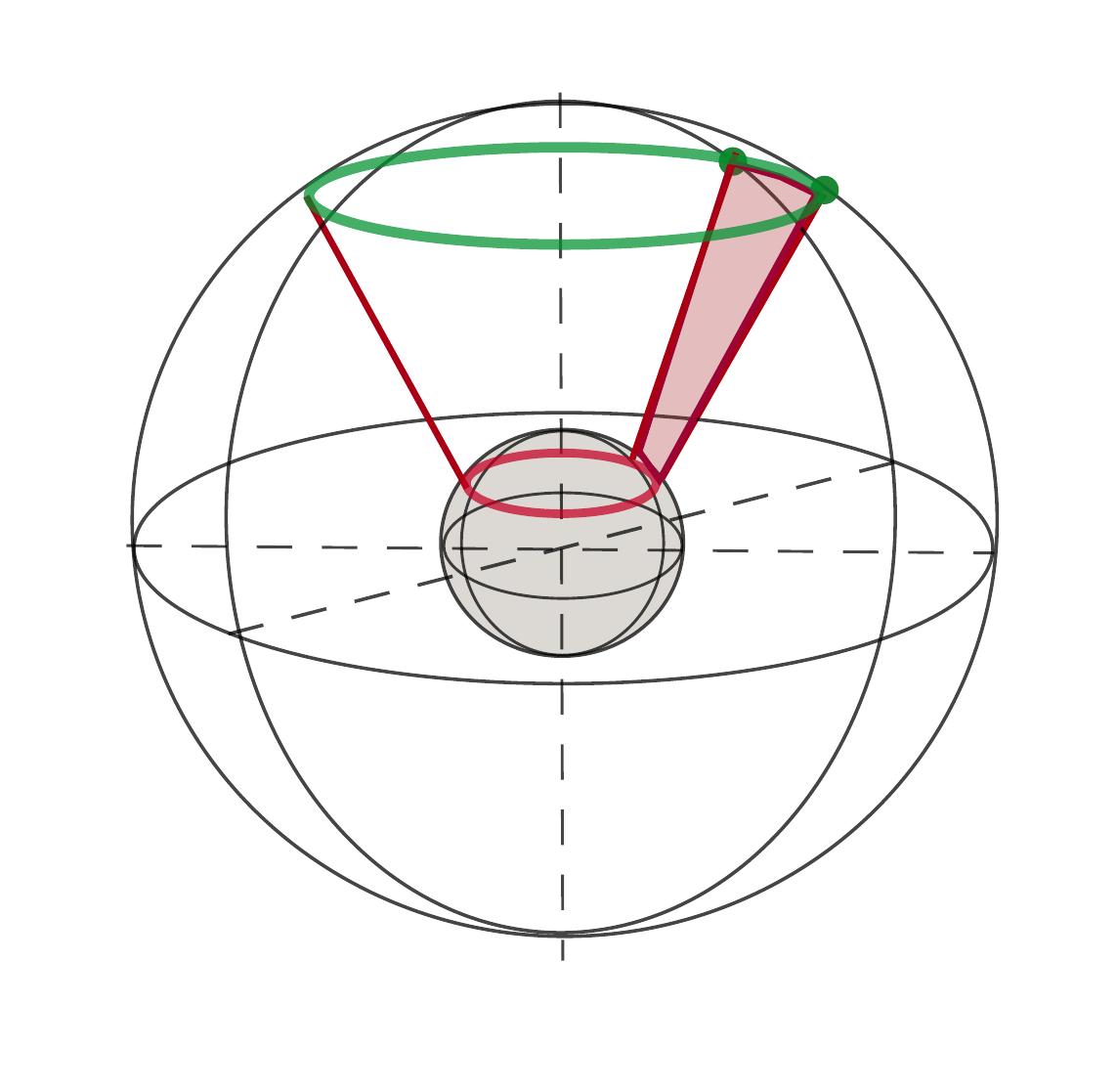}
    \\ B \hspace{70 mm} C
 \caption{Heavy quark moves on the boundary of ${\rm AdS}_5$ having topology ${\mathbb R}^1\times \mathbb{S}^3$ (${\mathbb R}^1$ and the third spatial coordinate, the angle $\psi$ are suppressed). It moves along a circle with fixed $\theta_0$ and varying $\phi$. A) We consider the Wilson loop  fixing $\theta=\theta_0$ and $0\leq \phi\leq 2 \pi$ (the small green circle).
 The blue surface is an extremal surface that ends  on trajectory of a heavy quark and touches the horizon of the black hole.
 The red surface is an analog of the disconnected holographic surface for planar BH calculations.
B) The string suspended from the boundary and touching the horizon.
  C) The straight string ending on the horizon. }
  \label{Fig:RS}
\end{figure}
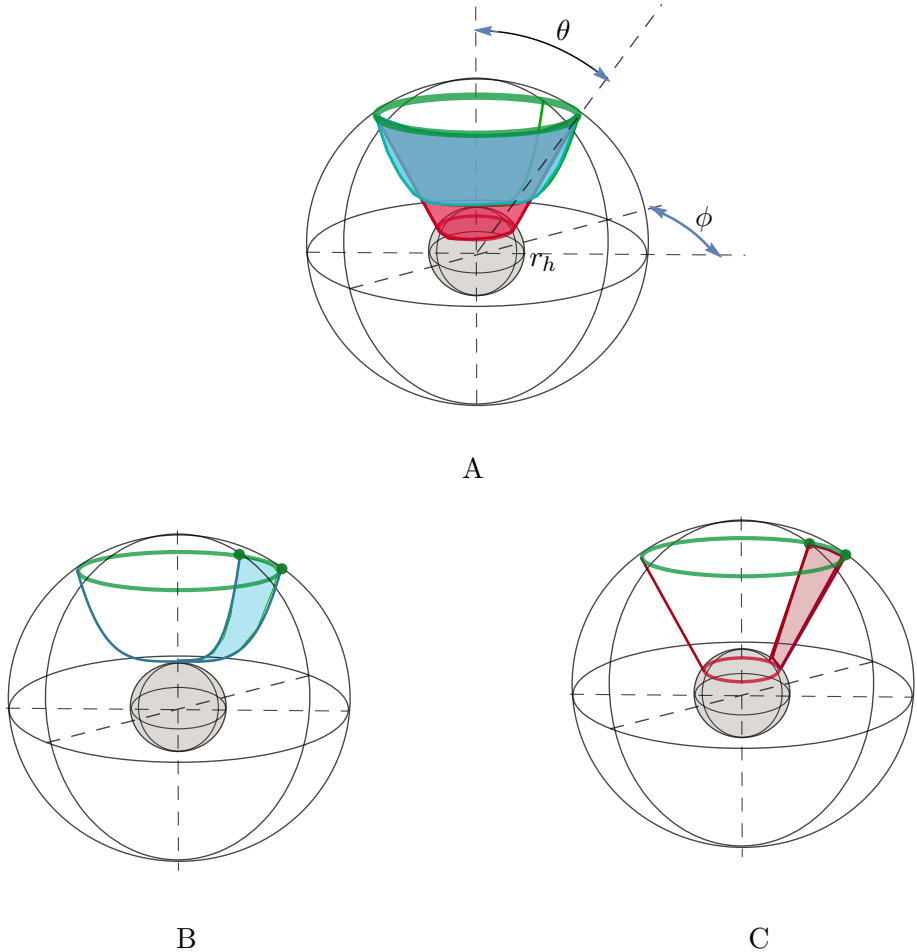

\newpage

\subsection{$D=5$ Kerr-AdS black holes}\label{Kerr5}

Our starting point is the following five-dimensional gravity model with a negative cosmological term:
\begin{eqnarray}\label{1}
S = - \frac{1}{16 \pi G_{5}}\int \dd^{5}x \sqrt{|G|}(R_{g} +12 \ell^{2}),
\end{eqnarray}
where $G_{5}$ is the five-dimensional Newton constant, $G = \det G_{\mu\nu}$, $G_{\mu\nu}$ being the spacetime metric,  and the cosmological constant is $\Lambda = - 6 \ell^{2}$.
The Einstein equations following from (\ref{1}) are given by
\be\label{EinEq}
R_{\mu\nu} = -4\ell^{2}G_{\mu\nu}.
\ee
Rotating black holes with an AdS aymptotics solve eqs. (\ref{EinEq}).
It is known that a rotating black hole in five dimensions is characterized by the mass and two angular parameters related to Casimir invariants of $SO(4)$.
The generic five-dimensional Kerr-AdS metric with two non-zero rotational parameters in the Boyer-Lindquist coordinates is \cite{HHT}
\bea\label{1.1GF}
ds^{2}& =& - \frac{\Delta_r}{\rho^{2}}\left(dt - \frac{a\sin^{2}\theta}{\Xi_{a}}d\phi - \frac{b\cos^{2}\theta}{\Xi_{b}}d \psi \right)^{2} + \frac{\Delta_{\theta}\sin^{2}\theta}{\rho^{2}}\left(adt - \frac{(r^{2} +a^{2})}{\Xi_{a}}d\phi \right)^{2}\nonumber\\
&+& \frac{\Delta_{\theta}\cos^{2} \theta}{\rho^{2}}\left( bdt  - \frac{(r^{2} +b^{2})}{\Xi_{b}}d\psi \right)^{2} + \frac{\rho^{2}}{\Delta_r}dr^{2} + \frac{\rho^{2}}{\Delta_{\theta} }d\theta^{2} \nonumber\\
&+&\frac{(1 + r^{2}\ell^{2})}{r^{2}\rho^{2}}\left(ab dt -  \frac{b(r^{2} + a^{2 })\sin^{2}\theta}{\Xi_{a}} d\phi  -\frac{a(r^{2} + b^{2 })\cos^{2}\theta}{\Xi_{b}}d\psi \right)^{2},
\eea
where  $0\leq\phi,\psi\leq 2\pi$, $0\leq\theta\leq\pi/2$, and the parameter $M$ is associated with the mass,
$a$, $b$ are related to the angular momentum and we also have
\bea
\Delta_r &=& \frac{1}{r^{2}}(r^{2} + a^{2})(r^{2} + b^{2})(1+ r^{2}\ell^{2}) -2M,\nonumber\\
\Delta_{\theta} &=& (1- a^{2}\ell^{2}\cos^{2} \theta -b^{2} \ell^{2}\sin^{2}\theta), \\ \label{Xi.ab}
\rho^{2}& =& (r^{2} + a^{2}\cos^{2}\theta + b^{2}\sin^{2}\theta),\nonumber \\
\Xi_{a}& = &(1 - a^{2} \ell^{2}), \quad \Xi_{b} = (1-b^{2}\ell^{2}).\nonumber
\eea
We note that we use Hopf coordinates for the spherical part of the metric (\ref{1.1GF}).
The horizon position is defined as a largest root $r_{+}$ to the equation $\Delta_{r} = 0$.
The rotational parameters $a$ and $b$ are constrained such that $a^{2}, b^{2}\leq \ell^{-2}$
and the angular momenta \cite{Gibbons:2004ai} are given by
\be
J_{a} = \frac{\pi M a}{2 \Xi^{2}_{a}\Xi_{b}}, \quad J_{b} = \frac{\pi M b}{2 \Xi^{2}_{b}\Xi_{a}}.
\ee
The Hawking temperature is defined as
\be\label{HawkT}
T_{H}= \frac{1}{2\pi}\left(r_{+}(1+ r^{2}_{+}\ell^{2})\Bigl(\frac{1}{r^{2}_{+} + a^{2}} + \frac{1}{r^{2}_{+} + b^{2}}\Bigr) - \frac{1}{r_{+}}\right).
\ee
The dependence of  $r_+$ on parameters $a$
 and $b$ is illustrated in Fig.\ref{Fig:roots}.

\begin{figure}[h!]
  \centering
\includegraphics[scale=0.225]{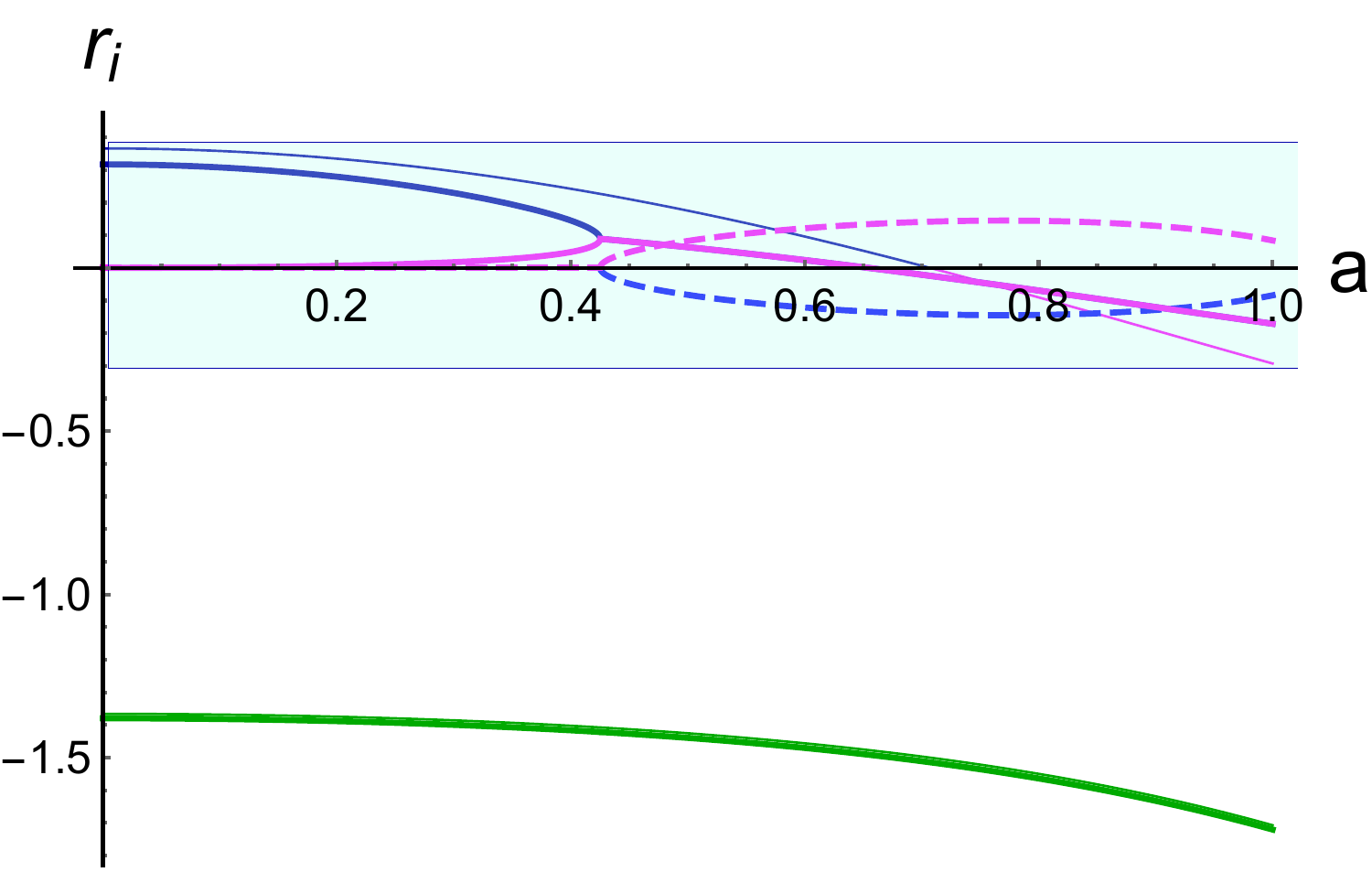}
\includegraphics[scale=0.25]{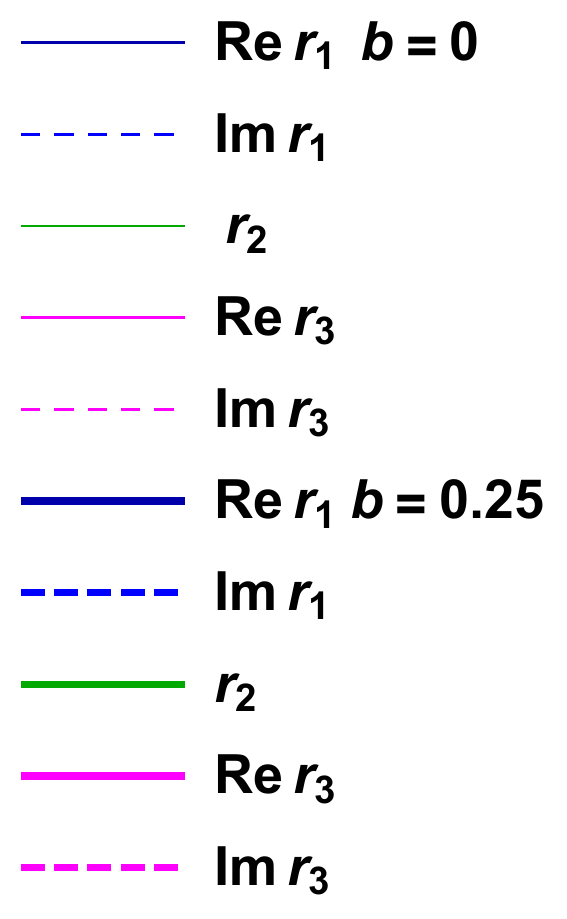}
 \includegraphics[scale=0.225]{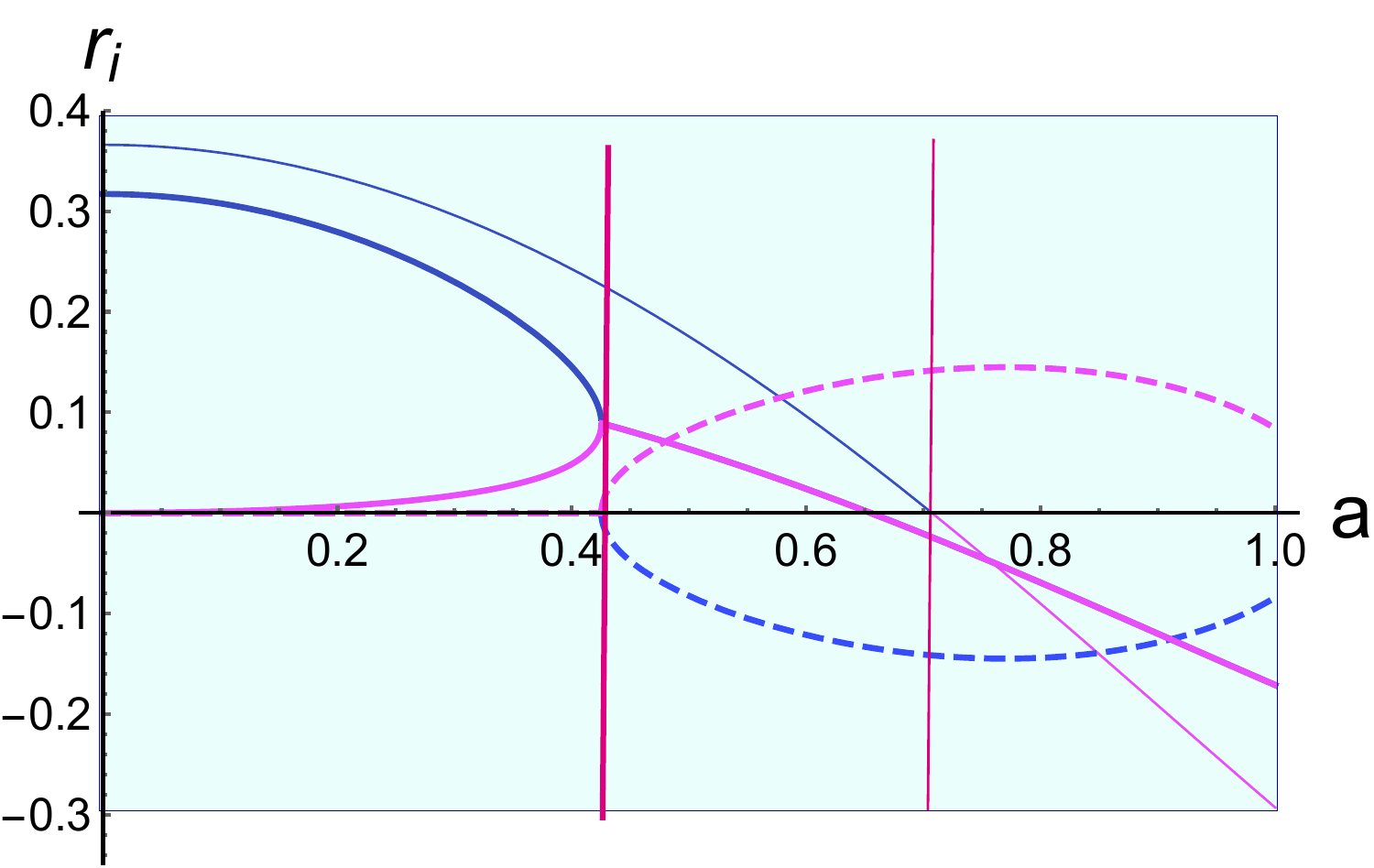}
 \\ A \hspace{70 mm} B\\
$\,$\\
\includegraphics[scale=0.25]{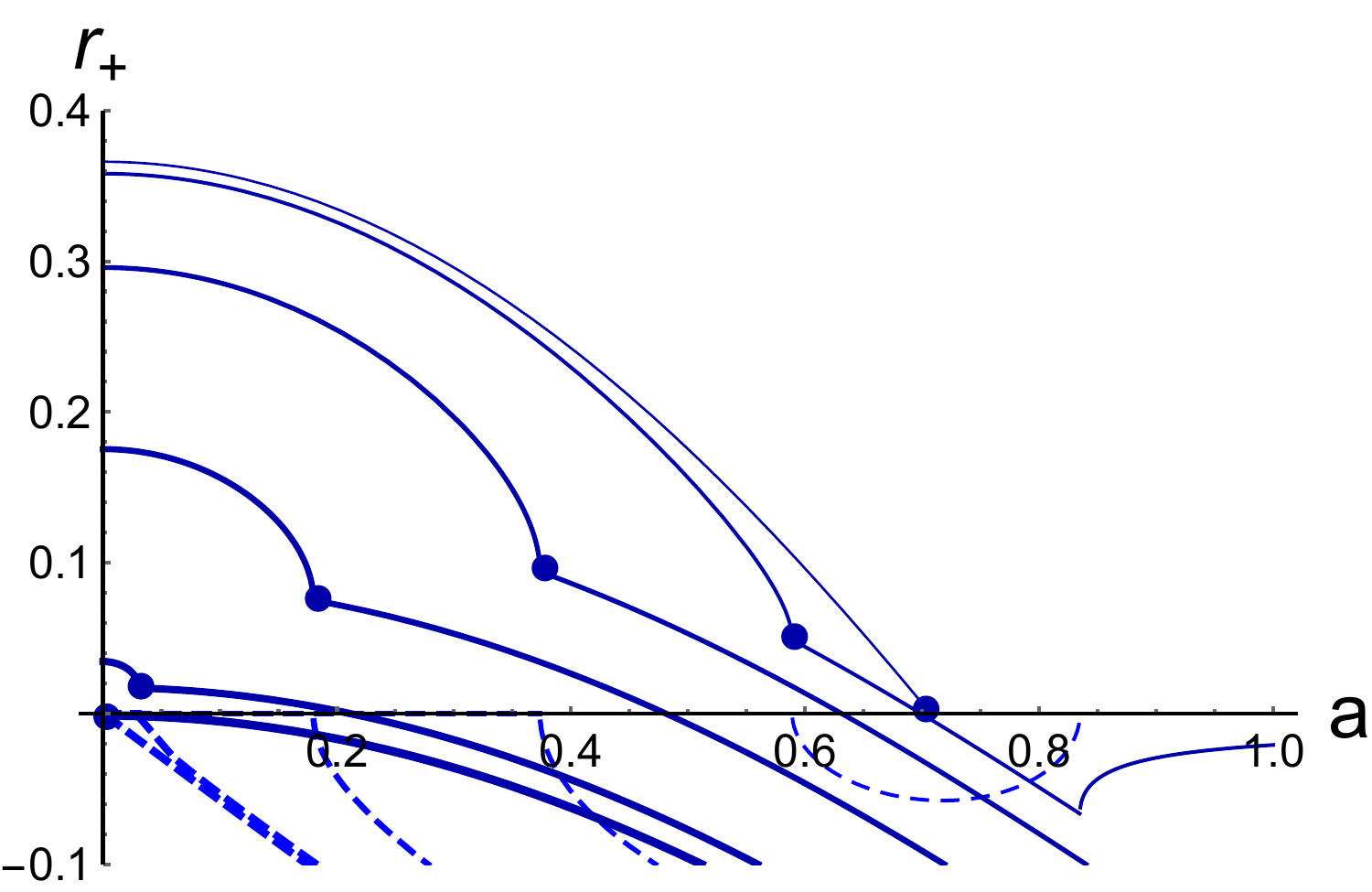}
\includegraphics[scale=0.25]{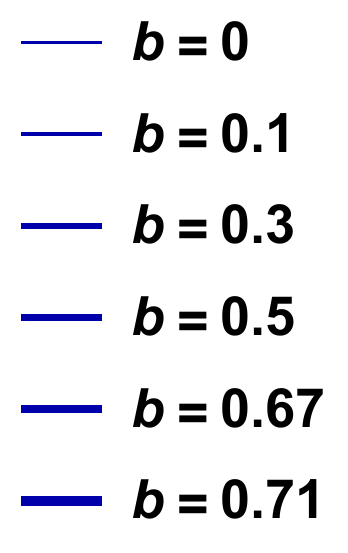} \qquad\includegraphics[scale=0.15]{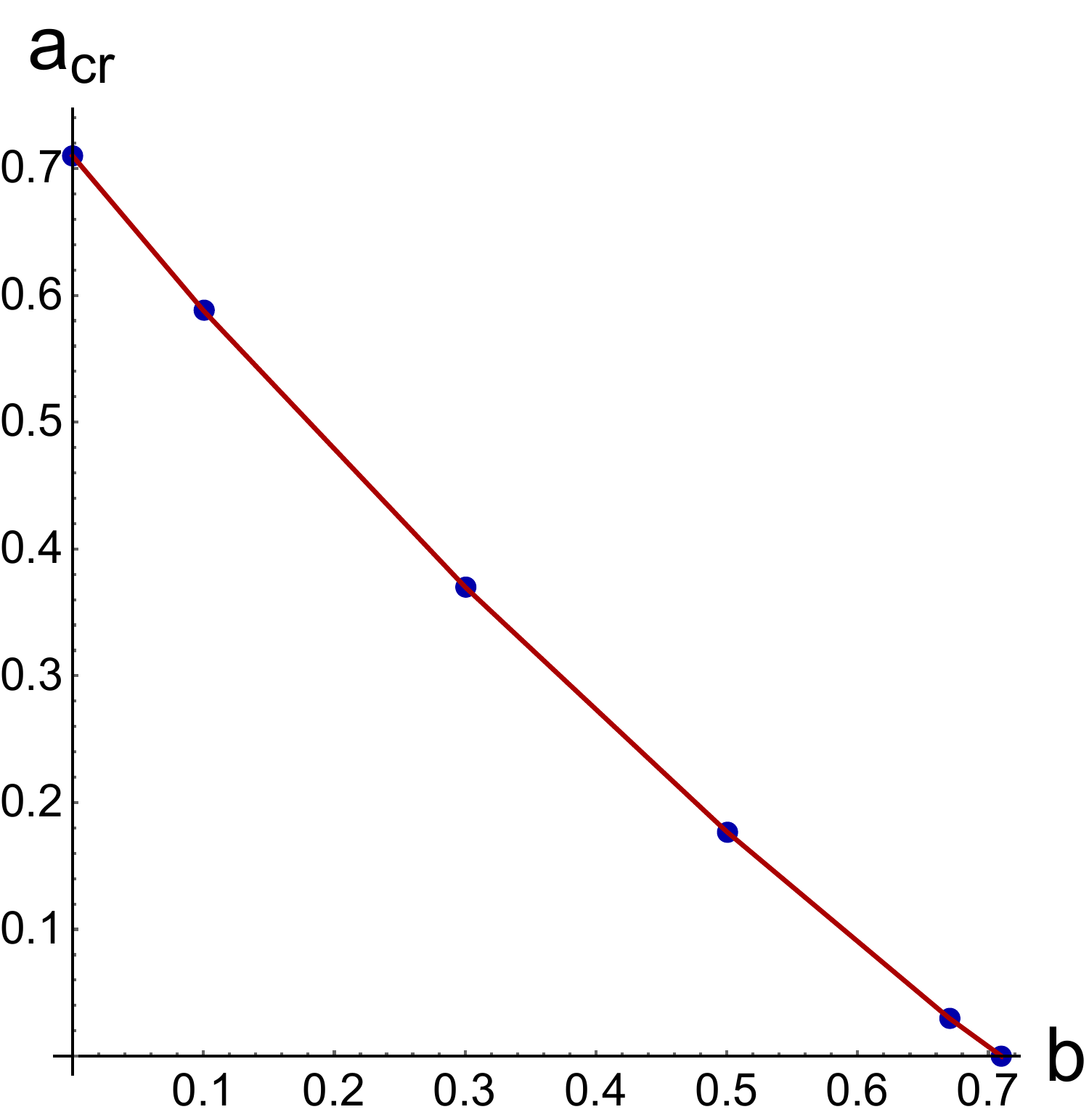}\\
C\hspace{70 mm}D
 \caption{A) 3 roots of equation $\Delta(r)=0$ vs $a$ for $b=0$ (thin lines) and $b= 0.2$ (thick lines).
  Dashed line shows the imaginary part of the first root $r_+$. B) First two roots  for the same values of $b=0,\,0.2$. C) The first root vs $a$ for variety of $b$. We see that only for $a<a_{cr}(b)$ the root $r_1$ is positive, $a_{cr}(0)=1/\sqrt{2}$. D) The dependence of critical points of $a_{cr}$ on  $b$. Here $M$ is taken to be $0.25$ and $\ell=1$.
  }
  \label{Fig:roots}
\end{figure}

\begin{figure}[h!]
  \centering
\includegraphics[scale=0.25]{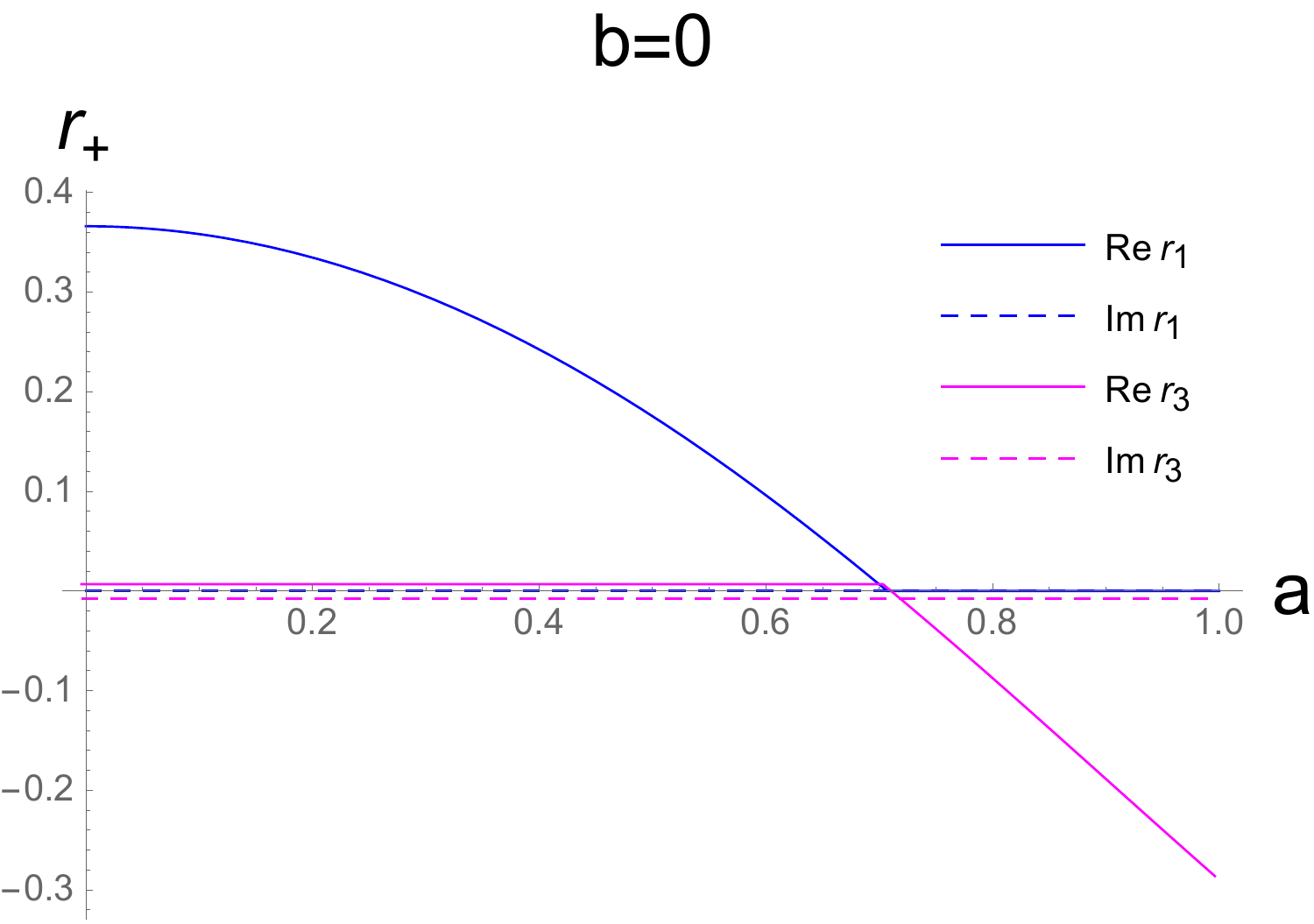}
\includegraphics[scale=0.25]{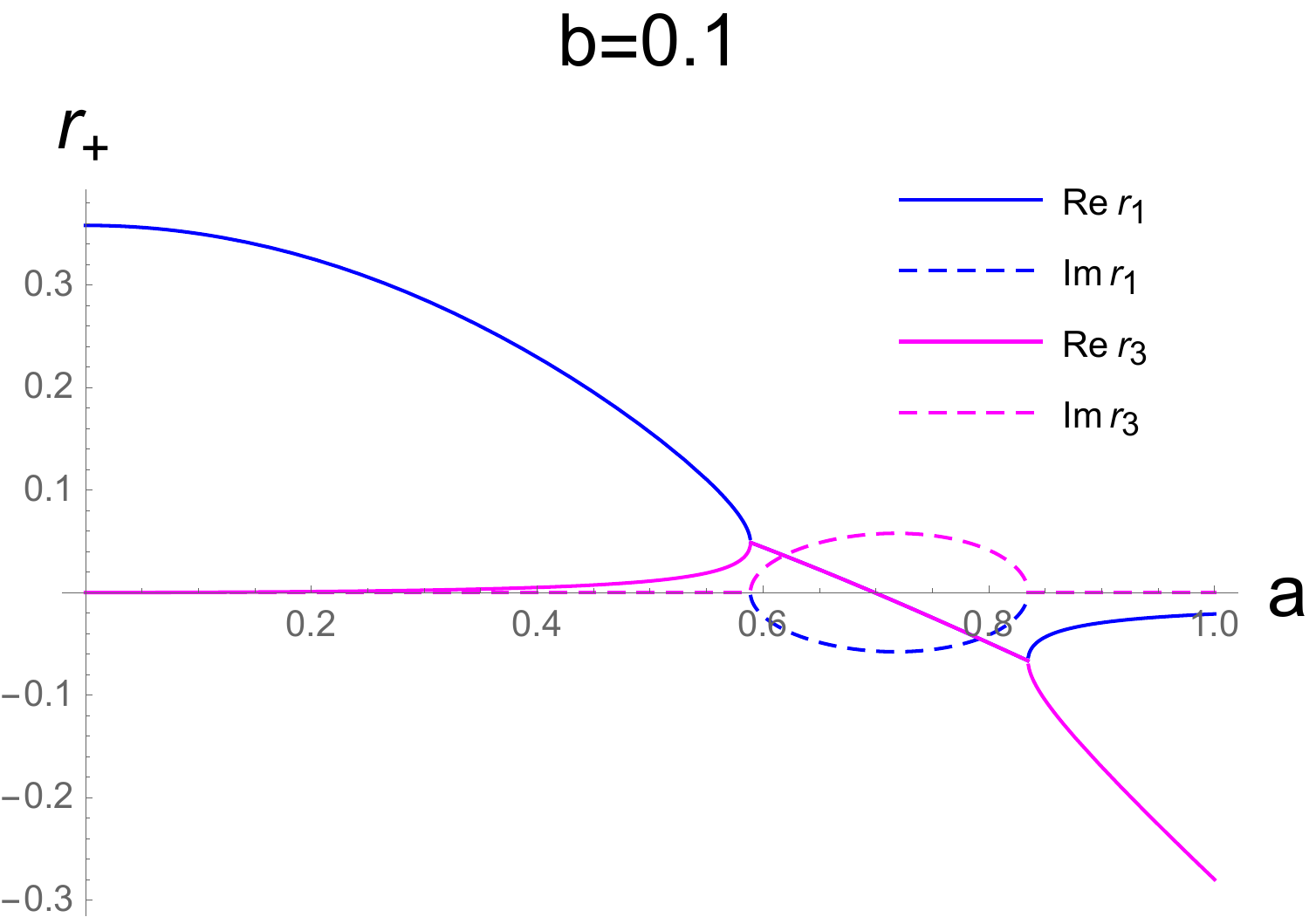}\\
A \hspace{70 mm} B\\$\,$\\
\includegraphics[scale=0.25]{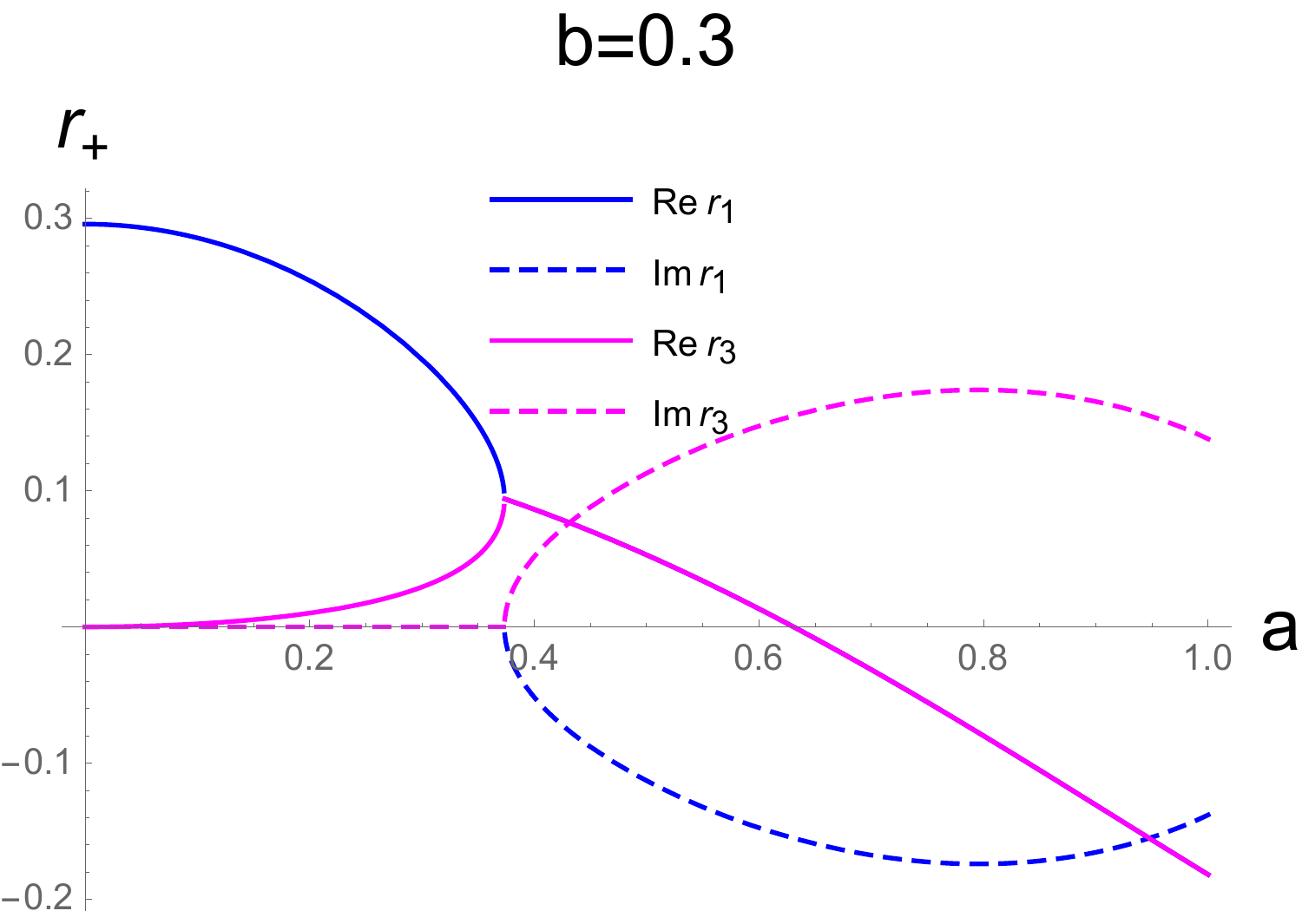}
\includegraphics[scale=0.25]{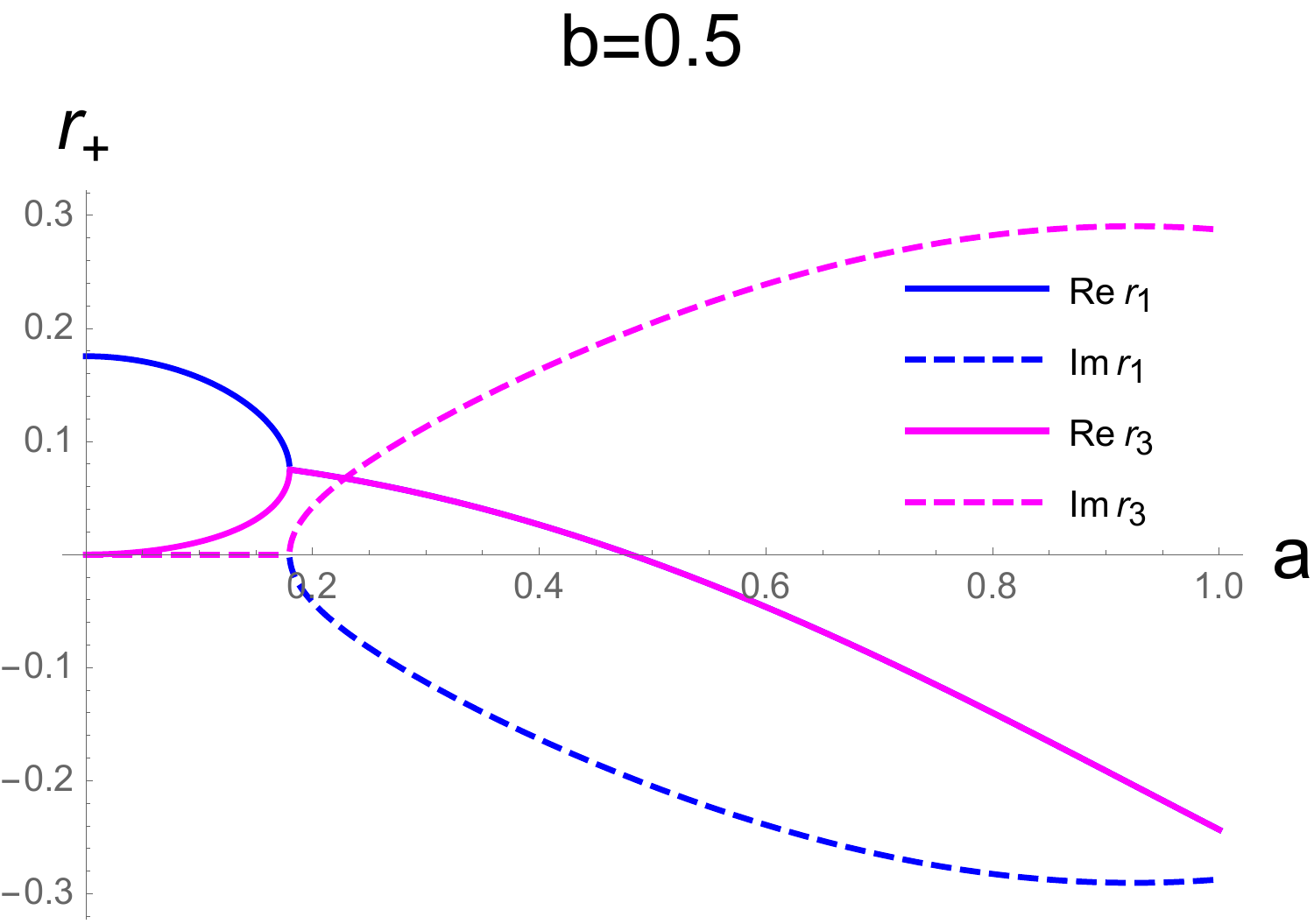}\\
C \hspace{70 mm} D\\ $\,$\\
\includegraphics[scale=0.25]{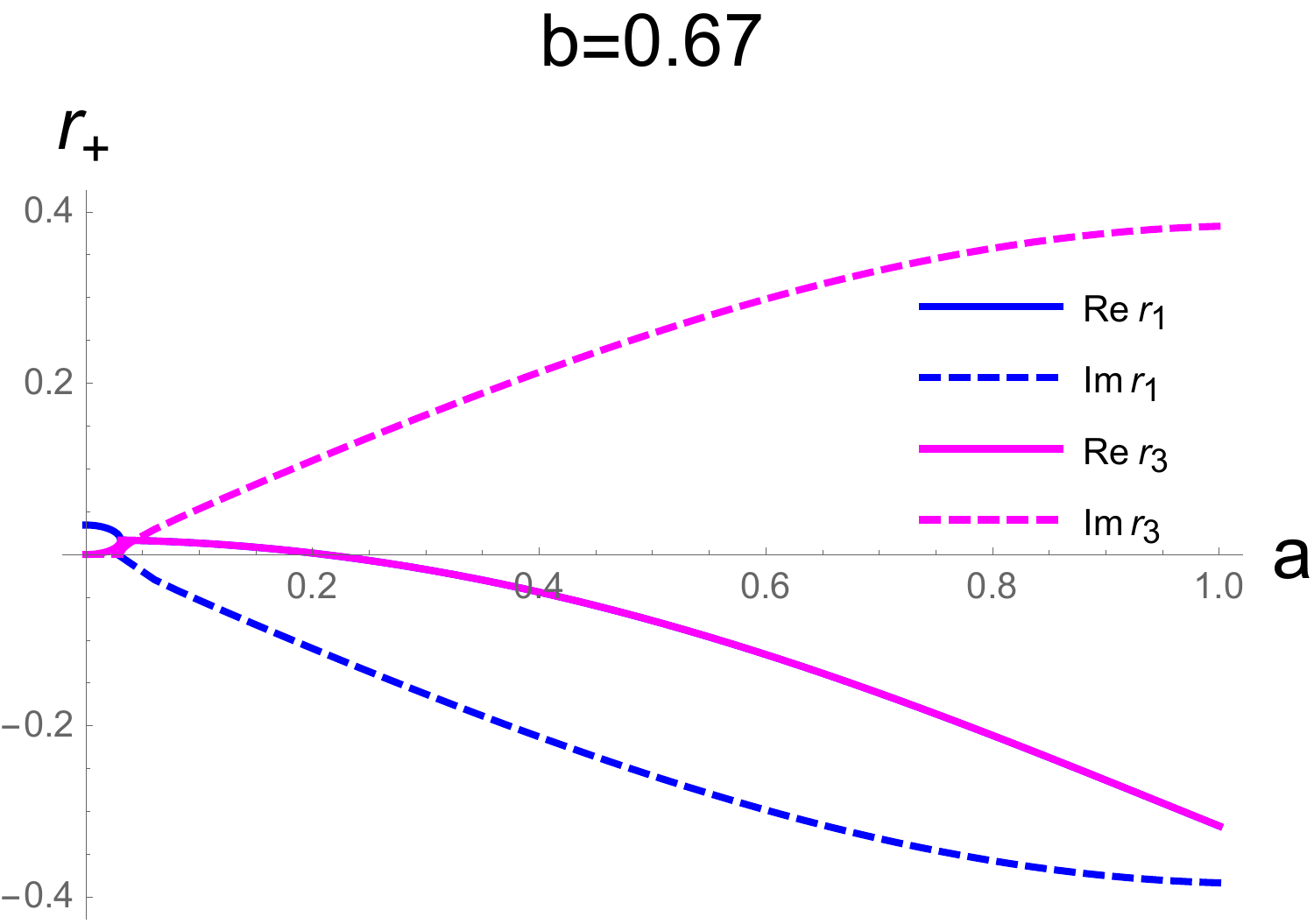}
\includegraphics[scale=0.25]{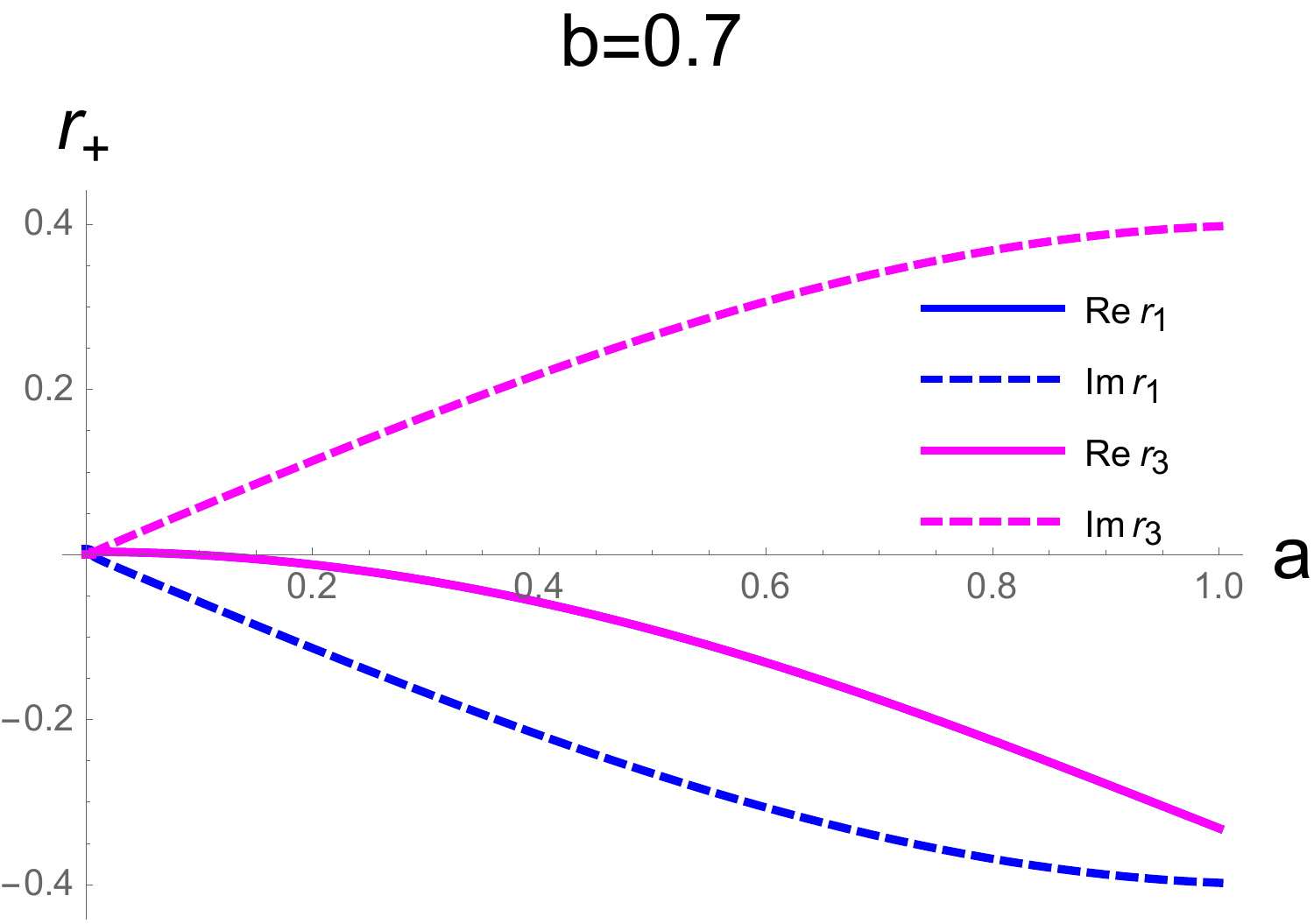}
\\ E \hspace{70 mm} F
 \caption{3 roots of equation $\Delta(r)=0$ vs $a$ for various values of $b$. We see that only the first root can be  positive for $a<b$ . Here $M$ is taken to be $0.25$ and $\ell=1$.
  }
  \label{Fig:roots}
\end{figure}

We note that in the Boyer-Lindquist coordinates, the metric is asymptotic to ${\rm AdS}_{5}$ in a rotating frame, with angular velocities
\be
\Omega^{\infty}_{\phi} = -a\ell^{2},\quad \Omega^{\infty}_{\psi} = -b\ell^{2}.
\ee
Let us consider  the case of the Kerr-AdS solutions with $a = b$, then the metric
(\ref{1.1GF}) takes the form
\bea\label{1.1SF2}
ds^{2}& =& - (1 + \rho^{2}\ell^{2} - \frac{2M}{\rho^{2}})dt^{2} + \frac{\rho^{2}}{\Delta_r}dr^{2} + \frac{\rho^{2}}{\Delta_{\theta} }d\theta^{2} \nonumber \\
&+& \frac{\sin^{2}\theta}{\Xi^{2}}\left(\rho^{2}\Xi + \frac{2a^{2}M}{\rho^{2}}\sin^{2}\theta\right)d\phi^{2}  + \frac{\cos^{2}\theta}{\Xi^{2}}\left(\Xi\rho^{2} + \frac{2a^{2}M\cos^{2}\theta}{\rho^{2}}\right)d\psi^{2}\nonumber \\
&+& \frac{2a\sin^{2}\theta}{\Xi}\left(\rho^{2}\ell^{2} - \frac{2M}{\rho^{2}}\right) dt d\phi + \frac{2a \cos^{2}\theta}{\Xi}\left(\rho^{2}\ell^{2} - \frac{2M}{\rho^{2}}\right)dtd\psi\\
&+& \frac{4Ma^{2}\sin^{2}\theta \cos^{2}\theta}{\Xi^{2}\rho^{2}}d\phi d\psi, \nonumber
\eea
where  for (\ref{Xi.ab}) we have
\bea \label{2.1a}
\Delta_r &=& \frac{1}{r^{2}}(r^{2} + a^{2})^{2}(1 +r^{2}\ell^{2}) - 2M, \\ \label{2.1b}
\Delta_{\theta} &= &(1 -a^{2}\ell^{2}), \quad  \rho^{2} = (r^{2}+ a^{2}), \quad \Xi  = (1- a^{2}\ell^{2}).
\eea
The 5d Kerr-AdS solution with a single rotational parameter ($a\neq 0$, $b=0$) reads
\be\label{1.1}
ds^{2} = - \frac{\Delta_{r}}{\rho^{2}}\left(\dd t - \frac{a}{\Xi_{a}}\sin^{2}\theta \dd\phi\right)^{2} + \frac{\rho^{2}}{\Delta_{r}}\dd r^{2} +  \frac{\rho^{2}}{\Delta_{\theta}}\dd\theta^{2}
+ \frac{\Delta_{\theta}\sin^{2}\theta}{\rho^{2}}[a\dd t - \frac{(r^{2} + a^{2})}{\Xi_{a}}\dd\phi]^{2} + r^{2}\cos^{2}\theta \dd\psi^{2},
\ee
with
\begin{eqnarray}
\Delta_{r}& =& (r^{2} + a^{2})(1+\ell^{2}r^{2}) -2M,\nonumber \\ \label{1.1a}
\Delta_{\theta}& = &1 - a^{2}\ell^{2}\cos^{2}\theta, \\
\Xi_{a} &= &1 -a^{2}\ell^{2}, \quad \rho^{2} = r^{2}+ a^{2}\cos^{2}\theta.\nonumber
\end{eqnarray}
For the single parameter 5d Kerr-AdS hole the horizon position is written down explicitly
\be\label{rhsinglea}
r_{+} = \frac{\sqrt{\sqrt{(1-a^{2}\ell^{2})^{2} + 8M\ell^{2}}- (1 - a^{2}\ell^2)}}{\sqrt{2}\ell}.
\ee
The case of single rotational parameter can be constructed  from 4d case as the stationary asymptotically flat higher dimensional black holes from the work by Myers and Perry \cite{MyPerry}.

The  transformations that convert (\ref{1.1GF}) from Boyer-Lindquist coordinates to asymptotically AdS coordinates $a\neq b\neq 0$ \cite{HHT} are
\bea
\Xi_{a} y^{2}\sin^{2}\Theta &= &(r^{2} + a^{2})\sin^{2}\theta,\nonumber \\
\Xi_{b} y^{2}\cos^{2}\Theta& = &(r^{2}+b^{2})\cos^{2}\theta,\nonumber\\ \label{yCAdS5ab}
\Phi &= &\phi + a \ell^{2}t, \\
\Psi&= & \psi  + b \ell^{2}t,\nonumber\\
T &= &t.\nonumber
\eea
It should be noted the coordinates (\ref{yCAdS5ab}) are difficult for direct representation of the 5d Kerr-AdS metrics,
except the case when we have the two non-zero rotational parameters which are equal by its magnitude:
\bea\label{globalAdS2}
ds^{2}& =& - (1+ y^{2}\ell^{2})dT^{2} + y^{2}(d\Theta^{2} + \sin^{2}\Theta d\Phi^{2} + \cos^{2}\Theta d\Psi^{2}) \\
&+& \frac{2M}{y^{2}\Xi^{3}}(dT - a \sin^{2}\Theta d\Phi - a\cos^{2}\Theta d\Psi)^{2}
+\frac{y^{4}dy^{2}}{y^{4}(1 + y^{2}\ell^{2}) - \frac{2M}{\Xi^{2}}y^{2} +\frac{ 2Ma^{2}}{\Xi^{3}}}.\nonumber
\eea
The position of the horizon of the extremal black hole in these coordinates \cite{HHT} reads
\be
y^{2} = \frac{1}{4\Xi}\left[4a^{2}\ell^{2} - 1 + \sqrt{1 + 8a^{2}\ell^{2}}\right].
\ee

\subsection{Thermodynamics of the 5d Kerr-AdS black holes}\label{TherKBH}

 Thermodynamics of the holographic background defines the first order phase transition on ($T$--parameters) phase
diagram. In the gravity language, the origin of this phase transition is related with is a non-trivial dependence of the Hawking temperature on the location of the horizon.
A typical example of such dependence is given by the Van der Waals curve.
 Here we are going to consider  the angular momentum of the rotating medium as physical parameters and find out how they affect on the phase diagram.
Technically this means that we have to understand a location of the phase transition
of the background in a two-dimensional space $(T-a)$ for the case of one rotational parameter  $a$ ($T$ is the Hawking temperature given by (\ref{HawkT})) and in
three-dimensional space $(T-a-b)$ for the case of two rotational parameter, $a$ and $b$.

\begin{figure}[h!]
  \centering
    \includegraphics[scale=0.125]{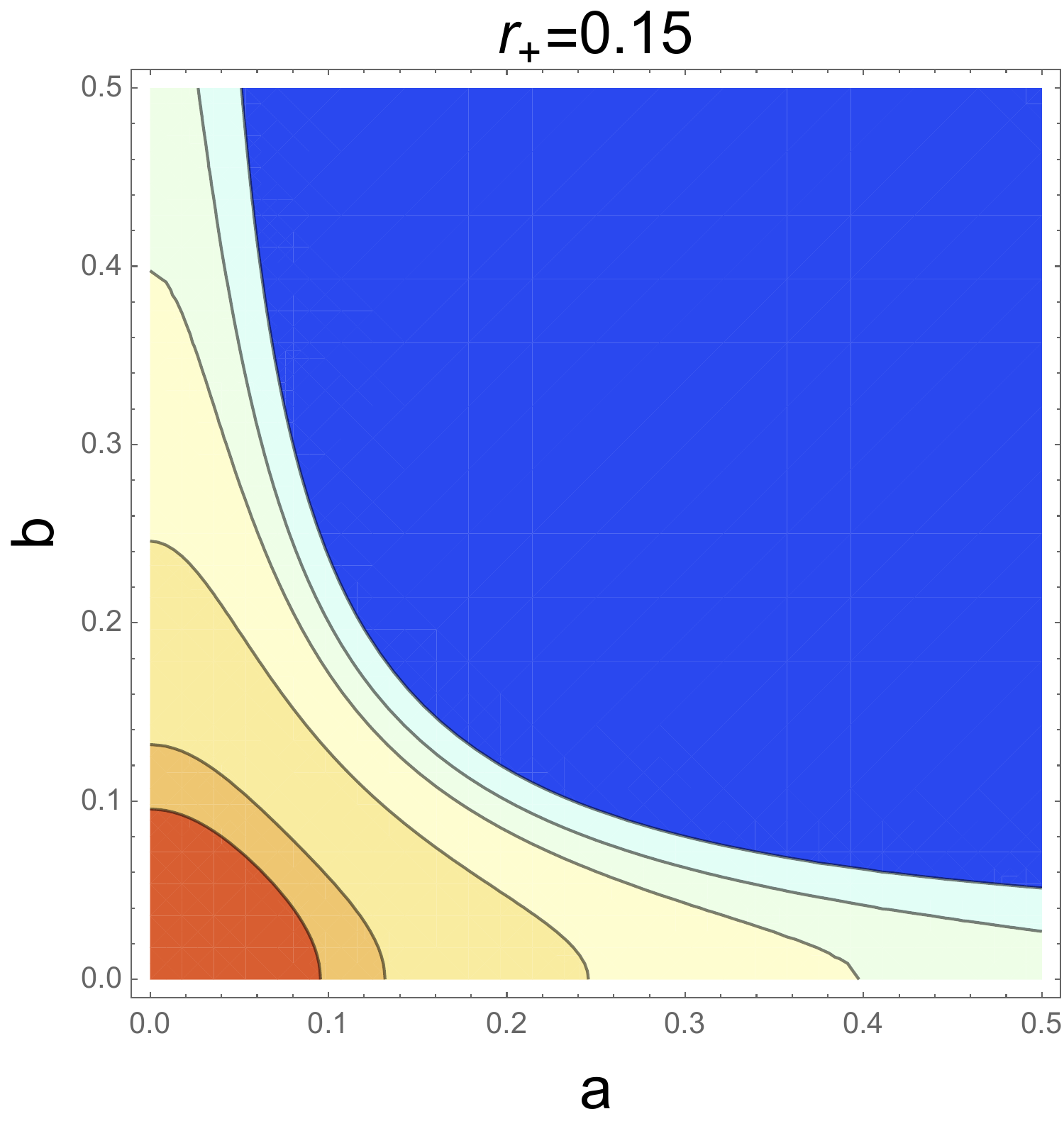}
  \includegraphics[scale=0.1]{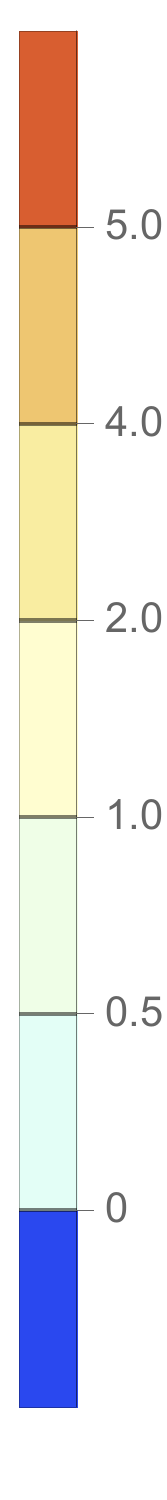}
  \includegraphics[scale=0.125]{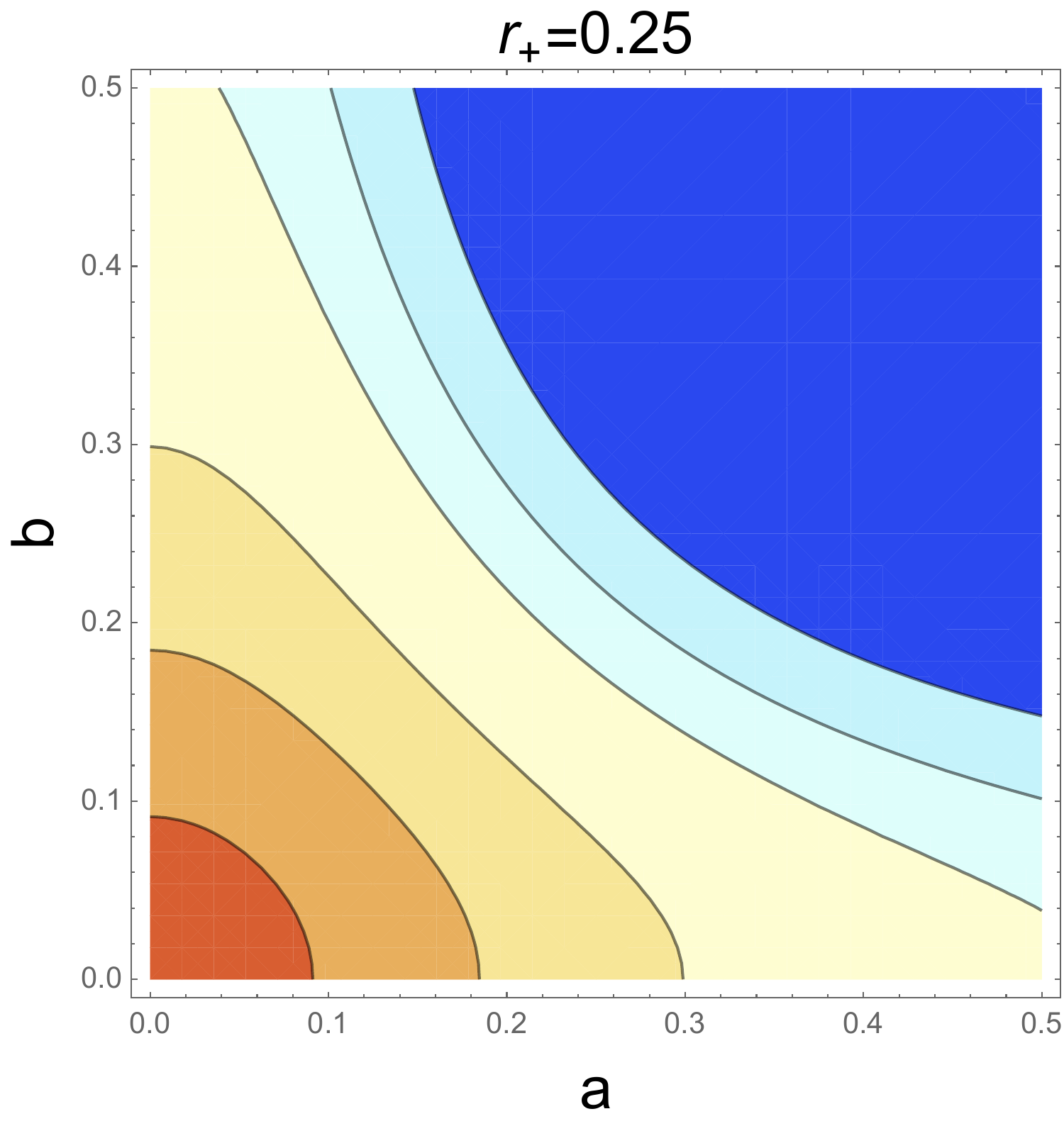}
  \includegraphics[scale=0.1]{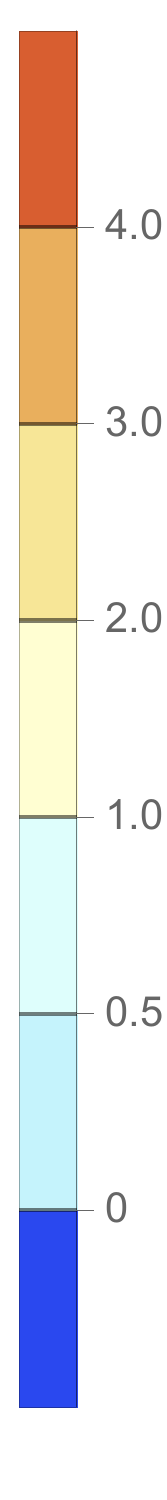}
  \includegraphics[scale=0.125]{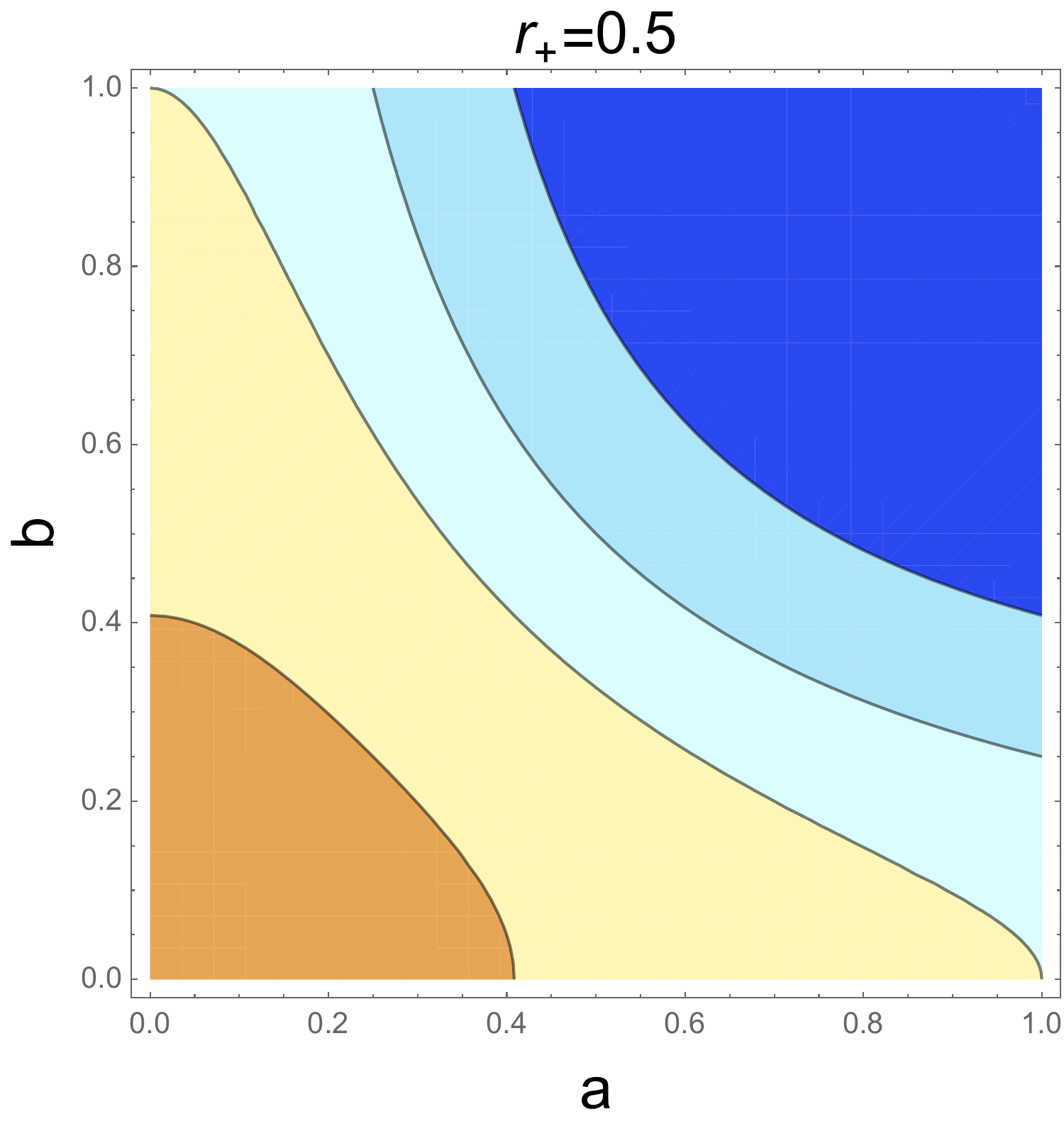}
    \includegraphics[scale=0.1]{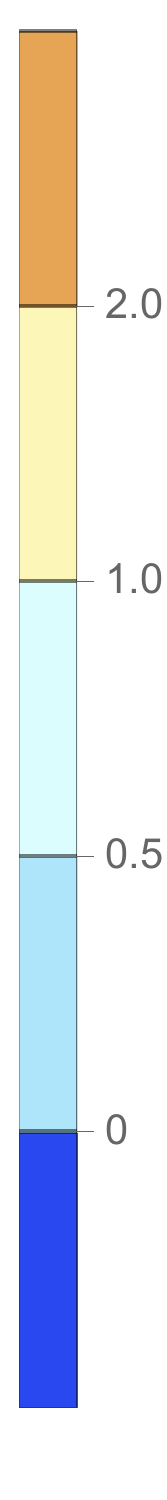}
      \includegraphics[scale=0.125]{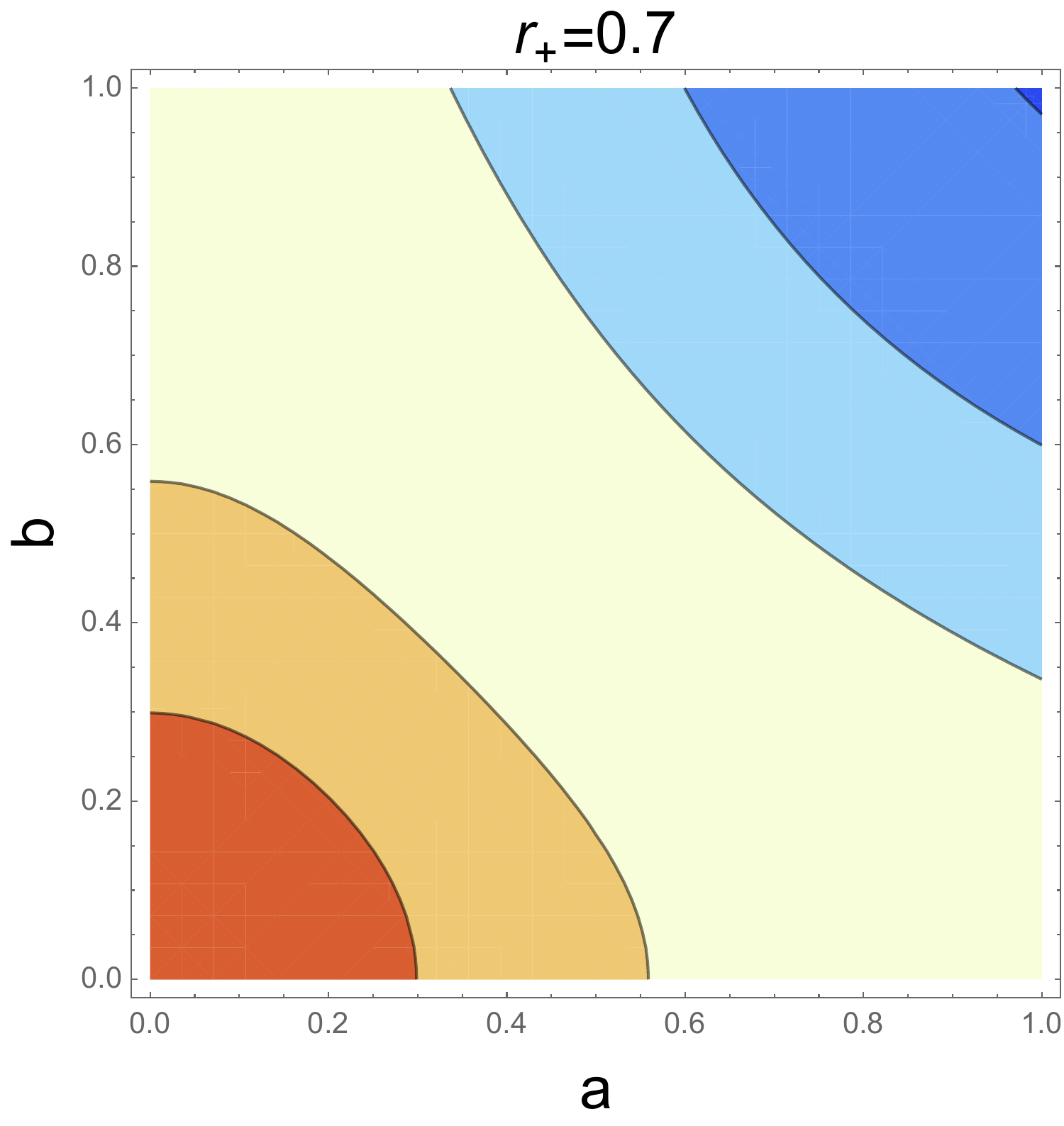}
    \includegraphics[scale=0.1]{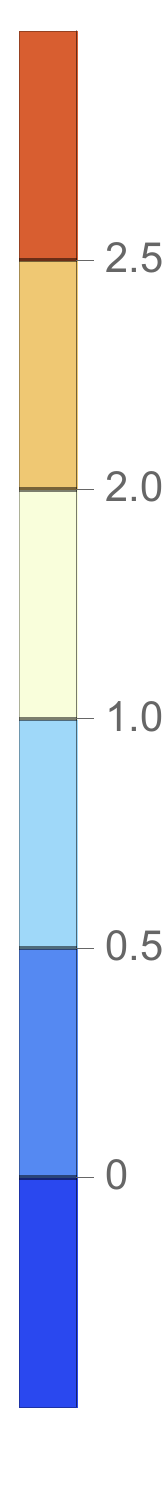}
  \caption{Contour plot of the Hawking temperature (\ref{HawkT})  multiplied by $2\pi\cdot 2.9$ depending on $a,b$ and $r_+$.
  }
  \label{Fig:PT}
\end{figure}

Below we illustrate that the Kerr-AdS black hole inherits the main thermodynamical property of the AdS black hole  -- the presence of the Hawking-Page phase transition.
For this purpose we trace the dependence of the Hawking temperature (\ref{HawkT}) on the rotational parameters $a$ and $b$ at fixed value of the horizon $r_{+}$.
 We present this in Fig.~\ref{Fig:PT}, using ContourPlot. The factor 2.9 ensures that the background phase transition for $a=b=0$ is at $T_{cr}(0,0)=0.160$ GeV, that is in agreement with lattice calculations and LHC and RHIC data.
From (\ref{HawkT}) it follows that the dependence of the horizon on the temperature is multivalued. To illustrate this we draw $T$ vs $r_{+}$ for different values of $a$ and $b$, see the left column of Fig.~\ref{Fig:TSF}. From these plots we observe that depending on the values of $a$ and $b$ the background allows to have the Van-der-Waals type of the temperature dependence on $r_{+}$ and for some cases the temperature becomes a three-valued function on the horizon $r_+$.  To show that the Hawking-Page transition takes place for the Kerr-AdS black hole, we have to consider the dependence of free energy $F$ on temperature $T$. To this purpose we find the dependence of the entropy $S$ on $T$
\be
S=\frac{ \pi^2 }{2 \Xi_a\cdot \Xi_b}\frac{(r_{+}^2+a^2)(r_{+}^2+b^2)}{r_{+}}
\ee
where $\Xi_{a}=1-a^2 \ell^2 $ and $\Xi_{b}=1- a^2 \ell^2 $.

From the middle column of Fig.~\ref{Fig:TSF} we see that for small enough values of $a$ and/or $b$, there is a region of temperatures at which  the entropy is three-valued.
If we will increase the value of one of the rotational parameters this region disappears and for the certain temperature the entropy becomes only one valued.

In the right column of Fig.~\ref{Fig:TSF} we show the influence of the rotating parameters on the dependence of the free energy on the temperature. For the free energy we use the following relation
\be
F(r) =  \int^{r_{+}}_{r}S(x)T'(x)dx.
\ee
 Again for the small values of $a$ and $b$ we observe that the free energy is multivalued and there are swallow tails in $(F-T)$ plane and the first order phase transition takes place.
  At some points, $(a_{cr},b_{cr})$, discontinuity of $F$ disappears and
  the first order phase transition ends.

 Finally, in Fig.~\ref{Fig:PT2} {\bf A}  we show how the rotating parameters affect on the value of the temperature of the phase transition. The higher the value of the rotating parameters ,  the lower temperature at which the phase transition takes place. So we see that it is quite justified to consider a case with one non-zero rotating parameter. In Fig.~\ref{Fig:PT2} {\bf B} we illustrate  the location of the transition plane $T_{cr}=T_{cr}(a,b)$.

\begin{figure}[t!]
  \centering
   \includegraphics[scale=0.14]{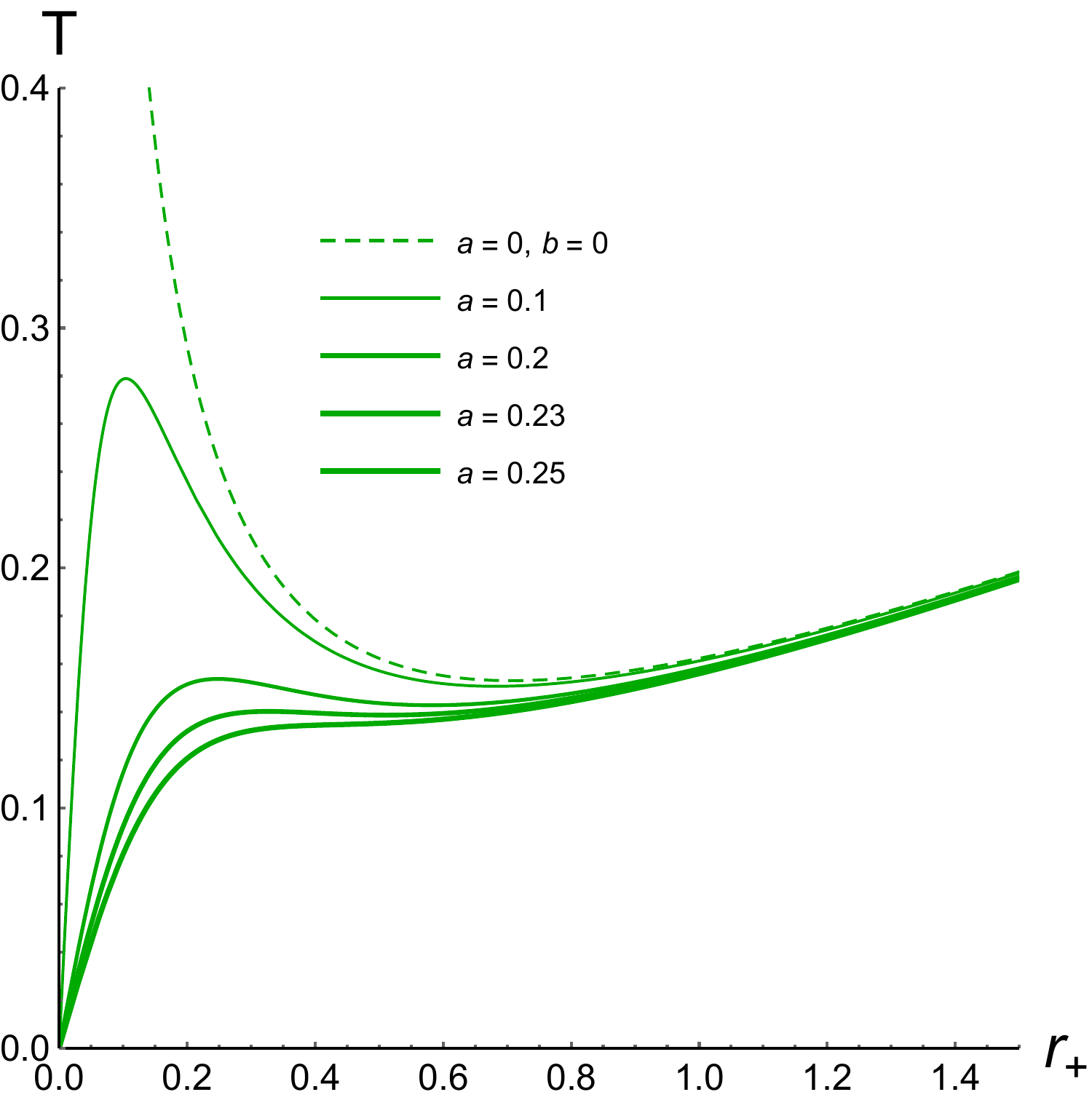}\qquad\includegraphics[scale=0.12]{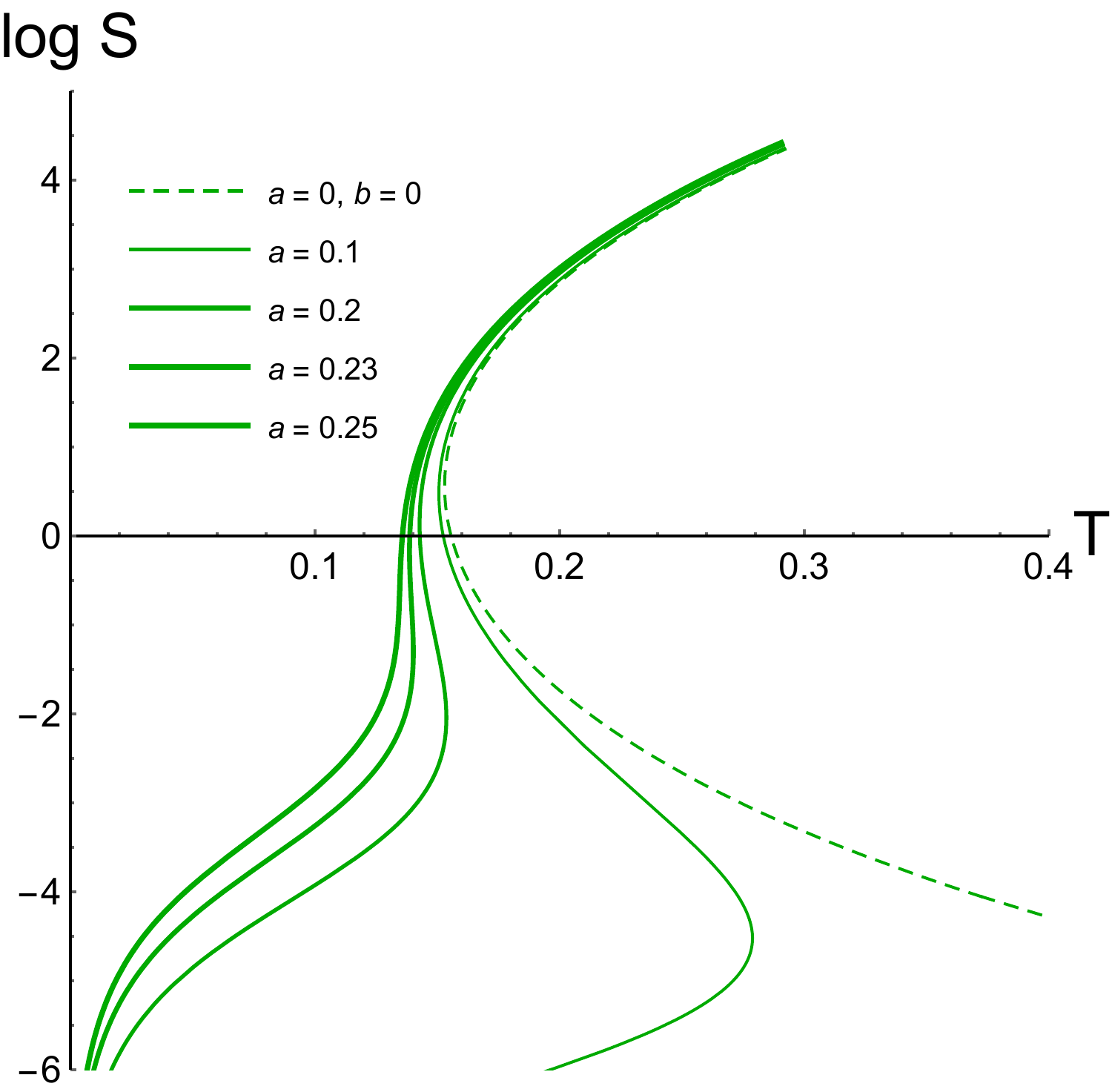} \qquad\includegraphics[scale=0.15]{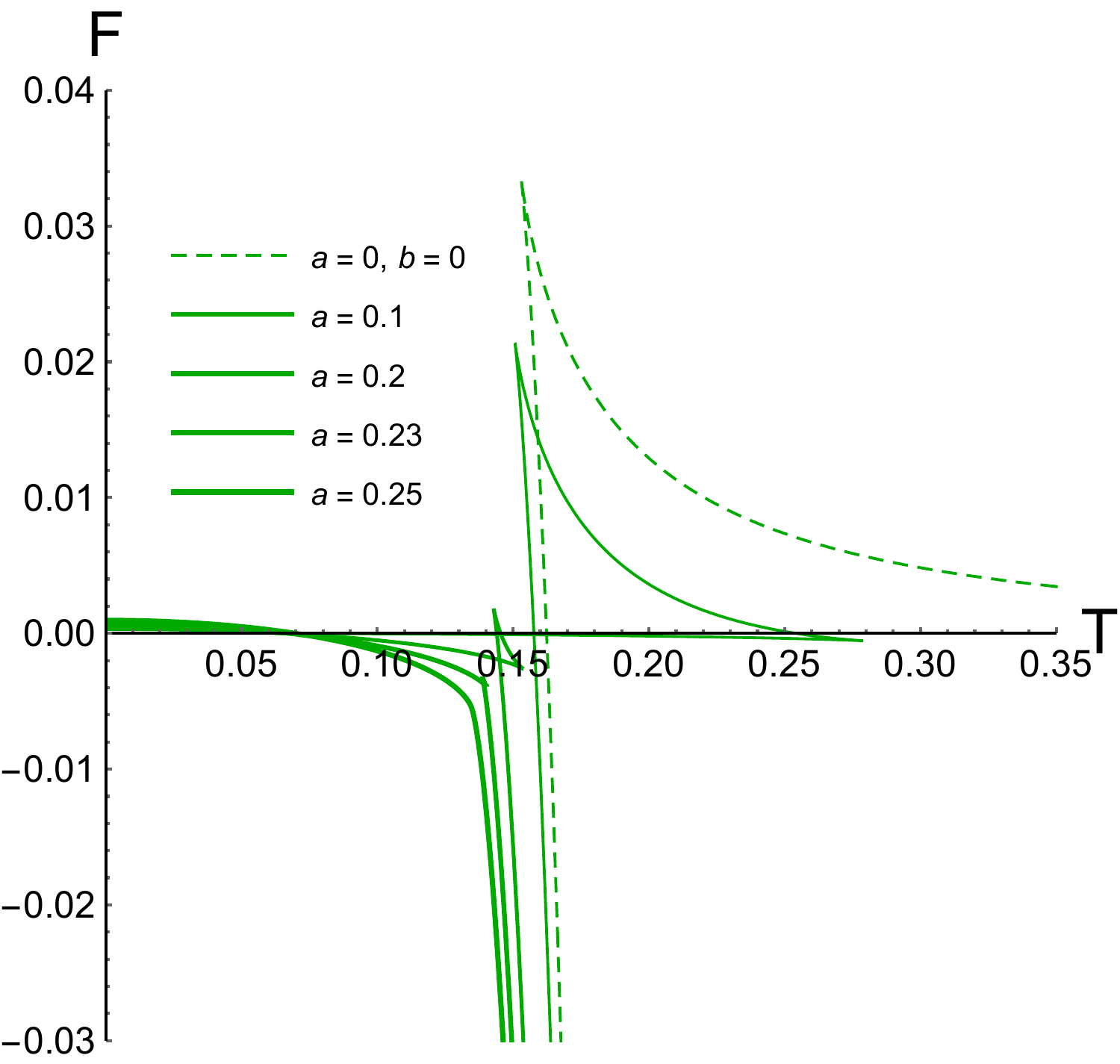}\\
 \includegraphics[scale=0.14]{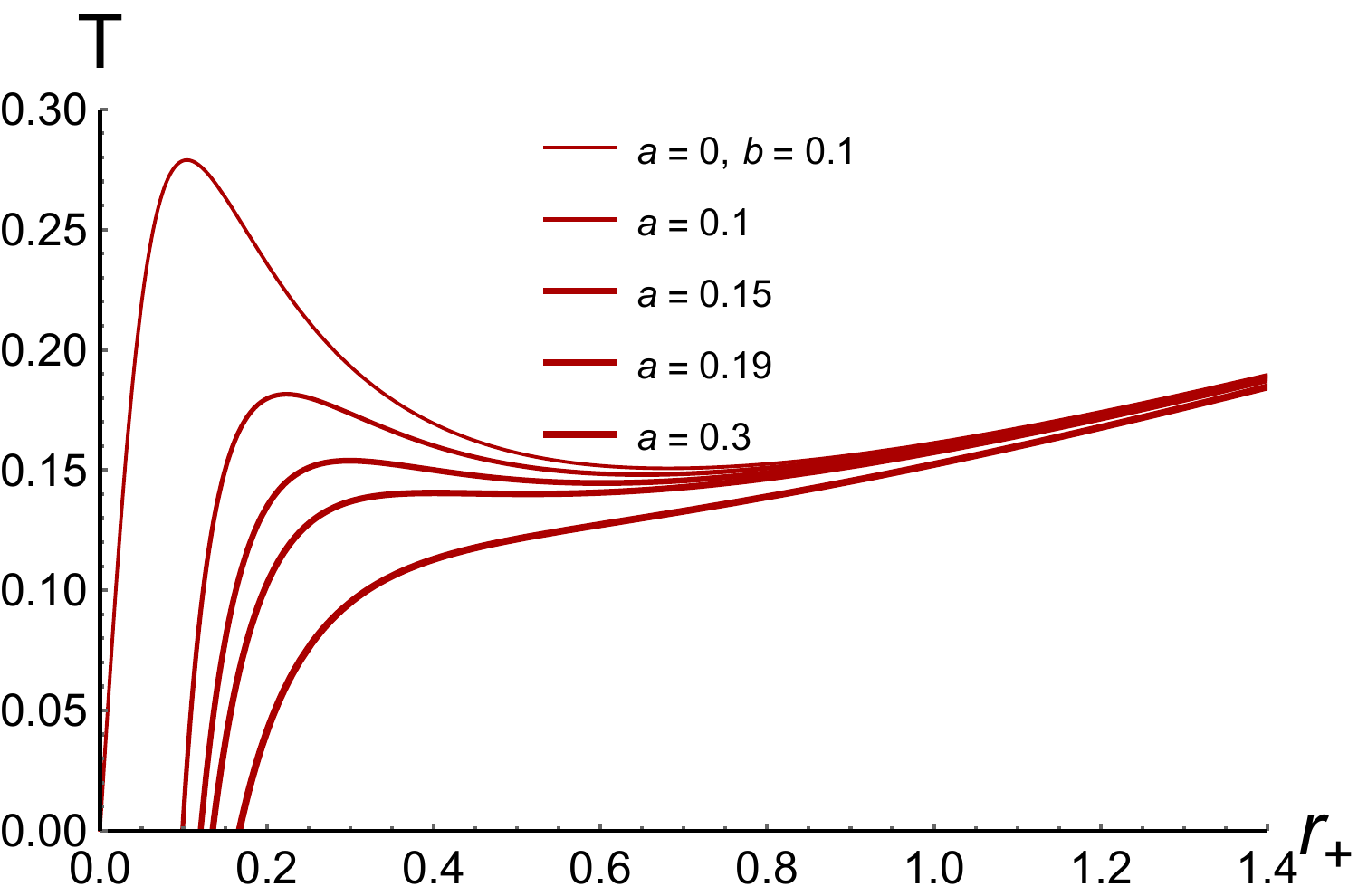}\qquad\includegraphics[scale=0.13]{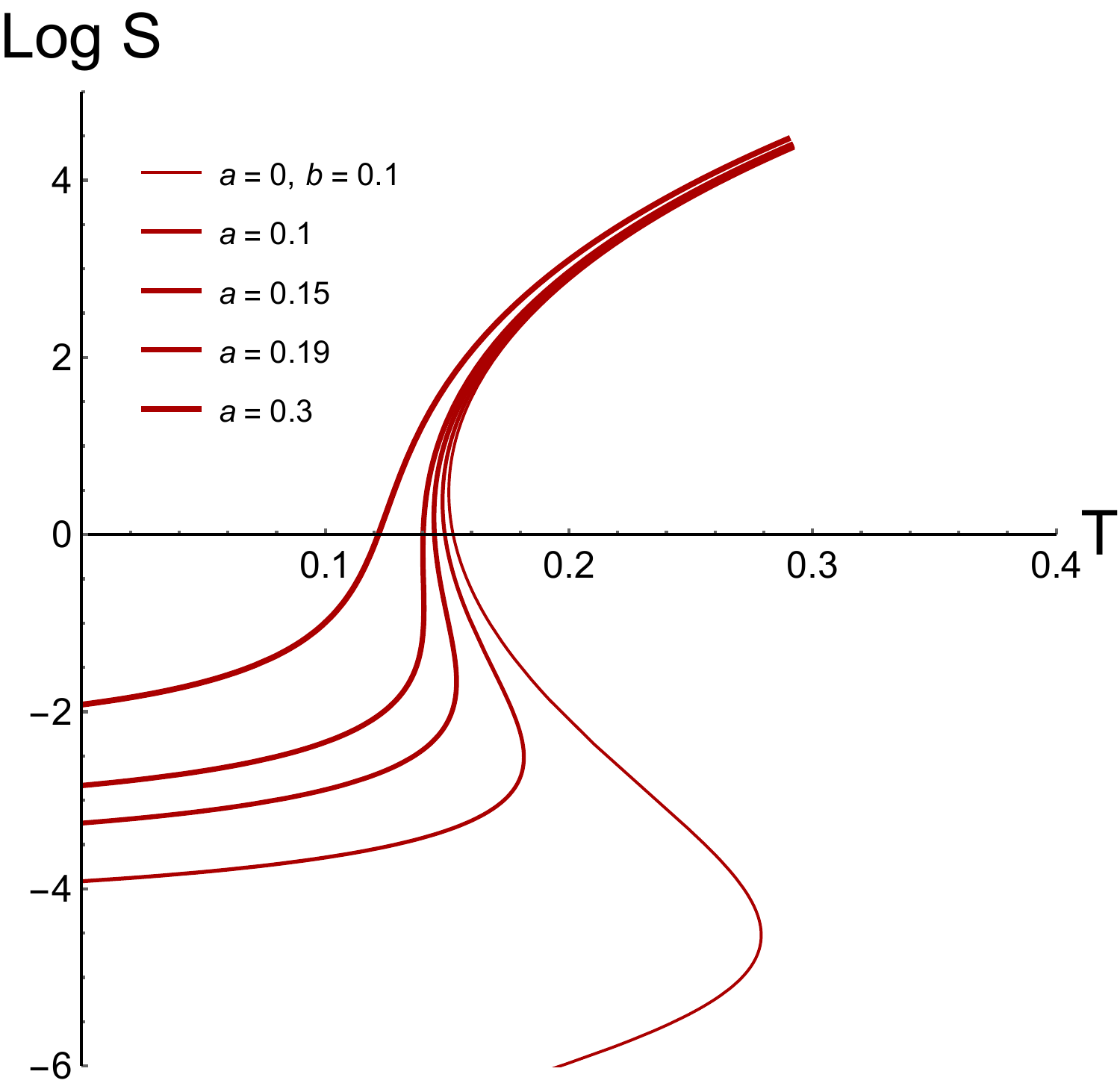}\qquad
  \includegraphics[scale=0.15]{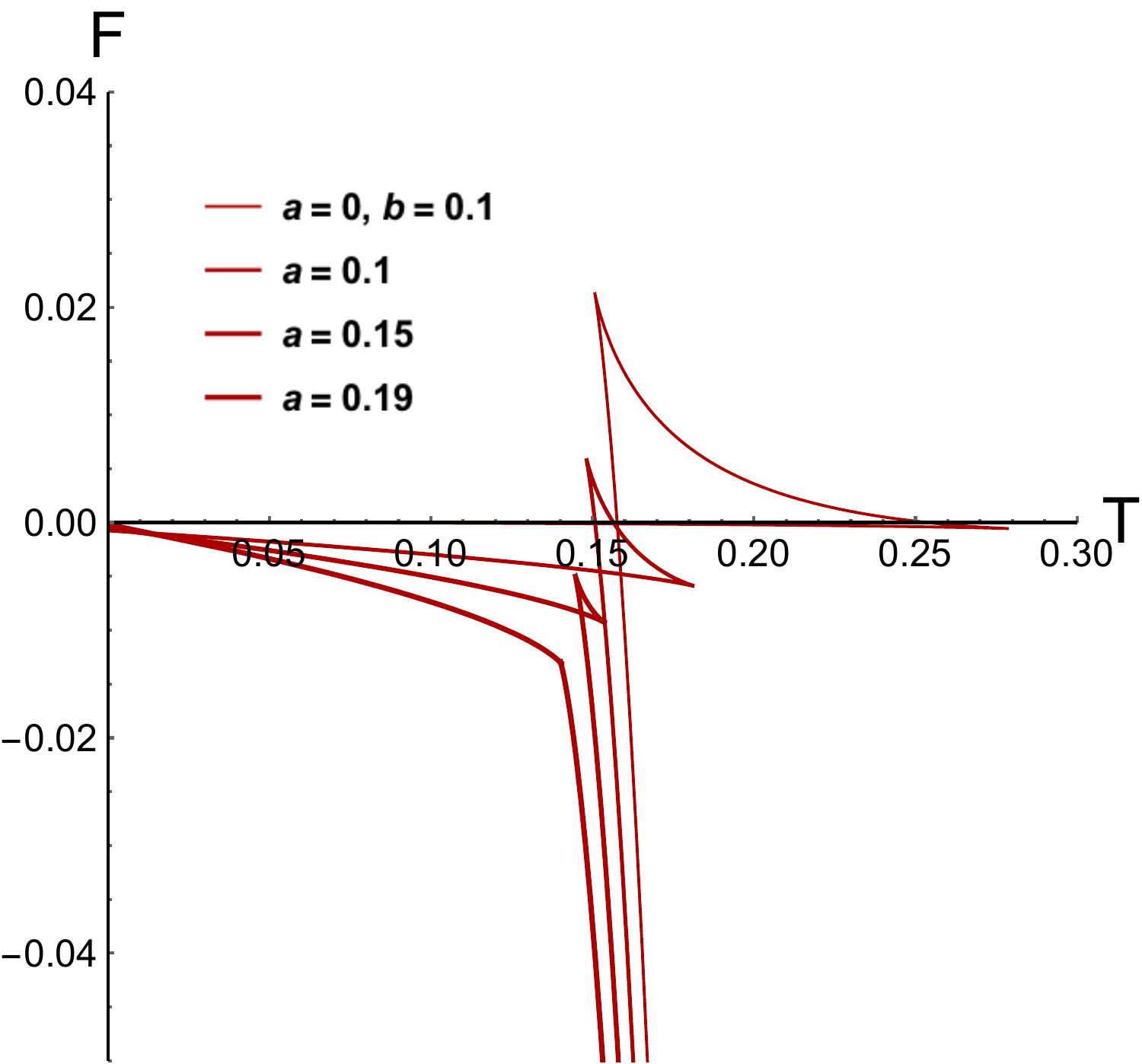}\\
   \includegraphics[scale=0.14]{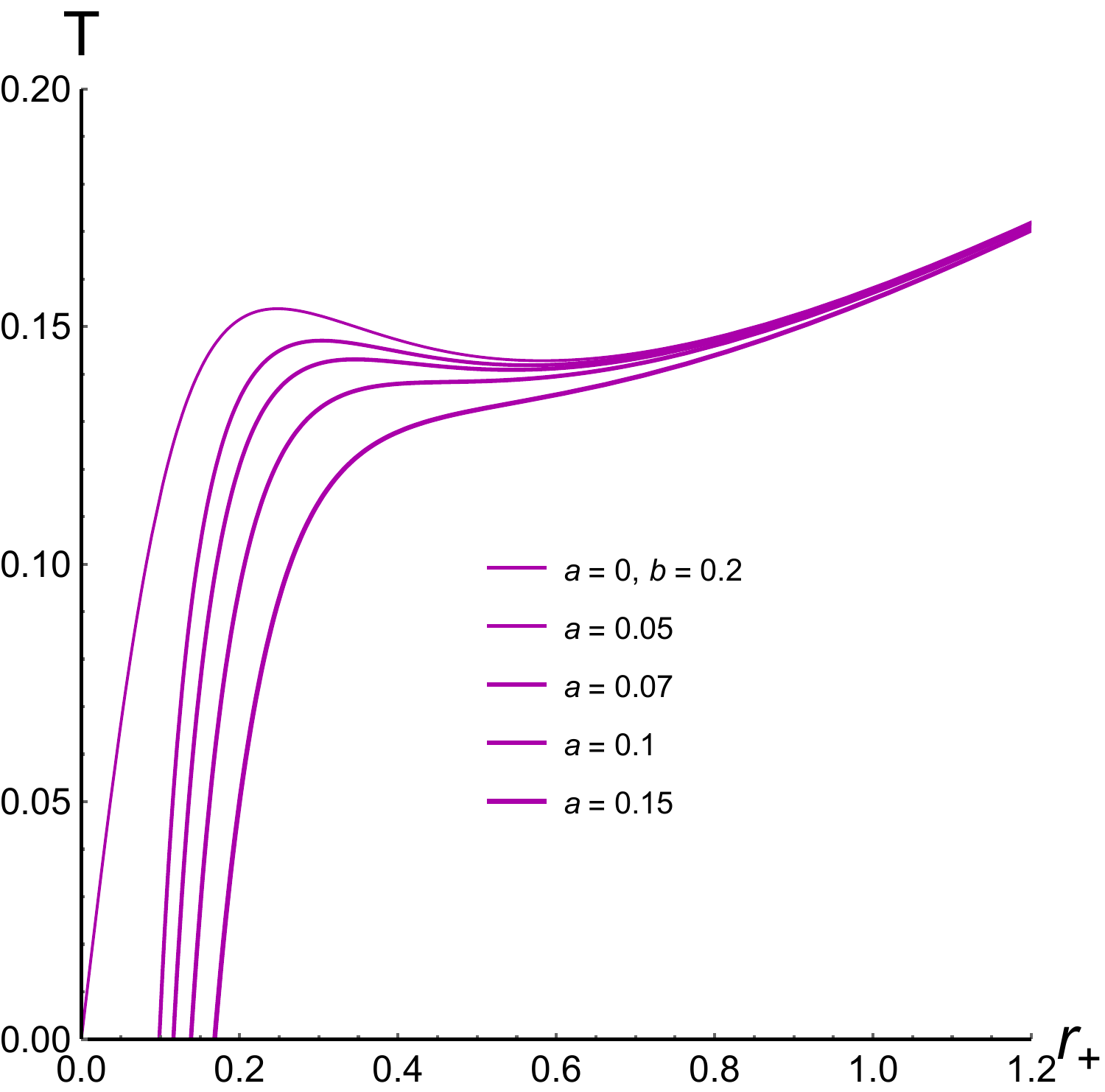}\qquad
    \includegraphics[scale=0.13]{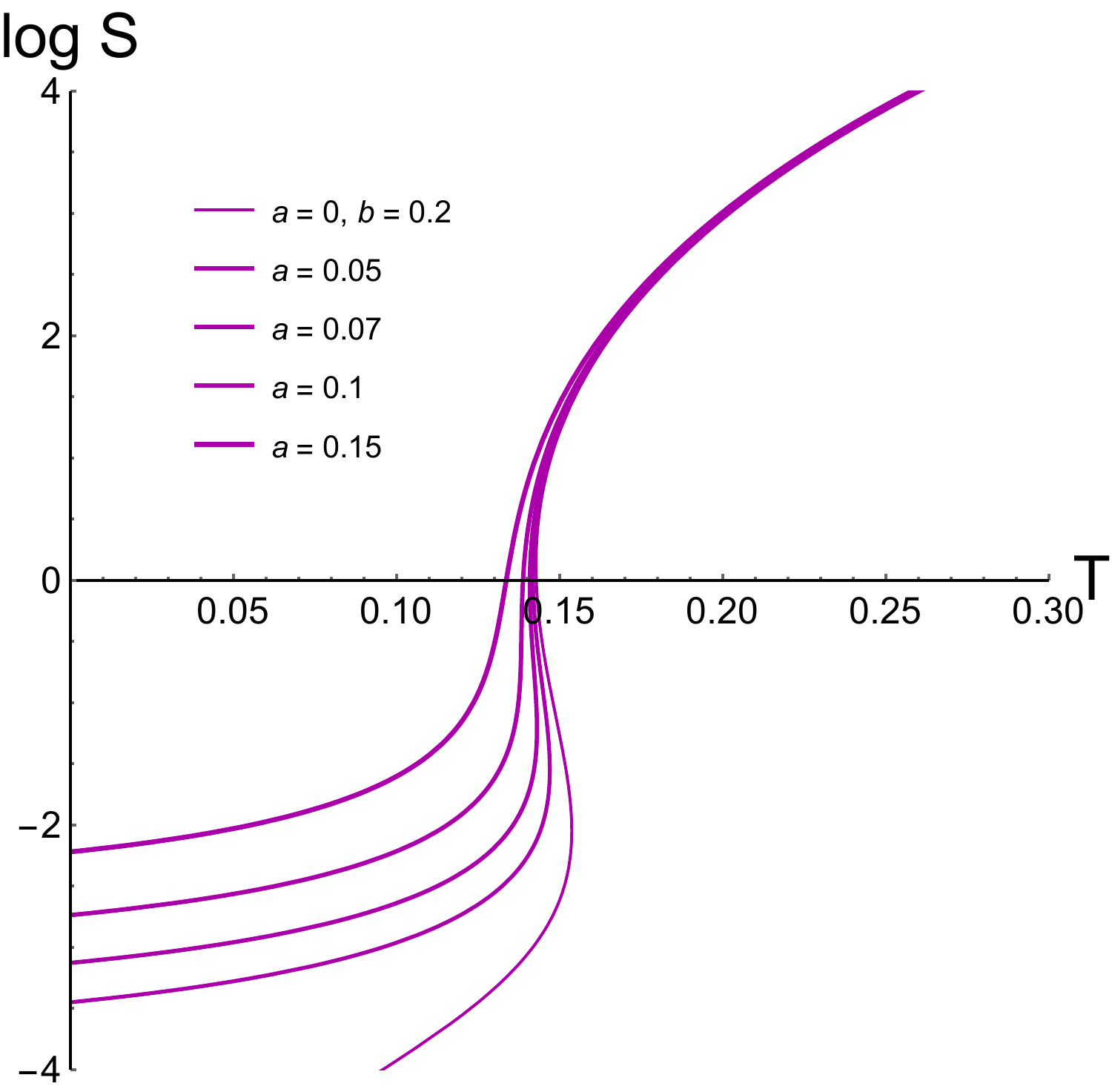}\qquad
  \includegraphics[scale=0.15]{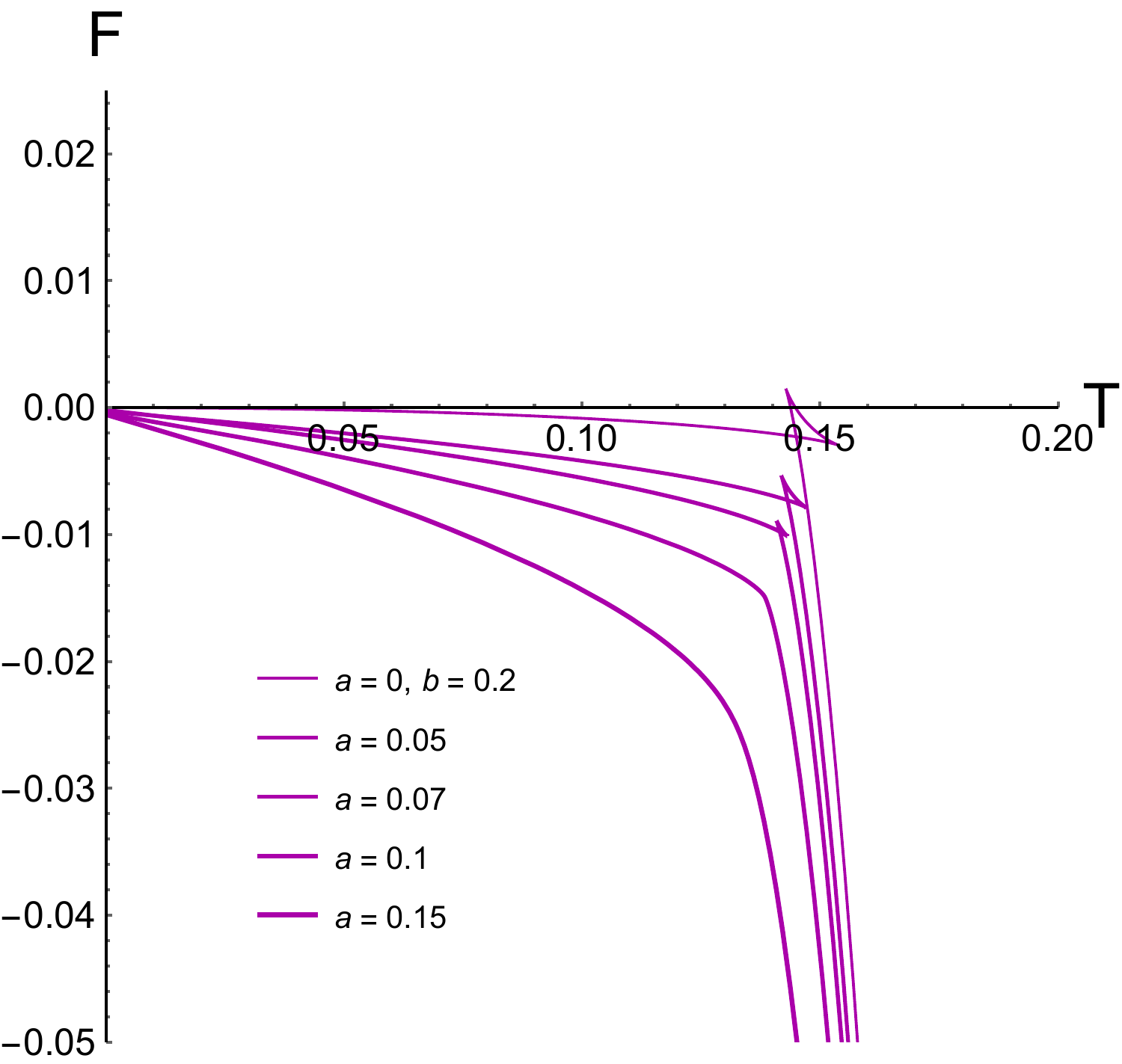}\\
     \includegraphics[scale=0.14]{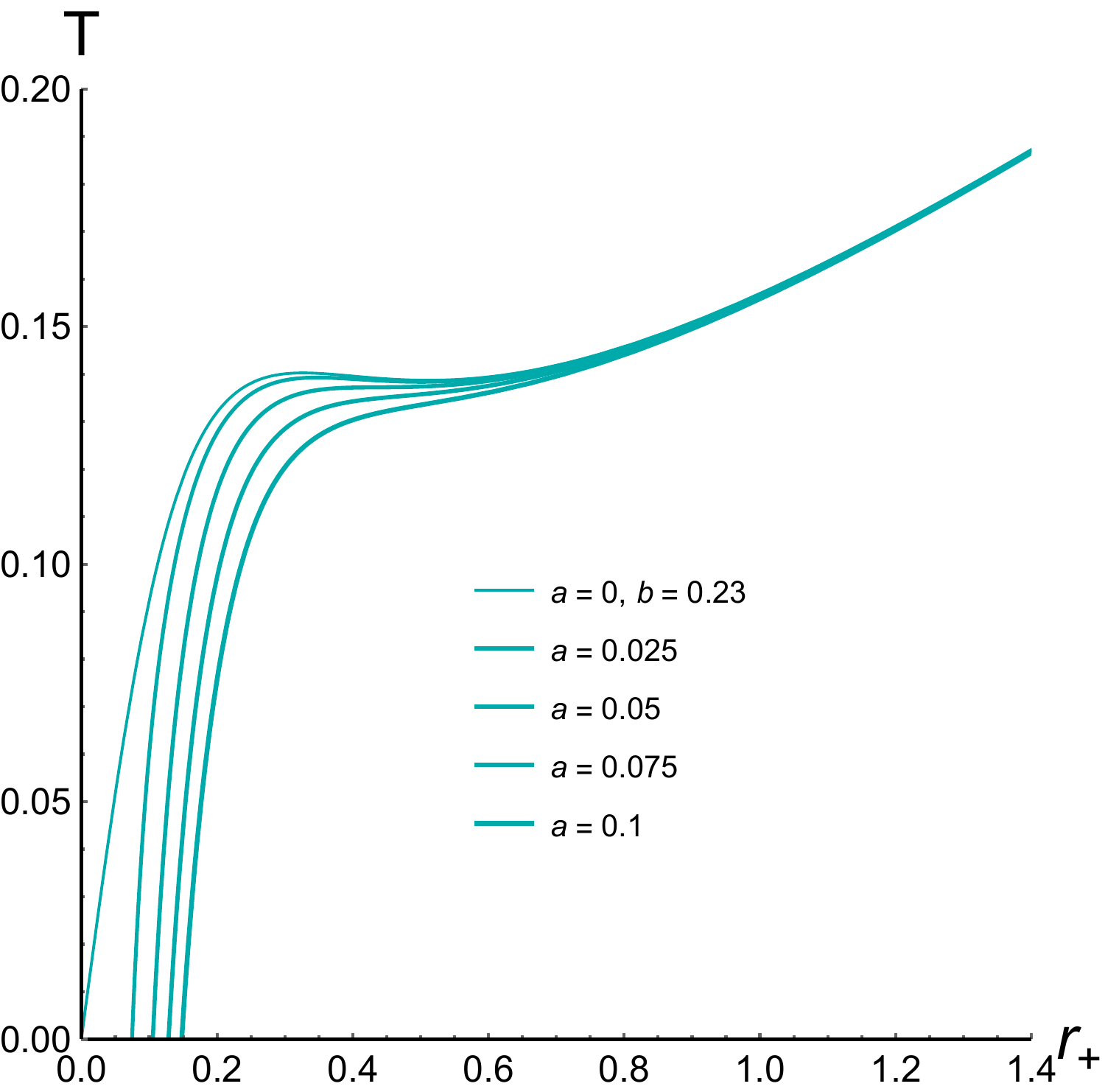}\qquad
    \includegraphics[scale=0.13]{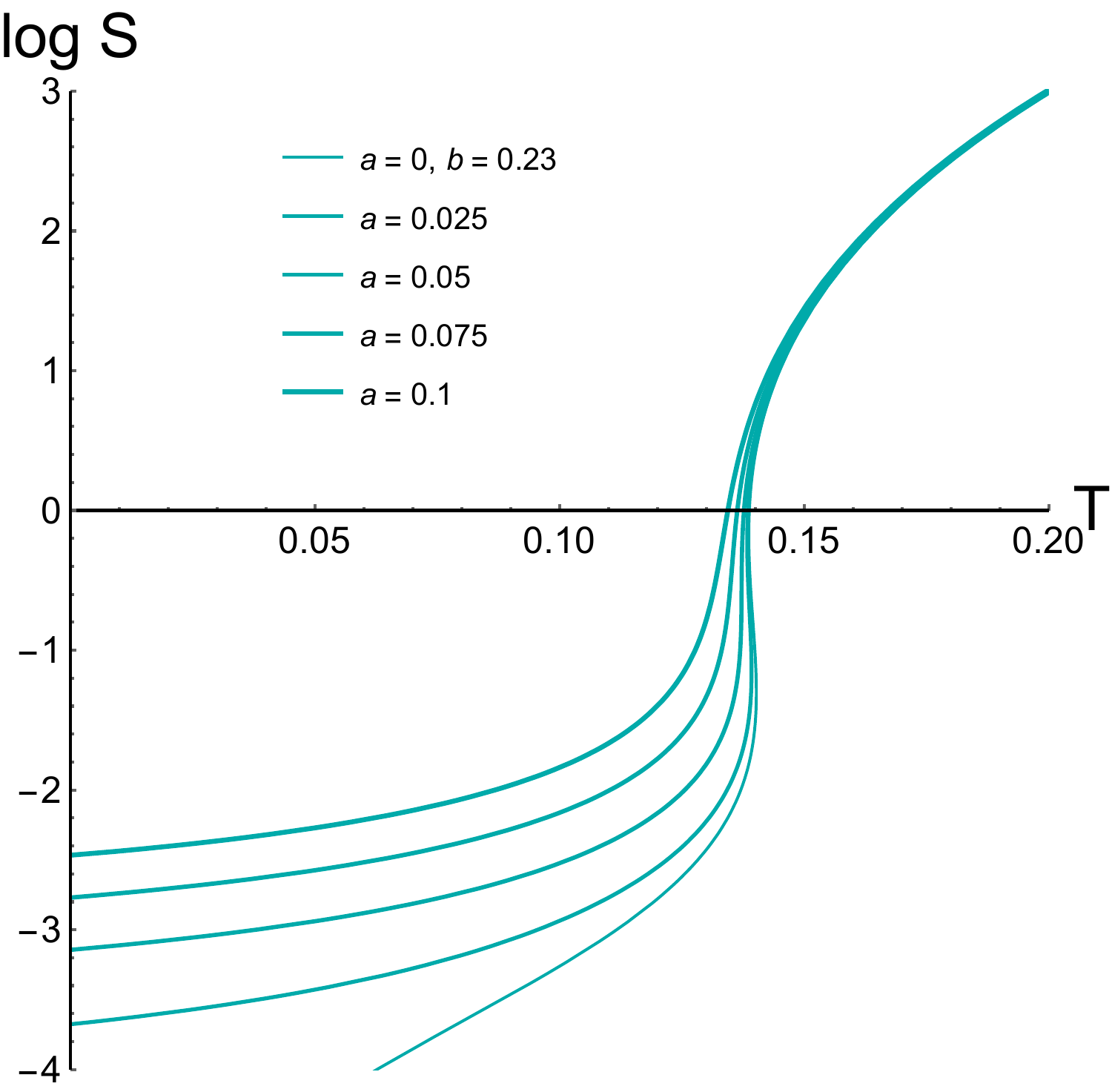}\qquad
  \includegraphics[scale=0.15]{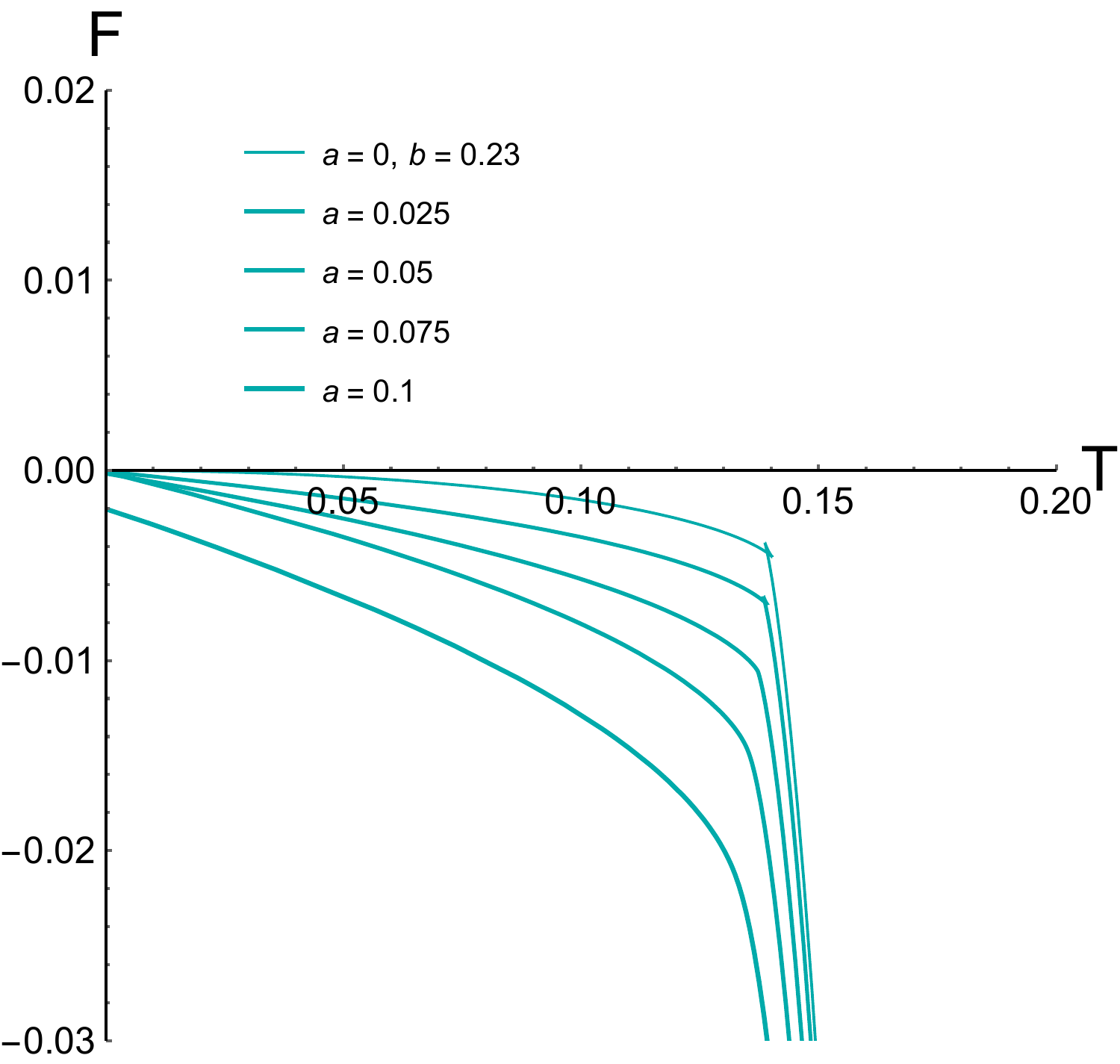}\\
     \includegraphics[scale=0.14]{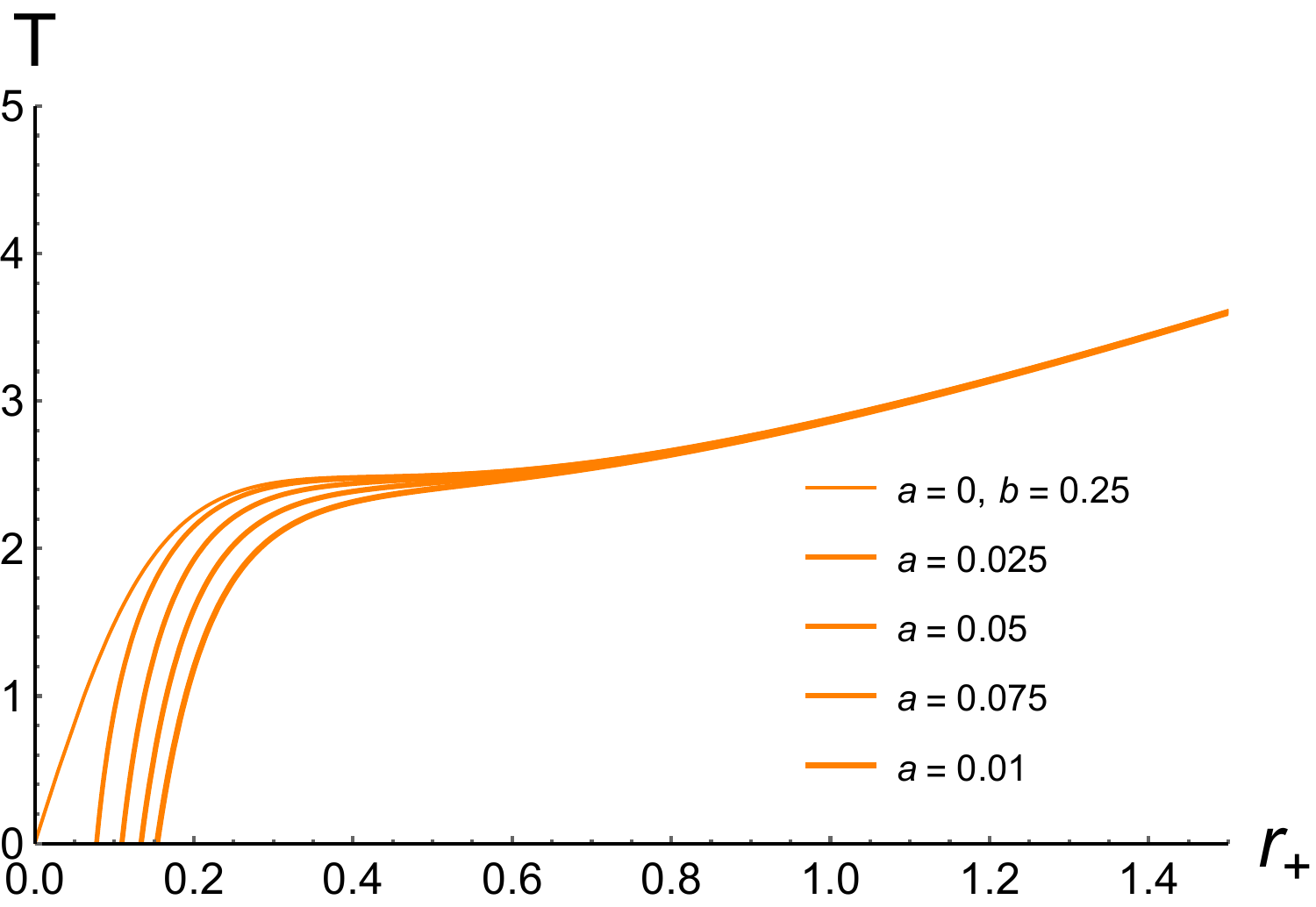}\qquad
    \includegraphics[scale=0.13]{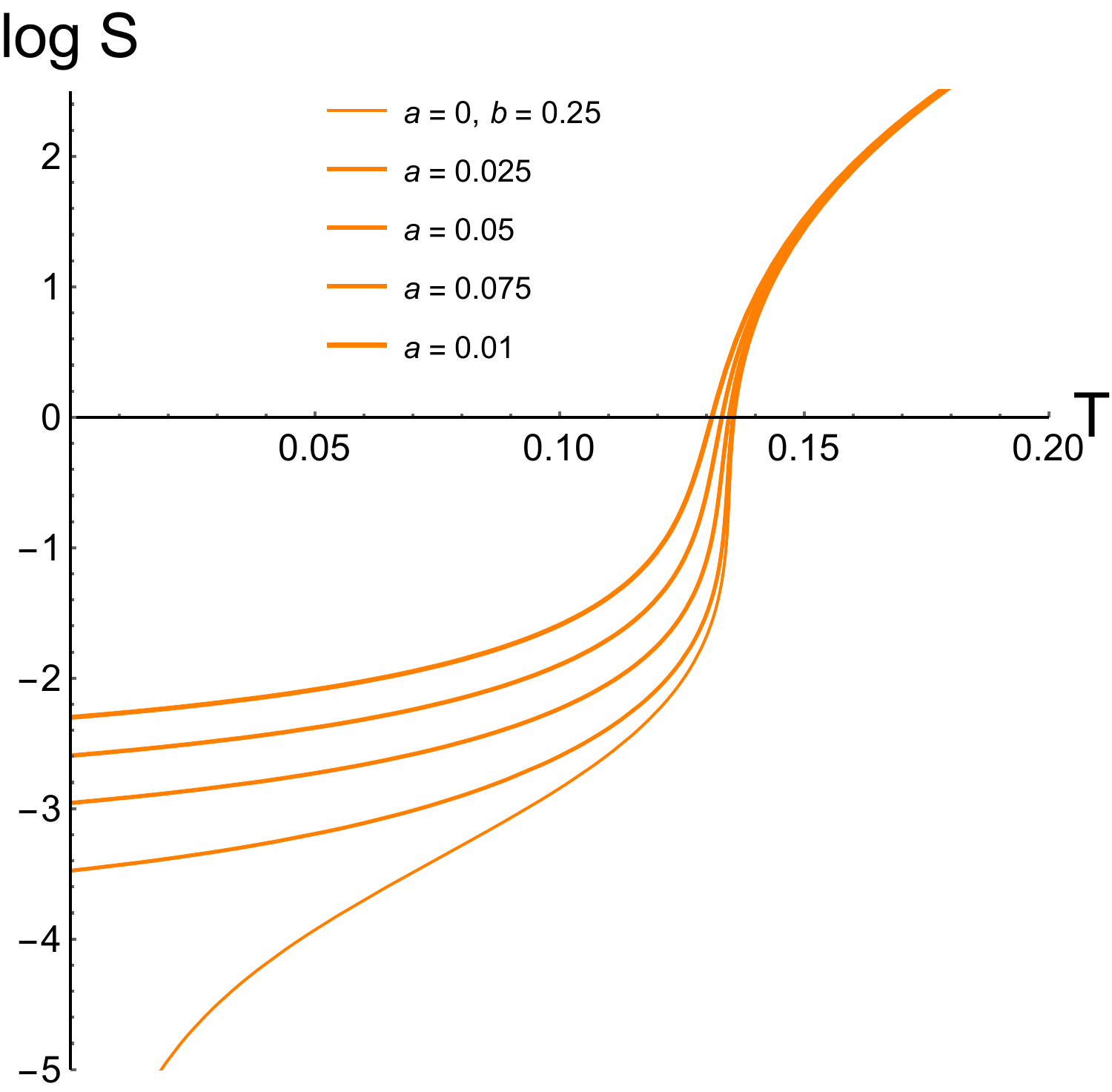}\qquad
  \includegraphics[scale=0.15]{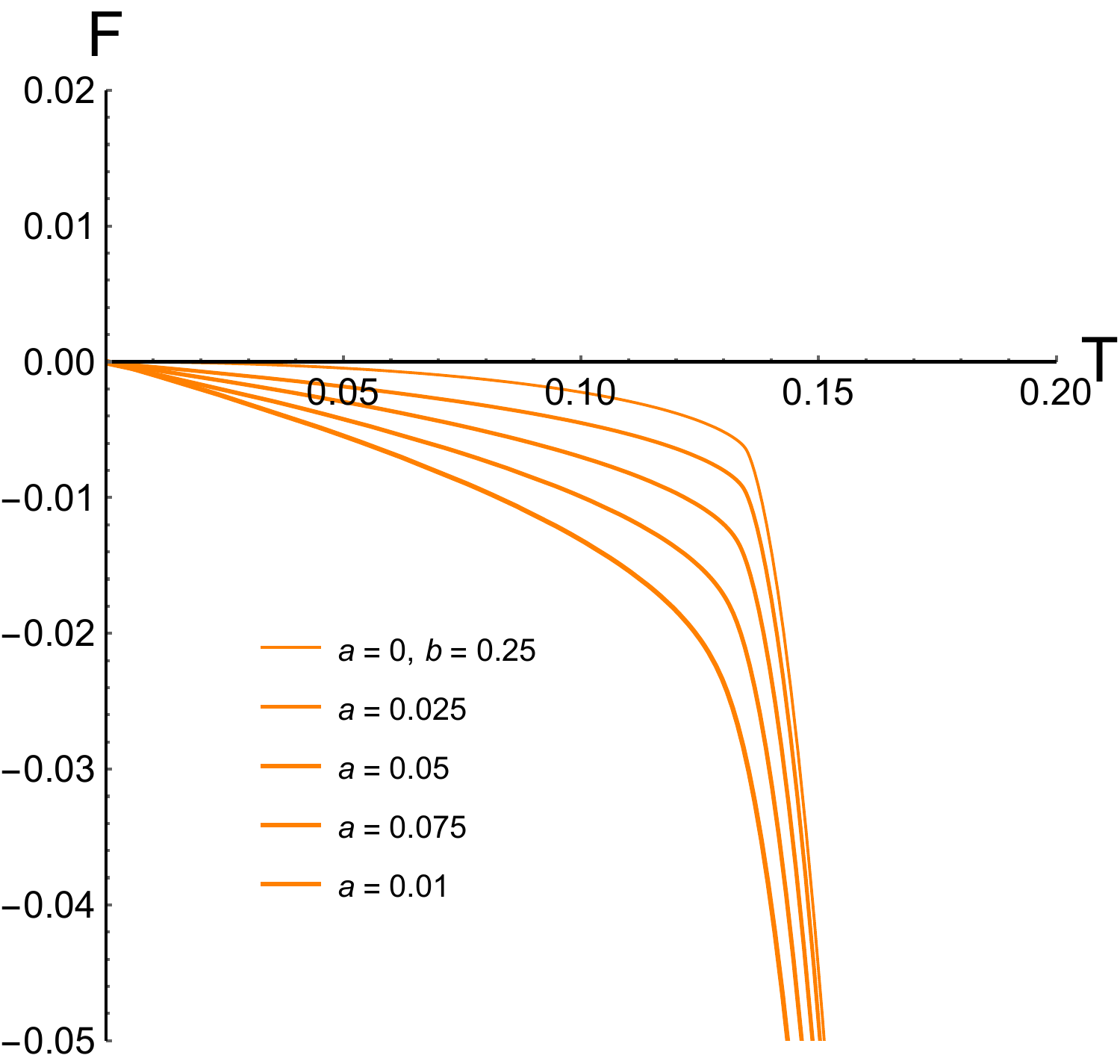}

  \caption{The plots in the first column show the dependence of temperature on $r_h $ for
  various values of
   $ a $ and $ b $.
  The plots in the second column show a logarithmic dependence of entropy on temperature.
  The plots in the third   column show the dependence of  free energy on temperature.}
  \label{Fig:TSF}
\end{figure}
$$\,$$
\newpage
\begin{figure}[h!]
  \centering
  \includegraphics[scale=0.27]{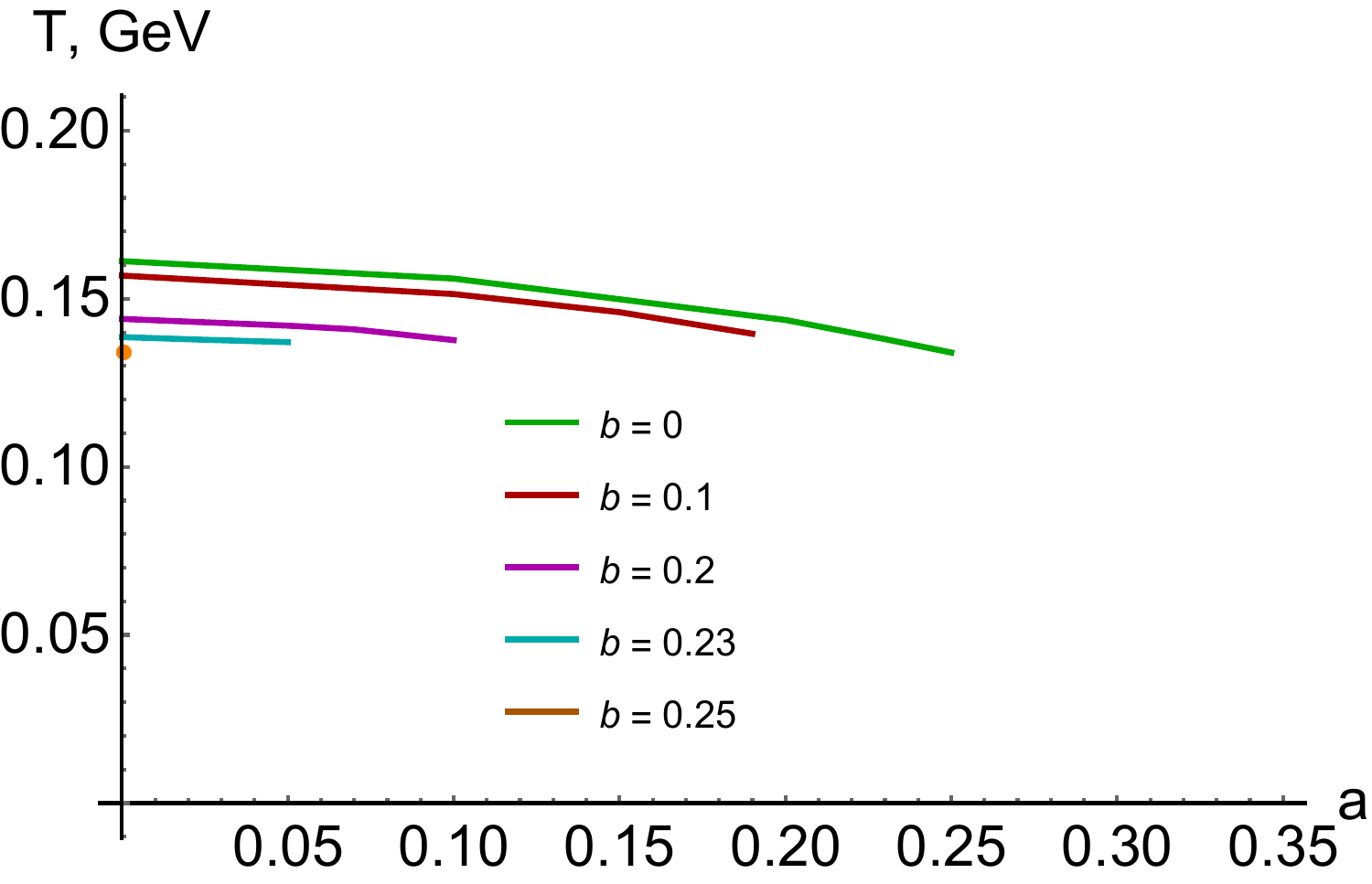} \quad  \includegraphics[scale=0.27]{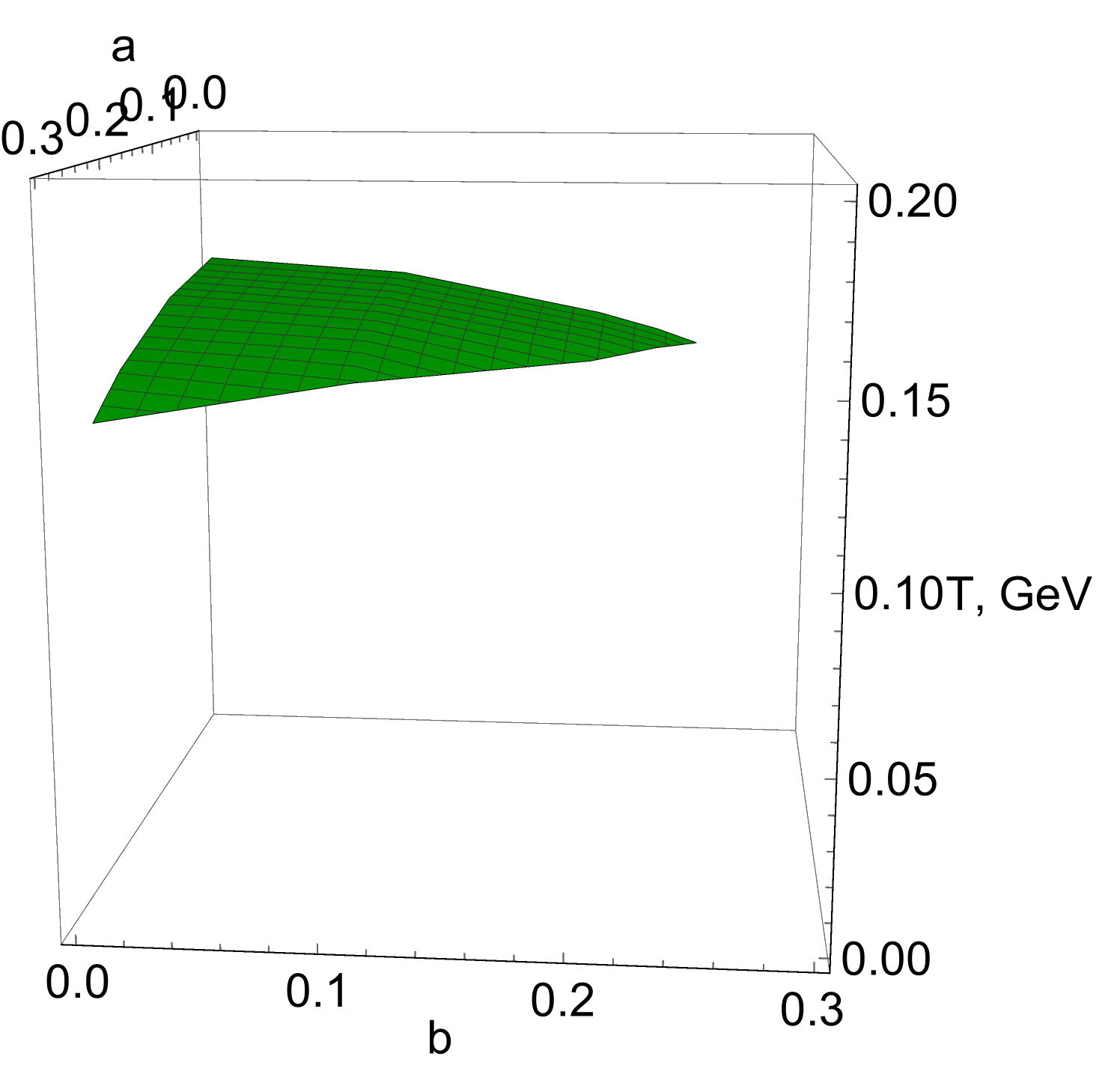} \\ A \hspace{70 mm} B
  \caption{Locations of the first order phase transition
  in terms of $a$  and $b$.
  A) Locations of the end points for fixed $b$ are indicated by dots. We see that all end points have the same temperature $T_{\rm CEP}=0.134$ GeV.  B)
  The location of the phase transition $ T_ {cr} = T_ {cr} (a ^ 2, b ^ 2) $ is shown by the green surface.   }
  \label{Fig:PT2}
\end{figure}
\newpage
\section{Drag force in 5d Kerr-AdS background with one rotational parameter}\label{HDF}
\subsection{Static curved string in the 5d Kerr-AdS background}\label{HDF1r}
Now we turn to discussion of a curved string  in the 5d Kerr-AdS background with one non-zero rotational parameter ($a\neq 0$, $b=0$).
We use the form of the metric in Boyer-Lindquist coordinates (\ref{1.1})-(\ref{1.1a}) \footnote{See the SageMath notebook \url{https://cocalc.com/share/565d762758a554b6adffa16d5562e097e0e565eb/Kerr-AdS-5D-string-b_zero.ipynb}
for details on the calculation.}. The string worldsheet is parametrized as $(\sigma^{0}, \sigma^{1}) = (t,r)$. The embedding is characterized by
\be\label{embed}
\theta = \theta(t,r), \quad \phi = \phi(t,r), \quad \psi = \psi(t,r).
\ee
Then non-zero components of the the induced metric $g_{\alpha\beta}$ (\ref{NGAIM}) look like
\bea\label{gttrrrt}
g_{tt}   &=& G_{tt} + 2G_{t\phi}\dot{\phi} + G_{\phi \phi}\dot{\phi}^{2}  + G_{\theta\theta}\dot{\theta}^{2} + G_{\psi\psi}\dot{\psi}^{2},\\
g_{rr} &= &G_{rr} + G_{\phi\phi}\phi^{'2} + G_{\theta\theta}\theta^{'2} + G_{\psi\psi}\psi^{'2},\\ \label{gttrrrt1}
g_{rt}& = &G_{t\phi}\phi' + G_{\theta\theta}\theta'\dot{\theta} +G_{\phi\phi}\phi'\dot{\phi} + G_{\psi\psi}\dot{\psi}\psi',
\eea
where $G_{\mu\nu}$ are components of the 5d Kerr-AdS metric (\ref{1.1})-(\ref{1.1a}) and we define $\dot{\,} = \frac{d}{dt}A$,  ${\,}^{\prime} =\frac{d}{dr}A$.

The equations of motion that follow from the Nambu-Goto action calculated with (\ref{gttrrrt})-(\ref{gttrrrt1}) seems to be quite difficult to work,
instead of them we will consider the linearized string equations of motion for  small values of the rotating parameter $a$.
For the ansatz of the curved string solution we take an expansion by order in $a$ as follows
 \bea\label{cs1}
\phi(t,r) &=& \Phi_{0} + \beta_{1}a\ell^{2}t  + \beta_{1}a\phi_{1}(r) + \mathcal{O}(a^{2}),\quad \nonumber\\
\theta(r)  &= &\Theta_{0} + a^{2}\theta_{1}(r) + \mathcal{O}(a^{4}),\\
\psi(t,r) &= &\Psi_{0} + \beta_{2 }a\ell^2 t +\beta_{2} a\psi_1(r) + \mathcal{O}(a^{2}), \quad \nonumber
\eea
where $\beta_{1}$ and $\beta_{2}$ are some constants such that $\beta^{2}_{1} + \beta^{2}_{2} = 1$.
We note that with $\beta_{2}=0$ we come to the case with fixed $\psi$, i.e. $\psi(t,r) = \Psi_{0}$.

Correspondingly, the induced metric (\ref{ind-metric}) with plugged (\ref{cs1}) into (\ref{gttrrrt})-(\ref{gttrrrt1}) takes the form
\bea\label{ind-metric3}
-g&=&\left(\left(a\Delta_r-a(r^2+a^2)\Delta_\theta\right)\frac{\sin^2\theta}{\Xi_{a}\rho^2}a\beta_{1}\phi'_{1}+
\right.\nonumber
\\ &&
\left.\ \ +\left(\Delta_\theta(r^2+a^2)^2-a^2\Delta_r\sin^2\theta\right)\frac{\sin^2\theta}{\Xi^2_{a}\rho^2}a^{2}\beta^{2}_{1}\ell^{2}\phi'_{1} +  a^{2}\beta^{2}_{2}\ell^{2}r^2 \cos^{2}\theta\psi'_{1}\right)^2 -\notag\\
&&\left(\frac{\rho^2}{\Delta_r}+\frac{\rho^2}{\Delta_\theta}a^{4}\theta'^2_{1}+\left(\Delta_\theta(r^2+a^2)^2-a^2\Delta_r\sin^2\theta\right)\frac{\sin^2\theta}{\Xi^2_{a}\rho^2} a^{2}\beta_{1}^{2}\phi'^2_{1} + a^{2}\beta^{2}_{2}r^{2}\cos^{2}\theta\psi'^{2}_{1}\right)\times\notag\\
&&\times\left(\left(a^2\Delta_\theta\sin^2\theta-\Delta_r\right)\frac{1}{\rho^2}+\left(a\Delta_r-a(r^2+a^2)\Delta_\theta\right)\frac{2\sin^2\theta}{\Xi_{a}\rho^2}a\beta_{1}\ell^{2}\right.\nonumber\\
&&\left.\qquad +\left(\Delta_\theta(r^2+a^2)^2-a^2\Delta_r\sin^2\theta\right)\frac{\sin^2\theta}{\Xi^2_{a}\rho^2}a^{2}\beta^{2}_{1}\ell^{4} + r^{2}\cos^{2}\theta a^{2}\beta^{2}_{2}\ell^{4}\right).
\eea
From the Nambu-Goto action (\ref{NGA})  with (\ref{ind-metric3}) we can find out the first integrals expanded in series by $a$ and the following relations hold
\begin{equation}\label{phipsiINT0}
\phi_{1}(r) = \mathfrak{p} \int^{r}_{r_{+}}  \frac{d\bar{r}}{\bar{r}^{4}h(\bar{r})}, \quad \psi_{1}(r) = \mathfrak{q} \int^{r}_{r_{+}}  \frac{d\bar{r}}{\bar{r}^{4}h(\bar{r})},
\end{equation}
where $\mathfrak{p}$ and $\mathfrak{q}$ are  some constants and  $h(r)$ is
\begin{equation}\label{hPSI}
h(r) = \ell^{2} + \frac{1}{r^2} - \frac{2M}{r^4},
\end{equation}
which is actually the blackening factor for the 5d AdS-Schwarzschild black hole.
It worth to be noted that the appearance of (\ref{hPSI}) is not surprising since we work with an  expansion with the small rotating parameter $a$ and the Kerr-AdS black hole asymptotes to the AdS-Schwarzschild black hole (\ref{adsSch}).

Substituting (\ref{phipsiINT0}) into the Nambu-Goto action (\ref{NGA}) with (\ref{ind-metric3}) we derive the linearized equation of motion for $\theta_{1}$ in the following form
\begin{equation}\label{psiexpeq}
\Upsilon' + \frac{2(r +2\ell^2 r^3)}{r^{4}h}\Upsilon + \frac{1+ \beta^{2}_{2}\mathfrak{q}^{2} - 4\ell^2\beta_{1} M - \beta^{2}_{1}\mathfrak{p}^{2} + (2\beta_{1}+\beta^{2}_{1}-\beta^{2}_{2})\ell^4r^4}{2r^{8}h^2}\sin(2\Theta_{0}) = 0,
\end{equation}
where we reduce the second order differential equation to a first one by the change of the variable $\Upsilon = \theta'_{1}$.

Taking into account the dimensions of the quantities
$[M] \sim r^2$, $[\ell] \sim \frac{1}{r}$, $[h] \sim \frac{1}{r^2}$, $[\theta_1] \sim \frac{1}{r^2}$ one sees that all terms in the LHS of the expression (\ref{psiexpeq}) have dimension  $\frac{1}{r^4}$.

The solution to eq. (\ref{psiexpeq}) can be represented in the following form
\bea\label{thetaprab0}
&&\theta'_{1}
=  \frac{C_{1}}{r^4h}-  \frac{\sin(2\Theta_{0})}{2 r^{4}h}(\beta^{2}_{1}-\beta^{2}_{2} + 2\beta_{1})\ell^{2}r - \frac{\ell\sin(2\Theta_{0})}{2\sqrt{2}\sqrt{1 + 8M\ell^2}r^{4}h}\tan^{-1}\left(\frac{\sqrt{2}\ell r}{\sqrt{1 -  \sqrt{1 + 8M\ell^2}}}\right)\times\nonumber\\
&&\times \frac{(4\ell^2 M +1-\sqrt{1 + 8M\ell^2}) (\beta^{2}_1 - \beta^{2}_2)- 2 (\beta^{2}_1\mathfrak{p}^{2} -1 - \beta^{2}_2\mathfrak{q}^{2})- 2\beta_{1}( \sqrt{1 + 8M\ell^2}-1)}{\sqrt{1-\sqrt{1 + 8M\ell^2}}}\\
&&+\frac{\ell\sin(2\Theta_{0})}{2\sqrt{2}\sqrt{1 + 8M\ell^2}r^{4}h}\tan^{-1}\left(\frac{\sqrt{2}\ell r}{\sqrt{1 + \sqrt{1 + 8M\ell^2}}}\right)\times\nonumber\\
&&  \frac{(4\ell^2 M +1 +\sqrt{1 + 8\ell^2 M} ) (\beta^{2}_1 - \beta^{2}_2)- 2 (\beta^{2}_1\mathfrak{p}^{2} -1 -  \beta^{2}_2\mathfrak{q}^{2})+2\beta_{1}(1+\sqrt{1 + 8M\ell^2})}{\sqrt{1+ \sqrt{1 + 8M\ell^2}}},\nonumber
\eea
where $C_{1}$ is a constant of integration. Remembering that in eq.(\ref{psiexpeq}) it appears a contribution of the AdS-Schwarzschild blackening function (\ref{hPSI})
one can use the relations for the horizon and the Hawking temperature by (\ref{rhschw})-(\ref{rhschwTR}) and represent the LHS of (\ref{thetaprab0})
 in terms of $r_{H}$
\bea\label{thetapr}
&&\theta'_{1}
=  \frac{\tilde{C}_{1}}{r^4h}-  \frac{\sin(2\Theta_{0})}{2 r^{4}h}(\beta^{2}_{1}-\beta^{2}_{2} + 2\beta_{1})\ell^{2}r \\
&+&\frac{\sin(2\Theta_{0})}{2(2\ell^2r^2_H +1)r^{4}h}\log\left(\frac{ r +r_{H}}{r- r_{H}}\right) \frac{\ell^4 r^4_H(\beta^{2}_1 - \beta^{2}_2)- (\beta^{2}_1\mathfrak{p}^{2} -1 - \beta^{2}_2\mathfrak{q}^{2})- 2\beta_{1} \ell^2 r^2_H}{ r_{H}}\nonumber\\
&+&\frac{\ell\sin(2\Theta_{0})}{2(2\ell^2 r^2_H +1 )r^{4}h}\tan^{-1}\Bigl(\frac{\ell r}{\sqrt{\ell^2 r^2_{H} +1}}\Bigr) \frac{(\ell^2r^2_H +1)^2 (\beta^{2}_1 - \beta^{2}_2)- (\beta^{2}_1\mathfrak{p}^{2} -1 - \beta^{2}_2 \mathfrak{q}^{2})+2\beta_{1} (r^2_{H}\ell^2 +1)}{\sqrt{r^2_{H}\ell^2 +1}}, \nonumber
\eea
where we also take into account that $\frac{1}{i}\tan^{-1}(\frac{x}{i}) = - \tanh^{-1}x$ and $\tanh^{-1}x = \frac{1}{2}\log\left(\frac{1+x}{1-x}\right)$.
Since the solutions to angular variables are known (\ref{phipsiINT0}), (\ref{thetapr}), we are able to write down the conjugate momenta with respect to (\ref{conjmom}) are
\begin{equation}\label{conjmom2}\begin{split}
&\pi^{r}_{\theta} =  h(r)r^{4}\theta'_{1}a^{2} + \mathcal{O}(a^{4}),\\
&\pi^{r}_{\phi} = h(r)r^{4}\phi'_{1}\sin^{2}(\Theta_{0})a + \mathcal{O}(a^{2}),\\
& \pi^{r}_{\psi} = h(r)r^{4}\psi'_{1}\cos^{2}(\Theta_{0})a + \mathcal{O}(a^{2}).
\end{split}
\end{equation}
Expanding (\ref{conjmom2}) with (\ref{phipsiINT0}), (\ref{thetapr}) near the boundary $r\to + \infty$ we get the following relations for the conjugate momenta
\bea\label{pirtheta}
\pi^{r}_{\theta} &=&\Bigl( \frac{2\tilde{C}_{1}}{\sin(2\Theta_{0})}-  (\beta^{2}_{1}-\beta^{2}_{2} + 2\beta_{1})\ell^{2}r -  \frac{(\beta^{2}_1 - \beta^{2}_2+2\beta_{1}) }{r}\\
&&+ \frac{\pi \ell}{2(2\ell^2 r^2_H +1 )}  \frac{(\ell^2r^2_H +1)^2 (\beta^{2}_1 - \beta^{2}_2)+2\beta_{1} (r^2_{H}\ell^2 +1) - (\beta^{2}_1\mathfrak{p}^{2} -1 -  \beta^{2}_2\mathfrak{q}^{2})}{\sqrt{r^2_{H}\ell^2 +1}}\Bigr)\sin(2\Theta_{0})\frac{a^{2}}{2}\nonumber\\
&+&{\cal O}(a^2)\nn\\
\pi^{r}_{\phi} &= &\mathfrak{p}\sin(\Theta_{0})^{2}a + \mathcal{O}(a^2),\\ \label{pirpsi}
\pi^{r}_{\psi} &= &\mathfrak{q}\cos(\Theta_{0})^{2}a + \mathcal{O}(a^2).
\eea
where we use $\log\Bigl(\frac{r+r_{H}}{r -r_{H} }\Bigl) \approx 2\frac{r_{H}}{r}$ and $\tan^{-1}\left(\frac{r\ell}{\sqrt{\ell^{2}r^{2}_{H} + 1}}\right) \approx \frac{\pi}{2} - \frac{\sqrt{\ell^{2}r^{2}_{H} + 1}}{\ell r}$.

The components of  the drag force can be found owing to (\ref{dragfp}) and (\ref{pirtheta})-(\ref{pirpsi}) as
\bea\label{dprtheta}
\frac{dp_{\theta}}{dt} &=&\Bigl( -\frac{2\tilde{C}_{1}}{\sin(2\Theta_{0})}+  (\beta^{2}_{1}-\beta^{2}_{2} + 2\beta_{1})\ell^{2}r + \frac{(\beta^{2}_1 - \beta^{2}_2+2\beta_{1}) }{r}\\
&&- \frac{\pi \ell}{2(2\ell^2 r^2_H +1)}  \frac{(\ell^2r^2_H +1)^2 (\beta^{2}_1 - \beta^{2}_2)+2\beta_{1} (r^2_{H}\ell^2 +1) - (\beta^{2}_1\mathfrak{p}^{2} -1 - \beta^{2}_2 \mathfrak{q}^{2})}{\sqrt{r^2_{H}\ell^2 +1}}\Bigr)\frac{\sin(2\Theta_{0})}{4\pi\alpha'}a^{2},\nonumber\\
\frac{dp_{\phi}}{dt} &= &-\frac{1}{2\pi\alpha'}\mathfrak{p}\sin(\Theta_{0})^{2}a + \mathcal{O}(a^2),\\ \label{dprpsi}
\frac{dp_{\psi}}{dt} &= &-\frac{1}{2\pi\alpha'}\mathfrak{q}\cos(\Theta_{0})^{2}a + \mathcal{O}(a^2).
\eea
We see that the component $\frac{d p_{\theta}}{dt}$ (\ref{dprtheta}) has a linearly divergent term with $r\to \infty$
\be\label{dplead}
\frac{d p_{\theta}}{dt}=\fB_1(\fb)\ell^{2}r\frac{\sin(2\Theta_{0})}{2\pi\alpha'}a^{2} +\ldots,
\ee
where $\fB_1(\fb)=\beta^{2}_{1}-\beta^{2}_{2} + 2\beta_{1}$ and  we use parametrization
$\beta_{1}=\sin \fb$, $\beta_{2}=\cos \fb$, so
$\fB_1(\fb)=2\sin (\fb)-\cos (2 \fb)$.
The term in RHS of \eqref{dplead} can be associated to an infinite mass of the heavy quark \cite{HKKKY}.  One can  renormalize it introducing a cut-off $r_{m}$.  By virtue of (\ref{rhschwTR}) we get from (\ref{dplead})
\be\label{dpdtTH}
\displaystyle{\frac{dp_{\theta}}{dt}}=\fB_1(\fb)\left(\ell^2 m_{\rm rest} + \frac{1}{4\pi\alpha'}\left(\pi T_{H} + \sqrt{\pi^2 T^{2}_{H}- 2\ell^2}\right) \right)\frac{a^{2}}{2}\sin(2\Theta_{0})
+...,
\ee
where we take for the cut for $r_{m} = 2\pi\alpha'm_{\rm rest}$.

It is interesting to note that $\fB_1(\fb)$ can have different signs. Here the sign "$-$" shows that the drag force is  opposite the quark movement.

With respect to values of parameter $(\fb)$  we have the following special cases

\begin{itemize}
\item $\fb=\pi/2$,  $\fB_1(\fb)=3$.\\

 The dependence of $\frac{dp_{\theta}}{dt}$ on the temperature is
\bea
\displaystyle{\frac{dp_{\theta}}{dt}}&=&\left(6\ell^2 m_{rest} + \frac{3}{2\pi\alpha'}\left(\pi T_{H} + \sqrt{\pi^2 T^{2}_{H}- 2\ell^2}\right) \right)\frac{a^{2}}{4}\sin(2\Theta_{0})+ \mathcal{O}(a^2)
\nn\\
\frac{d p_{\phi}}{dt}&=&-\frac{1}{2\pi\alpha'}\mathfrak{p}\sin(\Theta_{0})^{2}a + \mathcal{O}(a^2), \quad \frac{d p_{\psi}}{dt}=0.
\eea
This case corresponds to fixed $\psi$ and the conjugate momentum in the $\psi$-direction is equals to 0 and covers the result of \cite{NAS}, i.e. the leading term of the drag force in the $\theta$-direction is $3\ell^{2}m_{\rm rest}$.

\item $\fb=\pi/4 $, $\fB_1(\fb)=\sqrt 2$,\\

\bea
\frac{d p_{\theta}}{dt} &=&\Bigl(-\frac{2\tilde{C}_{1}}{\sin(2\Theta_{0})} +  \sqrt{2}\ell^{2}r+ \frac{\sqrt{2}}{r}\nonumber\\
&-& \frac{\pi \ell}{2(2\ell^2 r^2_H +1)}  \frac{\sqrt{2} (r^2_{H}\ell^2 +1) + 1- \frac{1}{2}(\mathfrak{p}^{2}
 - \mathfrak{q}^{2})}{\sqrt{r^2_{H}\ell^2 +1}}\Bigr)\frac{a^{2}\sin(2\Theta_{0})}{2}, \\
\frac{d p_{\phi}}{dt}&=&-\frac{1}{2\pi\alpha'}\mathfrak{p}\sin(\Theta_{0})^{2}a + \mathcal{O}(a^2),\\
\frac{d p_{\psi}}{dt}&=&-\frac{1}{2\pi\alpha'}\mathfrak{q}\sin(\Theta_{0})^{2}a + \mathcal{O}(a^2).
\eea
Here the relations seems to have the same form comparing as
the previous case  except the coefficients.
\item $0\leq\fb\leq \tan ^{-1}\left(\frac{\sqrt{3}-1}{\sqrt{2}
   \sqrt[4]{3}}\right) \approx 0.375$,
   $-1\leq \fB_1(\fb)\leq 0$.\\

   In this case, the drag force acts on the quark in the opposite direction of the joint motion of the quark
and rotating medium. Note that this case is not exhibited in \cite{NAS}.
\end{itemize}

Eqs. (\ref{pirtheta})-(\ref{dprpsi}) reflect the drag force for the fixed quark in the rotating QGP, so for the rotating parameter $a=0$ it just vanishes. One can complete this relation supposing that the quark also slowly moves with an angular velocity that has components $(\omega_{\theta}, \omega_{\phi}, \omega_{\psi})$. Then we have
\bea\label{dpthetadtmot}
\displaystyle{\frac{dp_{\theta}}{dt}}&=&-\frac{\omega_{\theta}r_{H}^2}{2\pi\alpha'} + \fB_1(\fb)\left(\ell^2 m_{\rm rest} + \frac{1}{4\pi\alpha'}\left(\pi T_{H} + \sqrt{\pi^2 T^{2}_{H}- 2\ell^2}\right) \right)\frac{a^{2}}{2}\sin(2\Theta_{0})
+ ...,\nonumber\\
\,\\  \label{dphidtmot}
\displaystyle{\frac{dp_{\phi}}{dt}} &=& -\frac{1}{2\pi \alpha'}(\omega_{\phi}r^{2}_{H} - \mathfrak{p}\sin(\Theta_{0})^2 a) +..., \\  \label{dpsidtmot}
\displaystyle{\frac{dp_{\psi}}{dt}} &=& -\frac{1}{2\pi \alpha'}(\omega_{\psi}r^{2}_{H} - \mathfrak{q}\sin(\Theta_{0})^2 a) + ...,
\eea
where $r_{H}$ is given by (\ref{rhschwTR}).

It is also instructive to obtain a relation to friction coefficients.  One can only calculate this from the first terms of (\ref{dpthetadtmot})-(\ref{dpsidtmot}), related to the slow motion of the quark in a  plasma. The second term in (\ref{dpthetadtmot}) can be centripetal force.  So we find  with  (\ref{rhschwTR})
\be
\mu = \left(\frac{\pi T}{\ell}\right)^2\frac{1}{2m\pi \alpha'},
\ee
where $p = \frac{m\omega}{\ell^2}$.

\subsection{Energy of the static string in the 5d Kerr-AdS background}\label{HEL1r}

The total energy of the string with the worldsheet parametrization $(\sigma^{0}, \sigma^{1}) = (t,r)$ is given by (\ref{enstrdef})
\be\label{enstrdef}
E  =  -\frac{1}{2\pi \alpha'}  \int dr \pi^{0}_{\,\,\,t},
\ee
where by virtue of (\ref{conjmom}) the conjugate momentum for a string in  (\ref{1.1}) reads
\be\label{pitt}
\pi^{0}_{\,\,\,t} = \sqrt{-g}(g^{tr}G_{t\phi}\phi' + g^{tr} G_{t\psi}\psi' + g^{tt}G_{tt} +g^{tt} G_{t\phi}\dot{\phi} +g^{tt}G_{t\psi}\dot{\psi}).
\ee
A single quark at rest is described by a static string solution with
\be
\theta(r,t)=\theta_{0},\quad \phi(t,r)= \phi_{0}, \quad \psi(t,r) =\psi_{0},
\ee so eq. (\ref{pitt}) with (\ref{gttrrrt}) is represented by
\be\label{estring}
\pi^{0}_{\,\,\,t}=  g^{tt}G_{tt} = G_{rr}G_{tt}.
\ee
Taking into account that for the 5d Kerr-AdS with one rotational parameter (\ref{1.1}) we have
\be\label{metricCGrrGtt}
G_{rr} = \frac{\rho^2}{\Delta_{r}}, \quad G_{tt} = -  \frac{\Delta_{r}}{\rho^2} + a^{2}\frac{\Delta_{\theta}\sin^2\theta}{\rho^2},
\ee
where $\Delta_{\theta}$ and $\Delta_{r}$ are defined by (\ref{1.1a}).
Plugging (\ref{metricCGrrGtt}) into (\ref{estring}) we get
\be\label{pitt2}
\pi^{0}_{\,\,\,t} = -\left(1 - a^{2}\frac{\Delta_{\theta}}{\Delta_{r}}\sin^2\theta\right).
\ee
We note that comparing to \cite{HKKKY}  in (\ref{pitt2}) we have the second term related to a contribution with a rotation parameter $a^2$, $\Delta_{r}$, which root defines a location of the horizon.
Therefore, there is a divergence related to a pole for $\Delta_{r} =0$. This can be regularized by introducing a cut-off. So to regularize this divergence we  perform an integration from $r_{H} +\epsilon$.
So for eq. (\ref{enstrdef}) with (\ref{pitt2}) we have
\be\label{enstrdep}
E  =\frac{1}{2\pi \alpha'} \int^{r_{m}}_{r_{H} +\epsilon} dr  \left(1 - a^{2}\frac{\Delta_{\theta}}{\Delta_{r}}\sin^2\theta\right),
\ee
where $r_{m}$ is a quark location.  We note that the relations (\ref{pitt2})-(\ref{enstrdep}) are exact and work for arbitrary parameter $a$.

It's valuable to consider an expansion of (\ref{enstrdep}) by order in small $a$. For (\ref{pitt2}) we get
\be\label{pitt3}
1 - a^{2}\frac{\Delta_{\theta}}{\Delta_{r}}\sin^2\theta\, \approx \,1 - a^{2}\frac{1}{r^{2} + \ell^2 r^4 - 2M}\sin^2\theta.
\ee
Plugging (\ref{pitt3}) into (\ref{enstrdep}) and integrating we get
 \be\label{Egeneq}
 E =\frac{1}{2\pi \alpha'} \left(r -\left(\frac{\tan^{-1}\left(\frac{\sqrt{2}\ell r}{\sqrt{1-\sqrt{1+8\ell^2 M}}}\right) }{\sqrt{1- \sqrt{1+ 8\ell^2 M}}}-\frac{\tan^{-1}\left(\frac{\sqrt{2}\ell r}{\sqrt{1+\sqrt{1+8\ell^2 M}}}\right)}{\sqrt{1+ \sqrt{1+ 8\ell^2 M}}}\right)
 \frac{\sqrt{2}a^{2}\ell \sin^2\theta}{\sqrt{1 + 8\ell^2 M}}\right)\Big|^{r_{m}}_{r_{H}+\epsilon}.
 \ee
 The case of the zero temperature limit corresponds to $M=0$, so expanding  (\ref{Egeneq}) near $M=0$ we get
 \bea\label{EgeneqExp}
 E|_{\,\,T=0} &=& \frac{1}{2\pi \alpha'} \left(r -\Big(-\frac{1}{\sqrt{2} \ell r}+\frac{i \pi }{4 \ell
   \sqrt{M}}- \frac{\tan^{-1}(\ell r)}{\sqrt{2}}\Big)\sqrt{2}a^{2}\ell \sin^2\theta \right)
   \Big|^{r_{m}}_{r_{H}+\epsilon}\\ \label{EgeneqExp2}
   &=& \frac{1}{2\pi \alpha'} \left(r_m +\Big(\frac{1}{ \ell r_m}+  \tan^{-1}(\ell r_m)-\frac{1}{\varepsilon} \Big)a^{2}\ell \sin^2\theta \right).
 \eea
where using the relation for the horizon (\ref{rhschw}) we have $r_{H} =0$ with $M= 0$ and define $\varepsilon = \ell \epsilon $. From eq. (\ref{EgeneqExp}) we see that the energy of the quark at rest in the zero temperature limit differs from that one \cite{HKKKY} by the contribution from the rotational parameter and an extra term $\frac{1}{\varepsilon}a^{2}\ell \sin^2 \theta$ related to the renormalization.

At zero temperature the renormalized energy equals to the (Lagrangian) mass $m$ of the quark. For the case of the zero temperature limit from (\ref{EgeneqExp2}) we get
\be
E_{ren}\Big|_{T=0} =\frac{1}{2\pi \alpha'} \left(r_m +\Big(\frac{1}{ \ell r_m}+ \tan^{-1}(\ell r_m)\Big)a^{2}\ell \sin^2\theta \right) =m.
\ee
Increasing the temperature the relation for the energy takes the form
 \be
E = \frac{1}{2\pi \alpha'}\left(m -r_{H}- \left(\frac{\sqrt{2}  \tanh^{-1}\left(\frac{r_{H}+\epsilon}{r_{H}}\right)}{ r_{H}} + \frac{\tan^{-1}\left(\frac{\ell (r_{H}+\epsilon)}{\sqrt{2(\ell^2 r^{2}_{H}+ 1)}}\right)}{\sqrt{2}\sqrt{\ell^2 r^{2}_{H} + 1}}\right)\frac{a^2 \ell \sin^{2}\theta}{(2\ell^2 r^2_{H} + 1)}\right),
\ee
where we also take into account the relations for the horizon (\ref{rhschw}), (\ref{usrel4Mlm2})-(\ref{usrel4Ml3}).
Following \cite{HKKKY} we consider (\ref{enstrdep}) as the static thermal mass $M_{\rm{rest}}$
 \be\label{Mrest}
M_{\rm{rest}} =m -\Delta m(T,a),
\ee
where
\bea
 &&\Delta m(T,a) =  \frac{\sqrt{\lambda}}{2\pi}\Bigl(\frac{\pi T_{H} + \sqrt{\pi^2 T^{2}_{H}- 2\ell^2}}{2\ell^2} +
 \Bigl(-\frac{ \tanh^{-1}\Bigl(1+\frac{2\ell^2\epsilon}{\pi T_{H}+\sqrt{\pi^2 T^2_{H} - 2\ell^2}}\Bigr)}{\pi T_{H} + \sqrt{\pi^2 T^{2}_{H}- 2\ell^2}}\\
 &&+\frac{\sqrt{2} \tan^{-1}\Bigl(\frac{ (\pi T_{H} + \sqrt{\pi^2 T^2_{H} - 2\ell^2}+2\epsilon\ell^2)}{\sqrt{2}\sqrt{\pi^2 T^{2}_{H}+\ell^2+ \pi T_{H}\sqrt{\pi^2T^2_{H} -2\ell^2}}}\Bigr)}{\sqrt{\pi^2 T^2_H + \ell^2 + \pi T_{H}\sqrt{\pi^2 T^2_{H} - 2\ell^2}}}
\Bigr)\frac{\sqrt{2} a^2 \ell^4 \sin^{2}\theta}{\pi^2 T^2_{H} + \pi T_{H}\sqrt{\pi^2 T^2_{H} - 2\ell^2}}\Bigr),\nonumber
\eea
with $\lambda = 1/\alpha'^2$.
 We are also able to write down thermal corrections to a quark mass without expansion for an arbitrary value of $a$.
 Integrating eq.~(\ref{enstrdep}) we obtain
 \bea\label{engenquark}
 E& =&\frac{1}{2\pi \alpha'}\left( r +\left(\frac{\tan^{-1}\Bigl(\frac{\sqrt{2}\ell r}{\sqrt{1 +a^2 \ell^2+\sqrt{(1-a^2\ell^2)^2 + 8\ell^2 M}}}\Bigr)}{\sqrt{1 + a^{2}\ell^2+\sqrt{(1 - a^{2}\ell^2)^2 + 8\ell^2 M}}}\right.\right. \nonumber\\
 &-&\left.\left.\left.\frac{\tan^{-1}\Bigl(\frac{\sqrt{2}\ell r}{\sqrt{1 +a^2 \ell^2-\sqrt{(1-a^2\ell^2)^2 + 8\ell^2 M}}}\Bigr)}{\sqrt{1 + a^{2}\ell^2-\sqrt{(1 - a^{2}\ell^2)^2 + 8\ell^2 M}}}\right)\frac{\sqrt{2}a^2\ell\Delta_{\theta}\sin^{2}\theta}{\sqrt{8\ell^2 M + (1 - a^{2}\ell^2)^2}}\right)\right|^{r_{m}}_{r_{+}+\epsilon}.
 \eea
  Taking into account the relation for the horizon (\ref{rhsinglea}) we get
 the relation for the energy at finite temperature takes the form
\bea
E &=&\frac{1}{2\pi \alpha'}\Bigl( r_{m}+ \Bigl(\tan^{-1}(\ell r_{m})- \tan^{-1}\left(\frac{r_{m}}{a}\right)\frac{1}{ a\ell}\Bigr)\frac{a^2\ell\Delta_{\theta}\sin^{2}\theta}{(1 - a^{2}\ell^2)}\\
&-& r_{+}-\Bigl(\frac{\tan^{-1}\left(\frac{\ell r_{+}}{\sqrt{(r_{+}^2\ell^2 +1)}}\right)}{(1-a^2\ell^2 +2r^2_{+}\ell^2)\sqrt{(r^2_{+}\ell^2 +1)}}
 +\frac{\sqrt{2}\tanh^{-1}\left(\frac{(r_{+}+\epsilon)}{ r_{+}}\right)}{(1-a^2\ell^2 +2r^2_{+}\ell^2)r_{+}}\Bigr)a^2 \ell \Delta_{\theta}\sin^2\theta \Bigr), \nonumber
\eea
or in terms of (\ref{Mrest}) we have
\bea
m &=& \frac{\sqrt{\lambda}}{2\pi}\Bigl( r_{m}+ \Bigl(\tan^{-1}(\ell r_{m})- \tan^{-1}\left(\frac{r_{m}}{a}\right)\frac{1}{ a\ell}\Bigr)\frac{a^2\ell\Delta_{\theta}\sin^{2}\theta}{(1 - a^{2}\ell^2)},\\
\Delta m(T,a)& =& r_{+}+ \Bigl(\frac{\tan^{-1}\left(\frac{\ell r_{+}}{\sqrt{(r_{+}^2\ell^2 +1)}}\right)}{(1-a^2\ell^2 +2r^2_{+}\ell^2)\sqrt{(r^2_{+}\ell^2 +1)}}
 +\frac{\sqrt{2}\tanh^{-1}\left(\frac{(r_{+}+\epsilon)}{ r_{+}}\right)}{(1-a^2\ell^2 +2r^2_{+}\ell^2)r_{+}}\Bigr)a^2 \ell \Delta_{\theta}\sin^2\theta,\nonumber\\
\eea
where the horizon is related to the temperature through (\ref{HawkT}).
It is interesting to note that without performing expansion by $a$ we do not meet with  the divergence \eqref{EgeneqExp2}. This is related with non-analiticity of RHS of \eqref{engenquark} on $a$.
The divergence in $\tanh^{-1}\left(\frac{(r_{+}+\epsilon)}{ r_{+}}\right)$ at $\epsilon\to 0$
is the same in two approaches.

\section{Drag force from the 5d Kerr-AdS metric with two equal rotational parameters}\label{HDF2r}

\subsection{Static curved string in the Boyer-Lindquist coordinates}\label{HDF2rBL}

Here we consider the drag force from the 5d Kerr-AdS metric with two rotational parameters that are equal in magnitude $a=b$ (\ref{1.1SF2})-(\ref{2.1a})  \footnote{See the SageMath notebook \url{https://cocalc.com/share/8a0fbfc77f4422e8e29b66770873d3110ffd95a6/Kerr-AdS-5D-string-a_eq_b.ipynb} for details on the calculation.}. As for the one non-zero rotational parameter we are going to solve the linearized string equation of motion for small values of $a$ and $b$.  For the string worldsheet coordinates we take
$(\sigma^{0}, \sigma^{1})= (t,r)$ and the embedding is characterized by (\ref{embed}) as for the case with one rotational parameter.

The Nambu-Goto action calculated through the determinant of the induced metric of a string worldsheet looks like (\ref{NGA}),
where the components of the induced metric are
\bea
g_{tt} &= & G_{tt} + G_{\theta\theta}\dot\theta^2 + G_{\phi\phi}\dot\phi^2 + G_{\psi\psi}\dot\psi^2 + 2 \left(G_{t\phi}\dot\phi + G_{t\psi}\dot\psi + G_{\psi\phi}\dot\psi\dot\phi\right) ,\\
g_{tr} &= & G_{\theta\theta}\dot\theta\theta' + G_{\phi\phi}\dot\phi\phi' + G_{\psi\psi}\dot\psi\psi' +  G_{t\phi}\phi' + G_{t\psi}\psi' + G_{\psi\phi}\dot\psi\phi' +G_{\psi\phi}\psi'\dot{\phi},\\
  g_{rr} &= & G_{\theta\theta}\theta'^2 + G_{\phi\phi}\phi'^2+ G_{\psi\psi}\psi'^2+G_{rr}+ 2 G_{\psi\phi}\psi'\phi'.
\eea
We expand  the transversal variables $\phi$, $\theta$ and $\psi$ by small $a$ as we do this in the previous case (\ref{cs1})
\bea\label{exp3}
\phi(t,r) &=& \Phi_{0} + a\beta_{1}\ell^{2}t  + a\beta_{1}\phi_{1}(r) + \mathcal{O}(a^{2})\\
\psi(t,r) &= &\Psi_{0} + a\beta_{2}\ell^2 t + a\beta_{2}\psi_1(r) + \mathcal{O}(a^{2}), \\
\theta(r)  &= &\Theta_{0} + a^{2}\theta_{1}(r) + \mathcal{O}(a^{4}),\quad
\eea
where as in the previous section $\beta_{1},\beta_{2}$ are some parameters with $\beta^{2}_{1}+\beta^{2}_{2}=1$.

Then we obtain the determinant of the induced metric as follows
\bea\label{indmaeqb}
- g& =& \Bigl( \frac{a^2 \ell^2 \rho^{2}}{\Xi^{2}}\left(\beta^{2}_{1}\sin^{2}\theta\phi'_{1} + \beta^{2}_{2}\cos^{2}\theta \psi'_{1}\right) + \frac{2Ma^{4}\ell^{2}}{\Xi^{2}\rho^{2}}(
\beta_{1}\sin^{2}\theta + \beta_{2}\cos^{2}\theta)(\beta_{1}\sin^{2}\theta \phi'_{1} + \beta_{2}\cos^{2}\theta \psi'_{1})\nonumber\\
& +& \frac{a^{2}}{\Xi}\left(\rho^{2}\ell^{2} - \frac{2M}{\rho^{2}}\right)(\beta_{1}\sin^{2}\theta\phi'_{1} + \beta_{2}\cos^{2}\theta\psi'_{1})\Bigr)^{2}\\
&- & \Bigl(-1 - \rho^{2}l^{2} + \frac{2M}{\rho^{2}}  + \frac{a^2 \ell^4\rho^{2}}{\Xi} (\beta^{2}_{1}\sin^{2}\theta +\beta^{2}_{2}\cos^{2}\theta)+\frac{2a^{4}\ell^{4}M}{\Xi^{2}\rho^{2}}(\beta_{1}\sin^{2}\theta + \beta_{2}\cos^{2}\theta)^{2}\nonumber\\
&+& 2\frac{a^2 \ell^2}{\Xi}\left(\rho^{2}l^{2} - \frac{2M}{\rho^{2}}
\right)(\beta_{1}\sin^{2}\theta +\beta_{2}\cos^{2}\theta )\Bigr)\nonumber\\
&\times& (\frac{\rho^{2}}{\Delta_r} + \frac{ a^4\rho^{2}}{\Delta_{\theta} }\theta'^2_{1} + \frac{a^{2}\rho^{2}}{\Xi}\left(\beta^{2}_{1}\sin^{2}\theta\phi'^{2}_{1} + \beta^{2}_{2}\cos^{2}\theta \psi'^{2}_{1}\right)
+  \frac{2Ma^{4}}{\Xi^{2}\rho^{2}}(\beta_1\sin^{2}\theta \phi'_{1} + \beta_2 \cos^2\theta \psi'_{1})^{2}),\nonumber
\eea
where $\Delta_{r},\Delta_{\theta},\Xi, \rho$ are given by (\ref{2.1a})-(\ref{2.1b}).

The first integrals for $\phi_{1}$ and $\psi_{1}$ can be found from the Nambu-Goto action  with (\ref{indmaeqb}), in lower order by $a$,  and we can write down
\be\label{phipsiINT}
\phi_{1}(r) =\mathfrak{p}\int^{r}_{r_{+}} \frac{d\bar{r}}{\bar{r}^{4}h(\bar{r})}, \quad \psi_{1}(r) = \mathfrak{q}\int^{r}_{r_{+}} \frac{d\bar{r}}{\bar{r}^{4}h(\bar{r})},
\ee where we also use the conserved quantities  $\mathfrak{p}$, $\mathfrak{q}$ and $h(r)$ is given by (\ref{hPSI}).
The equation of motion for $\theta'$, that follows from the Lagrangian with (\ref{indmaeqb}) after substitution (\ref{phipsiINT}), reads
\bea\label{thetaab}
\Upsilon' + \frac{2(r + 2\ell^{2}r^{3})}{r^{4}h}\Upsilon + \frac{ \ell^{4}r^{4}(\beta^{2}_{1}-\beta^{2}_{2} + 2(\beta_{1} -\beta_{2}))}{2r^{8}h^{2}} \sin(2\Theta_{0})\nonumber\\
+ \frac{-\beta^{2}_{1}\mathfrak{p}^{2} + \beta^{2}_{2}\mathfrak{q}^{2}-4M\ell^{2} (\beta_{1} - \beta_{2})}{2r^8h^2}\sin(2\Theta_{0})=0,
\eea
where we define $\Upsilon = \theta'_{1}$.

Owing to  (\ref{usrel4Mlm2})-(\ref{usrel4Ml3})  we can represent the solution to eq.(\ref{thetaab})$ (\theta'_{1}$) as follows\footnote{The constant $\tilde{C}_{1}$ differs from $C_{1}$ by to  including an imaginary constant.}
\bea
\theta'_{1}&=& \frac{\tilde{C}_{1}}{r^{4}h} - \frac{\sin(2\Theta_{0})}{2 r^{4}h}(\beta^{2}_{1} -\beta^{2}_{2}+ 2(\beta_{1}-\beta_{2}))\ell^{2}r\\
&+&\frac{\sin(2\Theta_{0})}{2(2\ell^2 r^{2}_{H}+ 1)r^{4}h}\log\Bigl(\frac{r+r_{H}}{r-r_{H}}\Bigr)\frac{\ell^4 r^4_H(\beta^{2}_1-\beta^{2}_{2})-2 (\beta_{1}-\beta_{2})\ell^2 r^{2}_{H} - (\beta^{2}_{1}\mathfrak{p}^{2}-\beta^{2}_{2}\mathfrak{q}^{2})}{r_{H}}\nonumber\\
&+&\frac{\ell \sin(2\Theta_{0})}{2(2\ell^2 r^{2}_{H}+ 1)r^{4}h}\times\tan^{-1}\Bigl(\frac{\ell r}{\sqrt{\ell^2 r^2_{H} +1}}\Bigr)\nonumber\\
&\times& \frac{(\ell^2r^2_H +1)^2 (\beta^{2}_1-\beta^{2}_{2})+2(\beta_{1}-\beta_{2})(r^2_{H}\ell^2 +1) - (\beta^{2}_{1}\mathfrak{p}^{2}-\beta^{2}_{2}\mathfrak{q}^{2})}{\sqrt{\ell^2 r^{2}_{H} + 1}},\nonumber
\eea
where $\tilde{C}_{1}$ is a constant.
The corresponding conjugate momenta can be calculated using (\ref{conjmom}). On the boundary $r \to +\infty$ they take the following form
\bea\label{twoabeqpi}
\pi^{r}_{\theta}&=& \Bigl(\frac{2\tilde{C}_{1}}{\sin(2\Theta_{0})} - (\beta^{2}_{1} -\beta^{2}_{2}+ 2(\beta_{1}-\beta_{2}))\ell^{2}r -\frac{\beta^{2}_1-\beta^{2}_{2}+2(\beta_{1}-\beta_{2})}{r} \\
&+&\frac{\pi \ell }{2(2\ell^2 r^{2}_{H}+ 1)}\frac{(\ell^2r^2_H +1)^2 (\beta^{2}_1-\beta^{2}_{2})+2(\beta_{1}-\beta_{2})(r^2_{H}\ell^2 +1) - (\beta^{2}_1\mathfrak{p}^{2}-\beta^{2}_2\mathfrak{q}^{2})}{\sqrt{\ell^2 r^{2}_{H} + 1}}\nonumber \Bigr)\sin(2\Theta_{0})\frac{a^{2}}{2}, \nonumber\\
\pi^{r}_{\phi} &= &\mathfrak{p}\sin(\Theta_{0})^{2}a + \mathcal{O}(a^2),\\
\pi^{r}_{\psi} &= &\mathfrak{q}\cos(\Theta_{0})^{2}a + \mathcal{O}(a^2).
\eea
 Thanks to (\ref{dragfp}) the drag force can be found as
 \bea\label{twoabeqdpdti0}
\frac{dp_{\theta}}{dt}&=& \Bigl(-\frac{2\tilde{C}_{1}}{\sin(2\Theta_{0})} + (\beta^{2}_{1} -\beta^{2}_{2}+ 2(\beta_{1}-\beta_{2}))\ell^{2}r + \frac{\beta^{2}_1-\beta^{2}_{2}+2(\beta_{1}-\beta_{2})}{r} - \frac{\pi \ell }{2(2\ell^2 r^{2}_{H}+ 1)} \nonumber\\
&\times&\frac{(\ell^2r^2_H +1)^2 (\beta^{2}_1-\beta^{2}_{2})+2(\beta_{1}-\beta_{2})(r^2_{H}\ell^2 +1) - (\beta^{2}_1\mathfrak{p}^{2}-\beta^{2}_2\mathfrak{q}^{2})}{\sqrt{\ell^2 r^{2}_{H} + 1}} \Bigr)\frac{\sin(2\Theta_{0})a^{2}}{4\pi \alpha'}, \,\qquad\\
\frac{d p_{\phi}}{dt} &= &-\frac{1}{2\pi \alpha'}\mathfrak{p}\sin(\Theta_{0})^{2}a + \mathcal{O}(a^2),\\
\frac{d p_{\psi}}{dt} &= &-\frac{1}{2\pi\alpha'}\mathfrak{q}\cos(\Theta_{0})^{2}a + \mathcal{O}(a^2).
\eea
As in the case with one non-zero rotational parameter, the component (\ref{twoabeqdpdti0}) has a divergent term with $r\to +\infty$ related to the infinite mass of the heavy quark
\be
\frac{dp_{\theta}}{dt} =\left(\mathfrak{B}_2(\mathfrak{b})\ell^{2}r + \ldots\right)\frac{\sin(2\Theta_{0})a^{2}}{4\pi \alpha'},
\ee
 where as in the previous section we introduce a parameter $\mathfrak{B}_2(\mathfrak{b}) =\beta^{2}_{1} -\beta^{2}_{2}+ 2(\beta_{1}-\beta_{2}) $ and  we use parametrization
$\beta_{1}=\sin \fb$, $\beta_{2}=\cos \fb$, so   $\fB_2(\fb)=-\cos (2\fb) + 2(\sin (\fb)-\cos ( \fb))$.
We see that it vanishes for the special choice of the parameters $\beta_{1}= \beta_{2}$.

For the special cases we have
\begin{itemize}
\item $\fb = \pi/2$,  $\mathfrak{B}_2(\mathfrak{b})=3$
\bea
\frac{d p_{\theta}}{dt}&=& \Bigl(-\frac{2\tilde{C}_{1}}{\sin(2\Theta_{0})}+3\ell^{2}r +\frac{3}{r} -\frac{\pi \ell}{2(2\ell^2 r^2_H +1 )}  \frac{(\ell^2r^2_H +2)^2 -1 - \mathfrak{p}^{2}}{\sqrt{r^2_{H}\ell^2 +1}}\Bigr)\frac{a^{2}\sin(2\Theta_{0})}{2},\nonumber\\
\,\\
\frac{d p_{\phi}}{dt} &= &-\frac{1}{2\pi \alpha'}\mathfrak{p}\sin(\Theta_{0})^{2}a + \mathcal{O}(a^2), \quad \frac{d p_{\psi}}{dt}=0.
\eea
\item $\fb = \pi/4$,  $\mathfrak{B}_2(\mathfrak{b})=0$
\bea \label{dpdtBLab}
\frac{d p_{\theta}}{dt}& =&\Bigl(-\frac{2\tilde{C}_{1}}{\sin(2\Theta_{0})}
- \frac{\pi \ell}{4(2\ell^2 r^2_H +1 )}  \frac{(-\mathfrak{p}^{2} +\mathfrak{q}^{2})}{\sqrt{r^2_{H}\ell^2 +1}}\Bigr)\sin(2\Theta_{0})\frac{a^{2}}{2},\\
\frac{d p_{\phi}}{dt} &= &-\frac{1}{2\pi \alpha'}\mathfrak{p}\sin(\Theta_{0})^{2}a + \mathcal{O}(a^2),\\ \label{dpdtBLab3}
\frac{d p_{\psi}}{dt} &=& -\frac{1}{2\pi\alpha'}\mathfrak{q}\cos(\Theta_{0})^{2}a + \mathcal{O}(a^2).
\eea
\end{itemize}
It should be noted that $\frac{dp_{\theta}}{dt}$ (\ref{dpdtBLab}) disappears for $\mathfrak{p}=\mathfrak{q}$ and $\tilde{C}_{1}=0$.
\subsection{Static curved string in global AdS coordinates}\label{GC}
Now we are going to consider the string in the 5d Kerr-AdS background with $a=b$ written in global AdS coordinates (\ref{globalAdS2}) \footnote{See the SageMath notebook \url{https://cocalc.com/share/0577fb7f5f67a290a1c863b0eb5f069b6df747b2/Kerr-AdS-5D-string-a_eq_b-AdS.ipynb} for details on the calculation.}. For the worldsheet coordinates we set
$\sigma^{0} = T$,  $\sigma^{1} = y$.
We suppose that the embedding is given by
\be\label{confgl2}
\Theta= \Theta(T,y), \quad \Phi=\Psi(T,y), \quad \Psi=\Psi(T,y).
\ee
The string action is given by (\ref{NGA}),
where the components of the induced metric (\ref{NGAIM}) are
\bea\label{gTTgl}
g_{TT} &= & G_{TT} + G_{\Theta\Theta}\dot\Theta^2 + G_{\Phi\Phi}\dot\Phi^2 + G_{\Psi\Psi}\dot\Psi^2 + 2 \left(G_{T\Phi}\dot\Phi + G_{T\Psi}\dot\Psi + G_{\Psi\Phi}\dot\Psi\dot\Phi\right) ,\\
g_{Ty} &= & G_{\Theta\Theta}\dot\Theta\Theta' + G_{\Phi\Phi}\dot\Phi\Phi' + G_{\Psi\Psi}\dot\Psi\Psi' +  G_{t\Phi}\Phi' + G_{t\Psi}\Psi' + G_{\Psi\Phi}\dot\Psi\Phi' +G_{\Psi\Phi}\Psi'\dot{\Phi},\\\label{gyygl}
  g_{yy} &= & G_{\Theta\Theta}\Theta'^2 + G_{\Phi\Phi}\Phi'^2+ G_{\Psi\Psi}\Psi'^2+G_{yy}+ 2 G_{\Psi\Phi}\Psi'\Phi'.
\eea
We take the following  expansion  for the transversal coordinates
\bea\label{expgl}
\Phi(T,y) &=& \Phi_{0} + a\beta_{1}\ell^{2}T  + a\beta_{1}\Phi_{1}(y) + \mathcal{O}(a^{2}),\\
\Theta(y)  &= &\Theta_{0} + a^{2}\Theta_{1}(y) + \mathcal{O}(a^{4}),\\ \label{expgl3}
\Psi(T,y) &= &\Psi_{0} + a\beta_{2}\ell^2 T + a\beta_{2}\Psi_1(y) + \mathcal{O}(a^{2}).
\eea
Plugging (\ref{expgl})-(\ref{expgl3}) into (\ref{gTTgl})-(\ref{gyygl}) we calculate the  determinant of the induced metric
\bea\label{degGL}
|g|&=&\Bigl(\ell^2 y^{2}(\beta^{2}_{1}\sin^{2}\Theta \Phi'_{1}+ \beta^{2}_{2}\cos^{2}\Theta \Psi'_{1})+ \frac{2\ell^2 a^{2}M}{\Xi^{2}y^{2}}(\beta^{2}_{1}\sin^{4}\Theta \Phi'_{1} + \cos^{4}\Theta\beta^{2}_{2}\Psi'_{1})  \nonumber\\  &-& \frac{2M}{y^{2}\Xi^{2}}(\sin^{2}\Theta\beta_{1}\Phi'_{1} +\cos^{2}\Theta\beta_{2}\Psi'_{1})+ \frac{2M a^{2}\sin^{2}\Theta \cos^{2}\Theta}{\Xi^{2}y^{2}}  \beta_{1}\beta_{2}\ell^{2}(\Phi'_{1}
+\Psi'_{1})\Bigr)^2a^4\nonumber\\
&-&(- (1 + y^{2}l^{2} - \frac{2M}{y^{2}\Xi^{2}}) + (\beta^{2}_{1}\sin^{2}\Theta + \beta^{2}_{2}\cos^{2}\Theta)a^{2} y^{2}\ell^{4}\nonumber\\
&+&(\sin^{2}\Theta\beta_{1} +\beta_{2}\cos^{2}\Theta)^{2}\frac{2a^{4}\ell^{4}M}{\Xi^2y^{2}} - 4 \frac{a^2M\ell^2 }{y^{2}\Xi^{2}}(\sin^{2}\Theta \beta_{1}+ \cos^{2}\Theta \beta_{2})\nonumber\\
&\times& (y^{2}a^{4}\Theta'^2_{1} +\frac{y^{4}}{y^{4}(1+y^{2}\ell^{2}) - \frac{2M}{\Xi^{2}}y^{2} + \frac{2Ma^{2}}{\Xi^{3}}}\nonumber\\
&+& (\beta^{2}_{1}\Phi'^2_{1}\sin^{2}\Theta+ \beta^{2}_{2}\Psi'^2_{1}\cos^{2}\Theta)a^{2}y^{2}+\frac{2a^{4}M}{\Xi^{2}y^{2}}(\sin^{2}\Theta\beta_{1}\Phi'_{1} + \cos^{2}\Theta\beta_{2}\Psi'_{1})^{2}),
\eea
with $\Xi$ given by (\ref{2.1b}).

As the in Boyer-Lindquist coordinates the variables $\Phi_{1}$ and $\Psi_{1}$ are cyclic,  so we can find the first integrals for $\Phi_{1}$ and $\Psi_{1}$ for the Nambu-Goto action  (\ref{NGA}) with (\ref{degGL})
\be\label{PhiPsiGL}
\Phi_{1}(y) = \mathfrak{p} \int^{y}_{y_{+}}  \frac{d\bar{y}}{\bar{y}^{2} + \ell^{2}\bar{y}^{4} - 2M}, \quad \Psi_{1}(y) =\mathfrak{q} \int^{y}_{y_{+}}  \frac{d\bar{y}}{\bar{y}^{2} + \ell^{2}\bar{y}^{4} - 2M}.
\ee
Substituting (\ref{PhiPsiGL}) into (\ref{degGL}) we derive the equation of motion for $\Theta_{1}$
\be\label{eqTHETAgAdS}
\Upsilon' + \frac{2(y+ 2\ell^2 y^3)}{-2M + y^2 + \ell^2 y^4}\Upsilon_{1}+ \frac{- \beta^{2}_{1}\mathfrak{p}^{2}+ \beta^{2}_{2}\mathfrak{q}^{2} - 4(\beta_{1} -\beta_{2})\ell^2 M +
\ell^4 y^4(\beta^{2}_{1} - \beta^{2}_{2})}{2(-2M +y^2 + \ell^2 y^4)^2}\sin(2\Theta_{0})=0,
\ee
with $\Upsilon = \Theta'_{1}$.
The solution to (\ref{eqTHETAgAdS}) leads to the following relation
\bea\label{bthetap}
\Theta'_{1}&=& \frac{\tilde{C}_{1}}{y^{4}\ell^{2} + y^{2} -2M} \\
&+&\frac{-2\ell^2 y^{2}_{H}(1+\ell^2 y^{2}_{H})(\beta_{1}- \beta_{2})+ (\beta^{2}_{1}-\beta^2_{2})\ell^4 y^4_H -\beta^{2}_{1}\mathfrak{p}^{2} + \beta^2_{2}\mathfrak{q}^2}{(2\ell^2 y^{2}_{H}+ 1)\ell y_{H}}\nonumber\\
&\times&\frac{\ell}{2(y^{4}\ell^{2} +y^{2} -2M)}\tanh^{-1}\Bigl(\frac{ y}{y_{H}}\Bigr)\sin(2\Theta_{0})\nonumber\\
&+&\frac{-2\ell^2 y^{2}_{H}(1+\ell^2 y^{2}_{H})(\beta_{1}-\beta_{2}) + (\beta^{2}_{1}-\beta^{2}_{2})(\ell^2 y^{2}_{H}+ 1)^2 - \beta^{2}_{1}\mathfrak{p}^{2} + \beta^2_{2}\mathfrak{q}^2}{(2\ell^2 y^{2}_{H}+ 1)\sqrt{\ell^2 y^{2}_{H}+ 1}}\nonumber\\
\nonumber\\
&\times&\frac{\ell}{2( y^{4}\ell^{2} + y^{2} -2M)}\tan^{-1}\Bigl(\frac{\ell y}{\sqrt{\ell^2 y^{2}_{H}+ 1}}\Bigr)\sin(2\Theta_{0})- \frac{(\beta^{2}_{1}-\beta^{2}_{2})\ell^2 y\sin(2\Theta_{0})}{2(-2M+y^{2} + \ell^2 y^4)},\nonumber
\eea
where $\tilde{C}_{1}$ is a constant of integration.
Using (\ref{conjmom}), (\ref{PhiPsiGL}), (\ref{bthetap}) 
and taking into account  (\ref{dragfp})  we can write down the drag force components
 \bea\label{twoabeqdpdti}
\frac{dp_{\theta}}{dt}&=& \Bigl(-\frac{2\tilde{C}_{1}}{\sin(2\Theta_{0})} + (\beta^{2}_{1} -\beta^{2}_{2})\ell^{2}y + \frac{\beta^{2}_1-\beta^{2}_{2}}{y}-\frac{\pi \ell}{2(2\ell^2 y^{2}_{H}+ 1)} \\
&\times&\frac{\left(-2\ell^2 y^{2}_{H}(1+\ell^2 y^{2}_{H})(\beta_{1}-\beta_{2}) + (\beta^{2}_{1}-\beta^{2}_{2})(\ell^2 y^{2}_{H}+ 1)^2 - \beta^{2}_{1}\mathfrak{p}^{2} + \beta^{2}_{2}\mathfrak{q}^2\right)}{\sqrt{\ell^2 y^{2}_{H}+ 1}}\Bigr)\frac{\sin(2\Theta_{0})a^{2}}{4\pi \alpha'}, \nonumber\\
\frac{d p_{\phi}}{dt} &= &-\frac{1}{2\pi \alpha'}\mathfrak{p}\sin(\Theta_{0})^{2}a + \mathcal{O}(a^2),\\
\frac{d p_{\psi}}{dt} &= &-\frac{1}{2\pi\alpha'}\mathfrak{q}\cos(\Theta_{0})^{2}a + \mathcal{O}(a^2).
\eea
Comparing  the result with this  for two rotational parameter $a=b$ performed in Boyer-Lindquist coordinates, we find that the relations for $\frac{dp_{\theta}}{dt}$ have  the common dependence on the radial coordinates ($y$ and $r$, correspondingly) and the horizons ($y_{H}$, $r_{H}$). At the same time the coefficients for these terms are different for the Boyer-Lindquist and AdS coordinates (we have $\beta^{2}_{1} -\beta^{2}_{2} +2 (\beta_{1}-\beta_{2})$ and  $\beta^{2}_{1} -\beta^{2}_{2}$, correspondingly).  This can be related the first coordinates are related to a rotating frame, while the AdS coordinates --  to a non-rotating frame.  However, one can reach an exactly same answer if we choose the parameters for the string dynamics as $\sin \fb =\beta_{1}$ $\cos \fb=\beta_{2}$ and take $\fb = \frac{\pi}{4}$.  Therefore, we have
 \bea\label{twoabeqdpdti2}
\frac{dp_{\theta}}{dt}&=& \Bigl(-\frac{2\tilde{C}_{1}}{\sin(2\Theta_{0})}  -\frac{\pi \ell}{4(2\ell^2 y^{2}_{H}+ 1)}\frac{\left( -\mathfrak{p}^{2} + \mathfrak{q}^2\right)}{\sqrt{\ell^2 y^{2}_{H}+ 1}}\Bigr)\frac{\sin(2\Theta_{0})a^{2}}{4\pi \alpha'}, \\
\frac{d p_{\phi}}{dt} &= &-\frac{1}{2\pi \alpha'}\mathfrak{p}\sin(\Theta_{0})^{2}a + \mathcal{O}(a^2),\\
\frac{d p_{\psi}}{dt} &= &-\frac{1}{2\pi\alpha'}\mathfrak{q}\cos(\Theta_{0})^{2}a + \mathcal{O}(a^2),
\eea
that matches with (\ref{dpdtBLab})-(\ref{dpdtBLab3}). Moreover,
 we can also observe a degenerate case if the parameters $\mathfrak{p}=\mathfrak{q}$ and the constant $\tilde{C}_{1} = 0$. Then the component (\ref{twoabeqdpdti2}) of the drag force  just equals to zero. This is in agreement with the calculations in the Boyer-Lindquist  coordinates and with the hydrodynamical results from Sect.~\ref{hydro} eq.~(\ref{pressure}) where we consider the case of equal angular velocities $\Omega_{a}=\Omega_{b}$.

\section{Conclusion and discussion}\label{Sum}
In this paper we have studied the drag force acting on a heavy quark moving in the rotating quark gluon plasma within the context of holographic duality.
We have considered a 5d Kerr-AdS black hole as a holographic dual of a 4d strongly coupled rotating QGP. We have focused on cases where the black hole solution has either only one non-zero rotational parameter ($a\neq 0$, $b=0$) or two non-zero rotational parameters that are equal ($a=b\neq 0$).
These cases are related to the presence of one and two Casimir invariants for $SO(4)$, correspondingly.

Following the holographic prescription we have associated the heavy quark with an end of a string suspended  on the boundary of the Kerr-AdS black hole into its interior.
We note that in this work the ansatz under consideration corresponds to the fixed quark in a rotating medium. We focused on the case when the rotating parameter is small.
We have solved the string equations of motion order by order in $a$.  Then we have found the corresponding conjugate momenta, which are related to the components of the drag force.

In the case of a single non-zero rotational parameter, we have performed calculations in Boyer-Lindquist coordinates. We have established that the relation for the leading term of the drag force is in agreement with the prediction of the work \cite{NAS}, which considered a lower dimensional holographic duality, namely, a 4d Kerr-AdS black hole for a 3d rotating quark-gluon plasma.
This is actually not surprising, since 4d rotating black holes have just one Casimir invariant corresponding to $SO(3)$.  We have also found corrections to the thermal quark rest mass. It worth to be noted that at zero temperature limit, the cut off term has a contribution from rotation comparing to the result \cite{HKKKY}.

 In the case of the Kerr-AdS black hole with two equal rotational parameters, we have calculated the drag force both in Boyer-Lindquist coordinates and in global AdS ones.
 We have observed that the results for the drag forces have the same dependence on the radial coordinates, however the coefficients may be different. This happens because the Boyer-Lindquist coordinates are related to a rotating frame, while the AdS coordinates --  to a non-rotating frame.
We have seen that there is a degenerate case where the drag force vanishes for certain values of the coefficients.
 This result matches with calculations from the hydrodynamical approach on 4-dimensional sphere.

To summarize our considerations about applied holography for Kerr-${\rm AdS}_{5}$ we can mention that
\begin{itemize}
\item  rotation has some influence  on the first order phase transition:
 \begin{itemize}
\item it decreases  the critical temperature;
\item the critical end point depends on some combination of the two rotation parameters.
\end{itemize}
\item The drag force contains two terms; one of them is related to the slow motion of the quark with respect to the fluid, the second one is interpreted as a centripetal force.
 \begin{itemize}
\item The friction coefficient depends on the temperature quadratically;
\item the centripetal term depends on the temperature linearly;
\item  the centripetal force vanishes in the case of two equal rotational parameters.
\end{itemize}
\item Comparing to \cite{HKKKY}, the energy of the quark at rest in the zero temperature limit has a contribution from the rotational parameter.
\end{itemize}

These phenomena can lead to measurable experimental signals in heavy-ion collisions. Besides the transport properties, rotation can affect the phase structure and phase transition of matter at energies relevant to LHC and RHIC.

A straightforward problem for future study is to investigate the drag force in the 5d Kerr-AdS black hole with two non-equal rotational parameters and to trace the influence of the second rotational parameter on energy loss.
Our considerations can also be generalized to higher dimensional cases \cite{MyPerry} with  compactifications (see also \cite{1707.03483}), as well as more complicate string configurations \cite{Fadafan}. Also it would be interesting to apply the exact results of \cite{Frolov} to the problem treated here.

Another interesting problem for some future work is the study of the drag forces in the rotating charged
Kerr-AdS metric \cite{Cvetic:2004hs}, which  describes a conformal plasma with non-zero chemical potential.  In this case the phase diagram for  $a=b$ in the space $(T,a,q)$
is very similar to the phase diagram presented in Fig.~\ref{Fig:PT} for two rotational parameters.

Note that  the Kerr-AdS black hole provides a good description of QGP at extremely high temperature, where the system tends to the conformal limit. Such a temperature is relevant to LHC and RHIC energies. Hoverer at low temperature, typical for FAIR and NICA, the conformal symmetry is broken  and  non-conformal holographic models are more relevant \cite{IA-NICA}.
The effects of vorticity in peripheral collisions at the NICA facility are widely discussed and it is expected to have hyperon polarization at NICA energies \cite{QGR1a,Csernai:2014ywa,Kharzeev:2015znc,1801.07610,QGPR3,QGPR2}.
Therefore, to describe holographically hyperon polarization at NICA energies, we have to deal with   5d deformed metrics with rotation.
To date, only one parametric deformed Kerr-AdS solution is known (solution with one rotation parameter) \cite{Sheykhi:2010pya,BravoGaete:2017dso,2003.03765,2010.14478}.
 In fact from the side of experimental data, it is not obvious that two parameters should be introduced to describe rotating QGP produced in HIC.  But it is certain that the deformation of Kerr-AdS solutions,
which would be relevant in the context of  NICA,  has to include the electromagnetic field.
From one side, this  is because here we deal with non-zero chemical potential and
 within the holography this is described by temporal component of the Maxwell field.
 From the other side, a huge magnetic field  is also expected \cite{1604.06231} to be present.
Study of perturbations of the electromagnetic field in deformed rotating metrics is also relevant in the context of the direct photons \cite{IA-NICA}.

We would like to mention that the properties of rotating strongly interacting matter have been also studied with lattice QCD \cite{1303.6292}.
As compared with chemical potential, rotation is a simpler task for lattice studies. It is also predicted a dependence of transport phenomena on rotation \cite{0706.1026,0906.5044,1010.0038}  and this is a subject for future investigation, to find  closer relation with our calculation. This is especially interesting, since lattice calculations can be done for two non-zero parameters of rotation and we plan to compare them with our future studies.

\section*{Acknowledgments}
The authors are grateful to Nata Atmaja  for useful clarifications.
The authors also thank for clarifying discussions Hristo Dimov, Oleg Teryaev and Sasha Zhiboedov. AG is grateful to the CERN Theory Division and the Paris Observatory, LUTh for kind hospitality, where a part of this work was done. This work is supported by RFBR Grant 18-02-40069 (I.A. and A.G.) and partially (I.A.) by the “BASIS” Science Foundation (grant No. 18-1-1-80-4). AG is also supported by the JINR grant for young scientists No. 20-302-02 and  the Landau-Heisenberg program. EG acknowledges support from CNRS 80 PRIME program TNENGRAV.

\newpage
\appendix
 \section{Some facts on Kerr-AdS black holes}\label{DetKerrAdS}
  \subsection{Special cases of 5d Kerr-AdS black holes}\label{PCKerrAdS}
Setting $a=b=0$ in (\ref{1.1GF}) for the Kerr-AdS metric, we get the Schwartzchild-AdS black hole:
\be\label{adsSch}
ds^{2} = - \frac{r^2 + r^4\ell^2 -2M}{r^{2}} dt^{2} + \frac{r^{2}}{r^2 + r^4\ell^2 -2M}dr^{2} + r^{2}d\Omega^2_{3},
\ee
where the horizon is defined by
\be\label{rhschw}
r_{H} = \frac{\sqrt{\sqrt{1 +8\ell^2 M }-1}}{\sqrt{2}\ell}.
 \ee
The Hawking temperature is given by
\be\label{rhschwT}
 T_{H} = \frac{4 r^2_{H}\ell^2 + 2  }{4\pi r_{H}} ,
\ee
so the location of the horizon is related to the temperature by
\be\label{rhschwTR}
r_{H} = \frac{1}{2\ell^2}\left(\pi T_{H} + \sqrt{\pi^2 T^{2}_{H}- 2\ell^2}\right).
\ee
From (\ref{rhschw}) one can write down some useful relations
 \bea\label{usrel4Mlm2}
 && 2\ell^2 r^{2}_{H}  = \sqrt{1+8\ell^2 M} -1,\quad 2\ell^2 r^{2}_{H}+ 1 = \sqrt{1+8\ell^2 M},\\  \label{usrel4Ml3}
&& 1+4\ell^2 M - \sqrt{1+8\ell^2 M} = 2\ell^4 r^4_H, \quad 4(\ell^2 r^{2}_{H}+ 1)^2 = (\sqrt{1+8\ell^2 M} + 1)^2.
\eea

In the so-called AdS coordinates (\ref{globalAdS2}), the 5d Kerr-AdS solutions with $M  =0$
takes the following form:
\be\label{Kerr-AdSab}
ds^{2}_{gAdS}= - (1+ y^{2}\ell^{2})dT^{2} + y^{2}(d\Theta^{2} + \sin^{2}\Theta d\Phi^{2} + \cos^{2}\Theta d\Psi^{2})
+\frac{dy^{2}}{(1 + y^{2}\ell^{2})},
\ee
which is a well known form of the global representation of the AdS solution, often called
\textit{static coordinates}.

 \subsection{Boundaries}
 The metric on the boundary for (\ref{1.1GF}) is
\be\label{rotb}
ds^{2}_{BL} = -dt^{2} + \frac{2a\sin^{2}\theta}{\Xi_{a}}dtd\phi + \frac{2b\cos^{2}\theta}{\Xi_{b}}dtd\psi + \frac{\ell^{2}}{\Delta_{\theta}}d\theta^{2} + \frac{\ell^{2}\sin^{2}\theta}{\Xi_{a}}d\phi^{2} +
\frac{\ell^{2}\cos^{2}\theta}{\Xi_{b}}d\psi^{2}.
\ee

From (\ref{Kerr-AdSab})  it is easy to see that  the 4d conformal boundary of 5d Kerr-AdS black hole is  $\mathbb{R}\times \mathbb{S}^{3}$ \cite{HReal, Gibbons:2004ai}, which is reached for $y \to \infty$:
\be\label{nonrotb}
ds^{2} = - \ell^2dT^{2} + d\Theta^{2} +\sin^{2} \Theta d\Phi^{2} + \cos^{2}\Theta d \Psi^{2}.
\ee
The boundary metrics (\ref{rotb}) and (\ref{nonrotb}) are related by
\be
ds^{2}_{BL,bnd} = \frac{y^{2}}{r^{2}} ds^{2}_{gAdS,bnd}.
\ee

 \section{Strings in a 5d Kerr-AdS black hole}
\subsection{Straight string solution in global AdS}\label{ststring}

Let us consider a  string motion in the 5d Kerr-AdS background written in the AdS coordinates. The general form of the metric in these coordinates is complicated.
However, under the assumption $M=0$, which corresponds to the absence of quark-gluon plasma, the Kerr-AdS metric reduces to the global AdS solution (\ref{Kerr-AdSab}),
which can be written as
\be\label{GlAdS2}
ds^{2}= -y^{2} h(y)dT^{2} + y^{2}(d\Theta^{2} + \sin^{2}\Theta d\Phi^{2} + \cos^{2}\Theta d\Psi^{2})
+\frac{dy^{2}}{y^{2}h(y)},\ee
where
\be
h(y)=\ell^{2}+\frac{1}{y^{2}}.
\ee
 We use the physical gauge with
\be(\sigma^{0},\sigma^{1})=(T,y).
\ee
 For the embedding we have $X^{\mu} = X^{\mu}(\sigma)$, so
\be\label{TPP}
  \Theta = \Theta(T,y), \quad \Phi = \Phi(T,y),\quad \Psi = \Psi(T,y).
 \ee
The induced metric is
\bea
g_{\alpha \beta} = \left[\begin{matrix}
  -y^2 h + G_{IJ}\dot{X}^{I}\dot{X}^{J} & G_{IJ}\dot{X}^{' \,I}X^{' \,J}  \\
G_{IJ}\dot{X}^{' \,I}X^{' \,J} &   \frac{1}{y^2 h}+  G_{IJ}X^{' \,I}X^{'\,J}
 \end{matrix}\right],
\eea
where the indices run as $I,J=(\Theta, \Phi,\Psi)$ and $\dot{\,} = \frac{d}{dT}$, ${\,'} = \frac{d}{dy}$.

  The Nambo-Goto action reads
\be\label{NGAss}
S_{NG} = \int dT dy \sqrt{|g|},
\ee
with the determinant of the induced metric:
\bea\label{gindM0}
g& =& -1 + h^{-1}\left(\dot{\Theta}^{2} + \dot{\Phi}^{2} \sin^{2}\Theta  + \cos^{2}\Theta \dot{\Psi}^{2}\right)- y^{4}h\left(\Theta^{'\,2} + \Phi^{'\,2} \sin^{2}\Theta  + \cos^{2}\Theta \Psi^{'\,2}\right)\nonumber\\
&+&y^{4}\Bigl((\dot{\Theta}^{2} + \dot{\Phi}^{2} \sin^{2}\Theta  + \cos^{2}\Theta \dot{\Psi}^{2})(\Theta^{'\,2} + \Phi^{'\,2} \sin^{2}\Theta  + \cos^{2}\Theta \Psi^{'\,2})\nonumber\\
&-&  (\dot{\Theta}\Theta' + \sin^{2}\Theta\dot{\Phi}\Phi' + \cos^2\Theta \dot{\Psi}\Psi')^{2} \Bigr) .
\eea
Assuming that the fluctuations $\dot{X}^{I}$ and $X^{' I}$ are small, we can write
\be\label{detgs}
\sqrt{|g|} \approx \sqrt{1 - h^{-1}\left(\dot{\Theta}^{2} + \dot{\Phi}^{2} \sin^{2}\Theta  + \cos^{2}\Theta \dot{\Psi}^{2}\right)+ y^{4}h\left(\Theta^{'\,2} + \Phi^{'\,2} \sin^{2}\Theta  + \cos^{2}\Theta \Psi^{'\,2}\right)}.
\ee
The equations of motion deduced from (\ref{NGAss}) and (\ref{detgs}) take the following
form:
\bea\label{Thetaeq}
 \frac{\partial}{\partial y}\left(h y^{4} \frac{\Theta'}{\sqrt{-g}}\right) - \frac{1}{h} \left(\frac{\partial}{\partial t} \frac{\dot{\Theta}}{\sqrt{-g}}\right)&=& \frac{\sin(2\Theta)}{2\sqrt{-g}}\left(hy^{4}(\Phi'^{2} -\Psi'^{2})- \frac{\dot{\Phi}^{2} - \dot{\Psi}^{2}}{h}\right) ,\\ \label{Phieq}
\frac{\partial}{\partial y}\left(h y^{4} \frac{\Phi'}{\sqrt{-g}}\right) - \frac{1}{h} \left(\frac{\partial}{\partial t} \frac{\dot{\Phi}}{\sqrt{-g}}\right)&=&0,\\ \label{Psi-eq}
\frac{\partial}{\partial y}\left(h y^{4} \frac{\Psi'}{\sqrt{-g}}\right) - \frac{1}{h} \left(\frac{\partial}{\partial t} \frac{\dot{\Psi}}{\sqrt{-g}}\right)&=&0.
\eea
Eqs. (\ref{Thetaeq})-(\ref{Psi-eq}) possess the static string solution, which reads
\be\label{STSTS}
\Theta = \Theta_{0}, \quad \Phi  = \Phi_{0},\quad \Psi = \Psi_{0},
\ee
where $\Theta_{0}$, $\Phi_{0}$ and $\Psi_{0}$ are some constants corresponding to a massive quark at rest.
We note that plugging the solution for the straight static string (\ref{STSTS}) into (\ref{gindM0}), we see that $-g$ is not positively defined. This fact was mentioned in \cite{HKKKY}.

The corresponding "time-dependent" solution in the Boyer-Lindquist coordinates with (\ref{yCAdS5ab}) is thus
\be\label{ststpsiM0}
\psi = \Psi_{0} \quad \phi =\Phi_{0} - a\ell^{2}t, \quad \theta = \arccos\left(\frac{y}{r}\cos\Theta_{0}\right),
\ee
where
\be\label{ststpsiM02}
y^{2} = \frac{r^{2}(r^{2} + a^{2})}{(1 -a^{2}l^{2}\sin^{2}\Theta_{0})r^{2} +a^{2}\cos^{2}\Theta_{0}}.
\ee

\subsection{Curved string in 5d Kerr-AdS with one rotational parameter(supplementary relations)}
 The determinant of the induced metric built on (\ref{gttrrrt})-(\ref{gttrrrt1}) is
\bea\label{ind-metric}
-g&=&\left(\left(a\Delta_r-a(r^2+a^2)\Delta_\theta\right)\frac{\sin^2\theta}{\Xi_{a}\rho^2}\phi'+\frac{\rho^2}{\Delta_\theta}\dot{\theta}\theta'\right.\nonumber \\
 &&\left.\ \ +\left(\Delta_\theta(r^2+a^2)^2-a^2\Delta_r\sin^2\theta\right)\frac{\sin^2\theta}{\Xi^2_{a}\rho^2}\dot{\phi}\phi' +  r^2 \cos^{2}\theta\dot{\psi}\psi'\right)^2\notag\\
&&-\left(\frac{\rho^2}{\Delta_r}+\frac{\rho^2}{\Delta_\theta}\theta'^2+\left(\Delta_\theta(r^2+a^2)^2-a^2\Delta_r\sin^2\theta\right)\frac{\sin^2\theta}{\Xi^2_{a}\rho^2}\phi'^2 + r^{2}\cos^{2}\theta\psi'^{2}\right)\times\notag\\
&&\times\left(\left(a^2\Delta_\theta\sin^2\theta-\Delta_r\right)\frac{1}{\rho^2}+\left(a\Delta_r-a(r^2+a^2)\Delta_\theta\right)\frac{2\sin^2\theta}{\Xi_{a}\rho^2}\dot{\phi}\right.\nonumber\\
&&\left.\quad +\left(\Delta_\theta(r^2+a^2)^2-a^2\Delta_r\sin^2\theta\right)\frac{\sin^2\theta}{\Xi^2_{a}\rho^2}\dot{\phi}^2+\frac{\rho^2}{\Delta_\theta}\dot{\theta}^2 + r^{2}\cos^{2}\theta\dot{\psi}^{2}\right),
\eea
where $\Delta_{r},\Delta_{\theta}, \Xi_{a},\rho$ are defined by (\ref{1.1a}).

Owing to (\ref{cs1})  the components of the induced metric can be represented in the following way:
\bea
g_{rt}  &=& \frac{a^2 \sin^{2}\theta}{\rho^{2}\Xi^2_{a}}\left(\beta_{1}\Xi_{a}(\Delta_{r} -\Delta_{\theta}(r^{2}+a^{2})) +\beta^{2}_{1} \ell^2(\Delta_{\theta}(r^{2} +a^{2})^{2} -a^{2}\Delta_{r}\sin^{2}\theta)\right)\phi'_{1}\nonumber\\
&+& a^2 \ell^2 \beta^{2}_{2}r^2 \cos^{2}\theta \psi'_{1}, \\
g_{tt}&=& \frac{a^{2}\Delta_{\theta}\sin^{2}\theta - \Delta_{r}}{\rho^{2}} + \frac{2\beta_{1}a^2 \ell^2 \sin^2\theta(\Delta_r - \Delta_{\theta}(r^2 + a^2))}{\rho^{2}\Xi_{a}}\nonumber\\
& +&\beta^{2}_{1}\frac{a^2 \ell^4 \sin^2 \theta}{\rho^2 \Xi^{2}_{a}}(\Delta_{\theta}(r^2 + a^2)^2 -a^2\Delta_{r}\sin^2\theta) +\beta^{2}_{2} a^{2}\ell^{4} r^{2}\cos^{2}\theta,\\
g_{rr}&=& \frac{\rho^{2}}{\Delta_{r}} + \frac{a^{2}\beta^{2}_{1}\sin^{2}\theta}{\Xi^2_{a}\rho^{2}}\left(\Delta_{\theta}(r^{2} +a^{2})^{2} - \Delta_{r}a^{2}\sin^{2}\theta\right)\phi'^{2}_{1} + a^{4}\frac{\rho^{2}}{\Delta_{\theta}}\theta'^{2}_{1} \nonumber\\
&+ &\beta^{2}_{2}a^2 r^2\cos^2 \theta \psi'^{2}_1.
\eea

\end{document}